\documentclass{article}

\usepackage{bookstyle}
\usepackage{graphicx}
\usepackage{cite}

\input{epsf}
\usepackage{epsfig}
\usepackage{amssymb}
\usepackage{amsfonts}
\usepackage{amsbsy}
\usepackage{amsmath}

\def\tD{\tilde{D}}
\def\tK{\tilde{K}}

\def\IC{\mathbb{C}}

\def\IZ{{\mathbb{Z}}}
\def\IR{{\mathbb{R}}}
\def\IP{\mathbb{P}}

\def\ICP{\mathbb{CP}}

\def\CM {{\cal M}}
\def\CK {{\cal K}}
\def\CN {{\cal N}}
\def\CR {{\cal R}}
\def\CD {{\cal D}}

\def\CL {{\cal L}}

\def\CO {{\cal O}}

\def\CG {{\cal G}}

\def\CS {{\cal S}}

\def\CK{{\cal K}}

\def\Re{{\rm Re \,}}
\def\Im{{\rm Im \,}}

\begin{document}

\author{Frederik Denef}
\address{Jefferson Physical Laboratory, Harvard University, \\
Cambridge, MA 02138, USA\\
and \\
Instituut voor Theoretische Fysica, KU Leuven, \\
Celestijnenlaan 200D, B-3001 Leuven, Belgium}
\title{Les Houches Lectures on Constructing String Vacua}


\frontmatter
\maketitle
\mainmatter

{\small \noindent {\bf Abstract:} These lectures give a detailed introduction to constructing and analyzing string vacua suitable for phenomenological model building, with particular emphasis on F-theory flux vacua. Topics include (1) general challenges and overview of some proposed scenarios, (2) an extensive introduction to F-theory and its relation to M-theory and perturbative IIB string theory, (3) F-theory flux vacua and moduli stabilization scenarios, (4) a practical geometrical toolkit for constructing string vacua from scratch, (5) statistics of flux vacua, and (6) explicit models.}

\section{Introduction}

The real world as we know it happens at energies well below the Planck scale, so it is very well described by effective field theory. There is a continuous infinity of consistent effective field theories. Remarkably, only a measure zero fraction of those seems to be obtainable from string theory. These effective field theories arise as low energy descriptions of certain ``vacua'' of string theory, which in some approximations schemes can be thought of as solutions to the equations of motion for the compactification space. Constraints on low energy effective particle spectra and interactions then typically arise from topological constraints on the internal degrees of freedom. Discreteness of the allowed values of parameters in the low energy effective action often arises from quantization effects, such as quantization of internal magnetic fluxes.

It is this remarkable selectivity of string theory which fuels the field known as string phenomenology. If the constraints from requiring the existence of a string theory UV completion are strong enough, then this can take us a significant step beyond the theorizing we can do based on field theory alone. This prompts the question: Are UV completion constraints merely academically fascinating, or are they strong enough to lead to experimental predictions?

In part triggered by developments in constructing string vacua meeting a number of rough observational constraints, a hypothetical picture has emerged in which the real world as we know it is just a tiny patch in a vast, eternally inflating multiverse, which effectively samples an gargantuan ``landscape'' of string vacua \cite{Susskind:2003kw}. If correct, this has profound implications for a number of paradigms in physics, including the notion of naturalness and how we should read some of the stunning fine tunings of parameters in nature, as explained by Nima Arkani-Hamed at this school. The string theory landscape picture is not uncontested \cite{Banks:2003es}, and prompts the question: Is it correct?

We are far from a systematic understanding of the Hilbert space(s) of string theory, or even the space of its approximate, semiclassical vacua with four large dimensions. In view of this, one might consider the above questions premature. Nevertheless, we do understand parts. This includes in particular AdS vacua which have a known dual CFT description, and, to some extent at least, approximate semiclassical vacua which can be constructed as solutions to the classical equations of motion in regimes where quantum corrections are small. The former class is not immediately useful yet as a description of the universe as we know it, as observations indicate our vacuum has a positive effective cosmological constant. We can however try to construct controlled vacua of the latter kind, and address within this class the general questions raised above, even if our vacuum might not be accessible in this way.

The main goal of these lectures is to provide a detailed introduction to the best studied and for phenomenological applications richest set of such approximate semiclassical vacua: type IIB flux compactifications. I will almost exclusively focus on the formal construction of these vacua in string theory and the development of general methods for their analysis, leaving out specific phenomenological applications. I felt this would best complement the existing literature, and be most likely to be useful in the long run. The path from string theory to the real world is long and twisted, and if we want to get beyond loosely string-inspired but further unconstrained effective field theories, there is no choice but to dive deep into the bowels of string theory itself.

I have tried to make the lectures more or less self-contained. In particular all of the geometrical tools needed to build actual models are introduced from scratch, assuming only basic knowledge of the differential geometry contained in section 2 of \cite{Greene:1996cy}. The framework in which I will be working is the F-theory description of IIB theory, because for many applications this is the most powerful and versatile approach, including for constructing semi-realistic string vacua. Nevertheless, as far as I know, no extensive elementary introduction bringing together all the basic ideas needed for applications of constructing F-theory vacua is available in the literature. I have therefore devoted a significant part of these lectures to explaining what F-theory is, how precisely it relates to M-theory and to the weak coupling limit of type IIB string theory, and how flux vacua and their number distributions over parameter space are obtained in this framework.

By the end of these lectures, you should be able to construct and analyze your own string vacuum.

The outline of this extended write-up is as follows:
\begin{enumerate}
\item In section \ref{sec:basics}, I outline some of the main challenges that arise when trying to construct controlled string vacua (of any kind, not just IIB) and give a brief overview of several of the scenarios that have been proposed. In particular I will explain why successful explicit constructions must be ``dirty'', involving many ingredients which might seem unnecessarily contrived at first sight.
\item In section \ref{sec:FtheoryandIIB}, I explain what F-theory is, give its detailed construction in M-theory, discuss ways to think about branes, fluxes and tadpoles in this framework, explain how in general perturbative type IIB orientifold compactifications arise as a particular weak coupling limit, and how in this limit localized D7-branes and their worldvolume gauge fields emerge. I have tried to be as explicit as possible, following an elementary physical approach rather than an algebraic geometrical one.
\item In section \ref{sec:IIB} we turn to the construction of F-theory flux vacua. First the four dimensional low energy effective action is given, both in the the general F-theory setting and in its perturbative IIB weak coupling limit. Next the effect of turning on fluxes is considered, how they induce an effective superpotential stabilizing the shape and 7-brane moduli and how they can produce strong warping effects. Fluxes leave the size moduli massless at tree level. The latter can get lifted by various quantum effects, which are discussed next. Finally, two concrete scenarios to achieve fully moduli stabilized vacua with small positive cosmological constant are reviewed, the KKLT \cite{Kachru:2003aw} and large volume \cite{Balasubramanian:2005zx} scenarios.
\item To find and analyze actual interesting concrete models of either of these two scenarios, a number of geometrical tools is needed. These are introduced from scratch in the hands-on geometrical toolbox which makes up section \ref{geometrictools}.
\item Another indispensable ingredient in constructing and analyzing these vacua are techniques for computing approximate distributions of flux vacua over parameter space. These techniques are introduced and explained in quite a bit of detail in section \ref{sec:statistics}, including a general abstract derivation of the continuum index approximation to counting zeros of ensembles of vector fields, which can then be applied to various flux vacua counting problems, including F-theory flux vacua. A summary is given of various results, and the section concludes with a short discussion of metastability, landscape population and probabilities in the context of flux vacua.
\item Finally, in section \ref{sec:explconstr}, we put all of these basic results together and outline how explicit models of moduli stabilized F-theory flux vacua can be obtained.
\end{enumerate}

The different sections can to a large extent be read independently, and readers only interested in certain aspects of the constructions can probably safely skip sections.

Finally, let me emphasize that these lecture notes are in no way supposed to be a comprehensive review of the subject. The references are not meant to reflect proper historical attribution and are seriously incomplete. They are merely intended to provide pointers to articles which can be used as a starting point for further reading.

\section{Basics of string vacua} \label{sec:basics}

\subsection{Why string vacua are dirty}

Even if you have never gotten your hands dirty constructing semi-realistic string vacua yourself, you probably have heard or read \cite{Susskind:2005bd} that they tend to have a certain Rube Goldberg \cite{wikiRubeGoldberg} flavor to them. In the following I will sketch what the challenges are to construct such vacua in a reasonably controlled way, and how meeting these challenges unavoidably requires adding several layers of complications.

\subsubsection{The Dine-Seiberg problem} \label{sec:DineSeiberg}

Perturbative supersymmetric string theories in flat Minkowski space and weakly curved deformations thereof only exist in ten dimensions. Observations on the other hand suggest only four large, weakly curved dimensions. The most obvious way out of this conundrum is to consider string theory on a space of the form
\begin{equation} \label{compact}
 M_{10} = M_4 \times X
\end{equation}
where $M_4$ corresponds to visible space and $X$ a compact manifold sufficiently small to have escaped detection so far. The size of the space $X$ could be of the order of the fundamental scale, it could be highly curved and even defy classical notions of geometry, and it could break supersymmetry at a very high scale. Currently available techniques to analyze such situations are limited. Therefore the sensible thing to do is to consider well controlled cases and hope that these either will be favored by nature too, or that at least we can draw valuable lessons from them.

The most obvious well controlled cases are provided by compactifications (\ref{compact}) for which $X_6$ is large compared to the fundamental scale, and for which supersymmetry is broken at a scale well below the compactification scale. The latter is most easily achieved by first constructing a compactification which preserves some supersymmetry, and then perturb this in a controlled way to break supersymmetry.

In this regime, we can use the long distance, low energy approximation to string theory, that is, ten dimensional supergravity, described by an effective action for the massless fields. We can also consider eleven dimensional supergravity, the low energy approximation to M-theory \cite{Witten:1995ex}, of which perturbative string theory is believed to be a particular weak coupling limit. Formally we can even go up one more dimension and imagine twelve dimensional F-theory \cite{Vafa:1996xn}, although this can be thought of more conservatively as a convenient geometrized description of type IIB string theory.

The following compactifications of string/M/F theory give rise to $\CN=1$ supersymmetry in four dimensional Minkowski space:
\begin{enumerate}
 \item heterotic or type I string theory on a three complex dimensional Calabi-Yau manifold
 \item type II string theory on a 3d Calabi-Yau orientifold
 \item M-theory on a $G_2$ holonomy manifold
 \item F-theory on a Calabi-Yau fourfold.
\end{enumerate}
All of these compactification manifolds are Ricci flat to leading order at large volume.

However, such compactifications immediately present a problem: at tree level, i.e.\ classically, they always come with moduli. Moduli are deformations of the compactification which do not change the 4d effective energy and therefore correspond to massless scalars in four dimensions. For example the size of the compactification manifold $X$ is always a modulus at tree level, due to the scale invariance $g_{\mu\nu} \to r g_{\mu\nu}$ of vacuum Einstein equations. Other possible moduli are
\begin{enumerate}
 \item The dilaton $e^\phi$, for all string theories. This is already a modulus of the ten dimensional theory in Minkowski space. It is the parameter controlling the worldsheet perturbative expansion of the theory.
 \item Axions: These arise when the supergravity theory under consideration has $p$-form potentials $C_p$ and $X$ has nontrivial harmonic $p$-forms (or equivalently nontrivial $p$-cycles $\Sigma_p$), as adding such a harmonic $p$-form to $C_p$ will not affect the field strength $F_{p+1} = d C_p$ and hence not affect the energy. On the other hand adding a generic harmonic form is not a gauge transformation either, so these are physical, massless modes in the 4d theory.
 \item Metric moduli:
 \begin{enumerate}
 \item CY complex structure (or shape) moduli, analogous to the complex structure modulus $\tau = \omega_2/\omega_1$ of the two-torus $T^2 = \IC / (\IZ \omega_1 \oplus \IZ \omega_2)$.
 \item CY K\"ahler (or size) moduli, analogous to the overall size of the $T^2$.
 \item $G_2$ structure (shape and size) moduli
 \end{enumerate}
 \item Brane deformation and/or vector bundle moduli. Including branes or bundles is often forced upon us, both by tadpole cancelation constraints and by the phenomenological desire to have gauge bosons and charged matter in the theory that could reproduce the Standard Model.
\end{enumerate}

In generic compactifications, there are thousands of these moduli. This is not good. Massless or very light scalars, if they couple at least with gravitational strength to matter, would be observed as long range ``fifth'' forces. No such forces have been observed. Moreover, it is difficult to allow for light scalars while preserving the successful predictions of standard cosmology.

Now, this would not seem to be such a big deal, since including quantum corrections, at least after breaking supersymmetry, are virtually guaranteed to give masses to the moduli, since there is in general nothing that protects scalars from becoming massive after supersymmetry is broken. However, one then runs into a universal problem of theoretical physics, which can be sloganized by saying that when corrections can be computed, they are not important, and when they are important, they cannot be computed. More concretely, the problem here is what is usually referred to as the Dine-Seiberg problem \cite{Dine:1985he}. The argument is very simple. Let $\rho$ be a modulus such as the volume $V_X$ or the inverse string coupling $e^{-\phi}$, with the property that $\rho \to \infty$ corresponds to the weakly coupled region where we trust our tree level low energy effective action. Then if, as expected, quantum corrections generate a potential $V(\rho)$ in the 4d effective theory, this potential will satisfy
\begin{equation}
 \lim_{\rho \to \infty} V(\rho) = 0 \, ,
\end{equation}
precisely because of our assumption that at $\rho \to \infty$, we can trust the tree level low energy effective action, which has zero potential for $\rho$, by definition, since $\rho$ is a modulus at tree level.

\begin{figure}
\centering
\includegraphics[width=\textwidth]{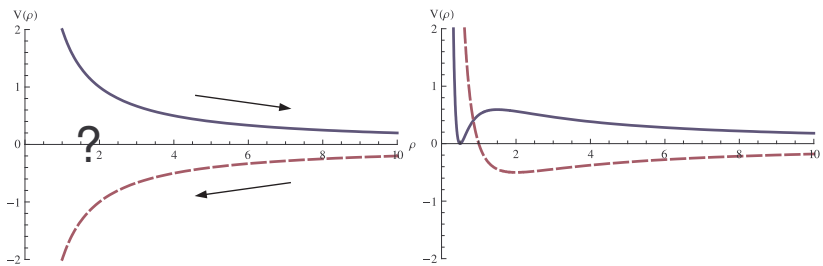}
\caption{On the left: two possible behaviors for the effective potential to first order. On the right: including higher order corrections.}\label{DineSeiberg}
\end{figure}

There are then two possibilities, as shown in figure \ref{DineSeiberg} on the left: either $V>0$ at large $\rho$, in which case the scalar has a runaway direction to $\rho = \infty$, or $V<0$ at large $\rho$, in which case the scalar is pulled to the strong coupling region. A local minimum can only arise if higher order corrections are included; one needs two more corrections for the first case and one more for the second, as illustrated in fig.\ \ref{DineSeiberg} on the right. But, tautologically, when these corrections are important enough to cause a significant departure from the first order shape of $V$, as is necessary to get a local minimum, one is no longer in the weakly coupled region, and in principle all higher order corrections might be important too. In the absence of extended supersymmetry ($\CN \geq 2$ in 4d), we generally lack the tools to compute more than a few orders in perturbation theory, so unavoidably we lose control.

On these grounds, Dine and Seiberg concluded in 1985 that the string vacuum we live in is probably strongly coupled. They may very well be right.

\subsubsection{Flux vacua and no-go theorems} \label{sec:nogo}

But new developments since then have changed the outlook somewhat. A crucial ingredient in these developments was the idea that by turning on $p$-form magnetic fluxes $F$ in the internal manifold $X$, many new string vacua could be designed \cite{Polchinski:1995sm}. Dirac quantization requires these fluxes $F$ to be integrally quantized, that is, integrals $\int_\Sigma F$ over closed $p$-cycles $\Sigma$ must be integral (for a suitable normalization of $F$). So fluxes are discrete degrees of freedom. The type of flux and the possible values of $p$ depend on the theory under consideration, but the basic idea is always the same. The crucial point is that turning on flux generates a moduli potential at \emph{tree level}, of the form\footnote{For simplicity of exposition we suppress for now the additional dilaton dependence which we get in string theory (which is different for the two types of flux one can turn on, RR and NSNS). We work in units in which the 10/11 dimensional Planck scale is set to one.}
\begin{equation} \label{fluxpot}
 V_F = \frac{m_p^4}{V_X^2} \int_X \sqrt{g} \, g^{mr} \cdots g^{ns} F_{m \cdots n} F_{r \cdots s} \, ,
\end{equation}
where $m_p$ is the four dimensional Planck mass and $V_X$ is the volume of $X$. The prefactor appears in the effective potential after rescaling the 4d metric as $g_{\mu\nu} \to \frac{m_p^2}{V_X} g_{\mu\nu}$ to remove the $V_X$-dependence of the Einstein-Hilbert term in the 4d effective action.

Since the internal metric depends on the moduli we originally had, we thus generate a tree level potential for the moduli. This is great, but unfortunately does not solve our problem. Under a rescaling $g_{mn} \to \lambda^2 g_{mn}$ of the internal metric, the potential scales as $V \to \lambda^{-2d+d-2p} V$, so if we parametrize the internal metric by $g_{mn} = r^2 g_{mn}^0$ where we normalize $g_{mn}^0$ such that $\int_X \sqrt{g^0} = 1$, we get
\begin{equation}
  V_F/m_p^4 = r^{-d-2p} \int_X \sqrt{g^0} \, (g^0)^{mr} \cdots (g^0)^{ns} F_{m \cdots n} F_{r \cdots s} \, .
\end{equation}
This is manifestly positive definite, so there is a runaway direction towards large $r$, except if $\int F^2 = 0$. This conclusion remains unchanged if we have different sets of fluxes with different values of $p$, since each term will be positive definite. Now, if $\int F^2=0$ and the geometry is nonsingular, then positive definiteness of the internal metric implies $F=0$, so there was no potential to begin with. Thus we conclude that in the regime in which classical geometry can be trusted, there can be no such flux vacua.

This analysis is a little too naive though. We neglected the possible backreaction of the fluxes on our originally Ricci-flat metric on $X$. If the Einstein-Hilbert action of $X$ becomes nonzero, this will give an additional effective potential term in four dimensions, which will also scale under $g_{mn} \to r^2 g_{mn}$:
\begin{equation} \label{Vextended}
 V/m_p^4 = \sum_p r^{-2p-d} \int_X \sqrt{g^0} F_p^2|_0  \, - \, r^{-2-d} \int_X \sqrt{g^0} R^0 \,.
\end{equation}
If the internal curvature is negative, the runaway gets only worse. So let us take it to be positive, as is the case for example for a sphere. If $F_p$ is nonzero only for $p \geq 2$, the curvature will in fact give the dominant contribution at large $r$. We are then in the case of the dashed line in figure \ref{DineSeiberg}, and we see that flux vacua with negative cosmological constant might be possible. Indeed, such flux vacua are abundant. A subset are the so-called Freund-Rubin vacua \cite{Freund:1980xh}, where $p=d$ and $X$ is taken to be an Einstein space; the simplest example is M-theory on AdS$_4 \times S^7$ with 7-form flux on the $S^7$. We will discuss these in more detail in the next subsection. However, such compactifications cannot be viewed as deformations of the compactifications to flat Minkowski space we started off with: The flux is what supports the internal manifold $X$; if we send $F$ to zero, $X$ collapses to zero size. You can see this from (\ref{Vextended}) from the fact that the minimum $r_* \to 0$ when scaling $F$ to zero keeping $g^0_{mn}$ fixed, or explicitly for example for AdS$_4 \times S^7$, where the size of the $S^7$ (and, pegged to it, the curvature radius of AdS$_4$) is proportional to a positive power of $N:=\int_X F_d$. Moreover, in examples studied so far, the KK scale is of order of the AdS scale. If this is so, then to get to our observed near-zero value of the cosmological constant, some quantum effect has to provide an additional contribution to the effective potential of the order of the KK scale, which tautologically means quantum corrections are not small compared to our leading order potential, and we lose control again. To see this correlation between the scales, note that
\begin{eqnarray}
 M^2_{\rm AdS} &=& \frac{V}{m_p^2} \sim r_*^{-2} \int_X \sqrt{g^0} R^0  \\
 M^2_{\rm KK} &\sim& \frac{1}{D^2} = r_*^{-2} \frac{1}{D_0^2} \, , \\
 \frac{M^2_{\rm AdS}}{M^2_{\rm KK}} &\sim& D_0^2 \int_X \sqrt{g^0} R^0 \, . \label{D0R0}
\end{eqnarray}
Here $r_*$ is the local minimum of $V(r)$, $D$ is the diameter\footnote{The diameter is the largest distance between two points. Its relation to the KK scale, i.e.\ the bottom of the eigenvalue spectrum of the Laplacian, is intuitively plausible but mathematically not trivial. See \cite{Acharya:2006zw} for further discussion.} of $X$ for the metric $g$, and $D_0$ is the diameter for the metric $g^0$. For the round sphere, (\ref{D0R0}) is of order 1, and it appears difficult to find examples where this scale ratio can be made very small. I don't known of a proof that this cannot be done though. If you looked a little harder, you  might well be able to find constructions with a large scale hierarchy. We will briefly return to these compactifications in the next subsection.

So far we discussed the cases with $F_1=F_0=0$. When only $F_1 \neq 0$ (i.e.\ a nonzero scalar gradient), things are qualitatively different. Now the two terms in (\ref{Vextended}) scale in the same way, so if the curvature can adjust itself so the two terms cancel at a given $r$, they will cancel at all $r$, and the potential will be zero. Such solutions do exist in type IIB theory. A simple example is $X = S^2 \times T^4$ with 24 7-branes transversal to the $S^2$, which source RR 1-form flux $F_1$. More generally, F-theory compactifications can be thought of as being of this kind. We will discuss these in detail further on. Still, such compactifications do not solve our problem. By construction, the overall scale modulus $r$ remains massless. On top of that, there will be other nonstabilized geometric moduli --- a typical F-theory compactification for example has thousands of moduli.

When $F_0 \neq 0$, the dominant term in (\ref{Vextended}) at $r \to \infty$ is the corresponding flux term. Together with other flux terms, this could in principle lead to the situation corresponding to the solid line in fig.\ \ref{DineSeiberg}, and could thus possibly even lead to metastable Minkowski or de Sitter vacua. We will return to this case shortly.

The above arguments are a baby version of a very general no-go theorem, proven by Maldacena and Nu\~nez \cite{Maldacena:2000mw} (see also \cite{deWit:1986xg} and recently \cite{Wesley:2008de}). The theorem can be stated as follows. We start with \emph{any} $D$ dimensional gravity theory whose gravitational dynamics is given by the standard Einstein-Hilbert action (without higher curvature corrections), coupled to arbitrary massless fields (scalars, $p$-forms, nonabelian gauge fields, \ldots) with positive kinetic terms, and with zero or negative potential (which could depend on the scalars). We then compactify this theory on a manifold $X$, with coordinates $y^m$, in the most general way, including a possible $y$-dependent ``warp factor'' $\Omega(y)$, to obtain a vacuum solution in $a<D$ dimensions:
\begin{equation}
 ds^2_D = \Omega(y)^2 (ds_a^2(x) + ds^2_X(y)) \, .
\end{equation}
By vacuum solution we mean a metric $ds_a^2(x) = \eta_{\mu\nu}(x) dx^\mu dx^\nu$ which is either Anti-de Sitter, Minkowski, or de Sitter. We assume that $X$ is compact and that the warp factor does not diverge anywhere.\footnote{This is somewhat stronger than the conditions under which the theorem was proven in \cite{Maldacena:2000mw}.}
The theorem now says that under these conditions, \emph{there are no compactifications down to de Sitter space, and none to Minkowski space except if $p=1$ or $p=D-1$, in which case we get Minkowski with $\Omega =$ constant.}

This is consistent with our simple analysis above. (The case $p=D-1$, which is related to $p=1$ by $F_1 = \star F_{D-1}$ was not considered in our analysis because we restricted to magnetic fluxes, which have all legs in the internal space.)

As promised above, we now return to the case $F_0 \neq 0$. This was studied separately in \cite{Maldacena:2000mw} for the particular string theory where it could occur, namely ``massive'' type IIA, with the conclusion that no compactifications to Minkowski or de Sitter can exist, dashing the hope left open by our simple scaling analysis earlier. A few AdS solutions with $F_0 \neq 0$ are known \cite{Tomasiello:2007eq}, but unfortunately they all have KK scales of the order of the AdS scale.

\subsubsection{Orientifold planes and type II flux vacua}

The Maldacena-Nu\~nez no-go theorem sounds like bad news, and in fact it is. Not because the assumptions of the theorem cannot be violated --- string theory violates them immediately, because its low energy effective action does have higher order curvature corrections, and because the theory contains singular negative tension objects (O-planes) --- but because it forces us to depart from the clean world of actions to second order in derivatives and smooth geometries, and to migrate to the dirty world of higher order corrections and orientifold singularities. As a result, control problems creep back in.

Still, let us proceed. We begin by returning to our simple scaling analysis, and add to (\ref{Vextended}) some of the extra contributions D-branes and O-planes provide. Space filling D$(3+k)$-branes wrapping a $k$-cycle in $X$ will give a contribution $\sim r^{-2d+k}$, while orientifold planes on a $k$-cycle give a similar contribution but (possibly) negative. Curvature of D$p$-branes typically gives a negative contribution to the energy density scaling like the energy of a D$(p-4)$-brane. Worldvolume fluxes on D$p$-branes give positive energy contributions scaling like those of lower dimensional branes.

Clearly now there are many more terms in the effective potential, include some more with negative signs, so we can expect to get minima more easily. We consider two special cases of interest. The first one is type IIB string theory on a $d=6$ Calabi-Yau orientifold with O3 and O7 planes, D3 and D7 branes, and NSNS ($H$) and RR ($F$) 3-form fluxes. Schematically this gives the potential
\begin{eqnarray}
 V(r,\phi)/m_p^4 &=& e^{4\phi} \left[ r^{-12} \, (e^{-\phi} \, T_3|_0 + \mbox{$\int$} F^2|_0 + e^{-2\phi}\mbox{$\int$} H^2|_0 ) \right. \nonumber \\
 && \, + \left. \, r^{-8} \, (e^{-\phi} \, T_7|_0 - e^{-2 \phi} \mbox{$\int$} R|_0) \right] \, .
\end{eqnarray}
Here we reinstated the dependence on the dilaton $e^\phi$, and $T_p|_0$ denotes the total tension from D$p$ and O$p$ branes in the metric $g^0_{mn}$ and with $\phi \equiv 0$. The structure of the potential suggests we can find nontrivial Minkowski flux vacua with $R=0$, provided the D7 and O7 tensions cancel, so $T_7 = 0$, and provided the contributions from fluxes, O3 and D3-branes cancel as well. Indeed such vacua turn out to exist \cite{Becker:1996gj,Dasgupta:1999ss,Giddings:2001yu}, and we will discuss them in great detail in section \ref{sec:IIB}.

Of course, by construction now, $r$ is still a modulus. In fact it turns out that all K\"ahler moduli remain unfixed by the flux potential. To stabilize those in this setup, one must resort to quantum corrections again, and the Dine-Seiberg problem kicks back in. Nevertheless an ingenious scenario for how this could be made to work in a reasonably controlled way, and moreover how a small positive cosmological constant could be achieved, was proposed by Kachru, Kallosh, Linde and Trivedi (KKLT) \cite{Kachru:2003aw}. We will return to this in section \ref{sec:IIB}.

The second case we consider is type IIA on a Calabi-Yau orientifold with O6 planes, order 1 units of RR flux $F_0$ and NSNS flux $H$, and $N$ units of RR flux $F_4$. Setting O(1) quantities to 1, this generates a potential of the form
\begin{equation}
  V(r,\phi)/m_p^4 \sim e^{4\phi} \left[ N^2 r^{-14} + r^{-12} e^{-2\phi} - r^{-9} e^{-\phi}  - r^{-8} e^{-2 \phi} \mbox{$\int$} R + r^{-6} \right] \, .
\end{equation}
The first negative term is the O6 contribution. Setting $R = 0$, this has minima for large $N$ at
\begin{equation} \label{rphisol}
 r \sim N^{1/4}, \qquad e^{\phi} \sim N^{-3/4} \, ,
\end{equation}
that is, large volume and weak string coupling. In \cite{DeWolfe:2005uu}, a more refined analysis was done and it was shown that such flux vacua indeed exist in type IIA string theory, at least at the level of the 4d effective theory, with all geometric moduli fixed. In \cite{Acharya:2006ne} this was promoted to full ten dimensional solutions in the approximation of smeared O6 charge. Although you cannot see this from the simple considerations we made, the minima turn out to be always AdS minima, but of a different nature than those of Freund-Rubin type we mentioned earlier. From the scaling (\ref{rphisol}), we see that $m_p^2 = e^{-2 \phi}r^6 \sim N^3$, $V/m_p^4 \sim -N^{-9/2}$, and consequently
\begin{equation}
 M^2_{\rm AdS} = \frac{V}{m_p^2} \sim N^{-3/2}, \qquad M_{\rm KK}^2 = \frac{1}{r(N)^2 D_0(N)^2} \sim \frac{N^{-1/2}}{D_0(N)^2} \, .
\end{equation}
Hence, unlike in the Freund-Rubin case, provided the diameter $D_0(N)$ of the unit volume normalized metric does not grow with $N$ (or grows less fast than $N^{1/2}$), we automatically get a hierarchy of KK and AdS scales in the large $N$ limit. This removes the immediate obstruction to controlled lifting to positive cosmological constant we mentioned for Freund-Rubin type vacua. Which is not to say that lifting is now straightforward or that there are no other control issues with these compactifications. We will come back to this in the next subsection.

In conclusion, we arrived at classical moduli stabilization scenarios in type IIB and IIA string theory which might have a chance of producing some reasonably controlled vacua. But they are not the simple smooth exact classical solutions we might have hoped for. The constructions we have at this point need many different ingredients. Calling them Rube Goldberg contraptions would be excusable. But as I hope I have made clear, it is the failures of simpler ideas\footnote{and perhaps the strategy to start from the highly supersymmetric string vacua we do know and control.} that has driven us this far.

\subsection{A brief overview of some existing scenarios}

We now turn to a brief overview of some of the constructions that have been proposed, and of their virtues and drawbacks. I will not try to be complete, far from it; the idea is to just give a flavor of what has been done and what the issues are. Several concepts mentioned may be foreign to you; some of the material will become more clear further on in the lectures. Much more can be found in the reviews \cite{Grana:2005jc,Douglas:2006es}. The references below are very incomplete and only meant to give you some pointers to the relevant clusters of papers.

\subsubsection{IIB orientifolds / F-theory} \label{sec:IIBFFFF}

These are variants of the KKLT scenario mentioned in the previous subsection. We will deeply get into the details of these scenarios in the next sections. Some key references for the basic setup are \cite{Becker:1996gj,Gukov:1999ya,Dasgupta:1999ss,Giddings:2001yu,Kachru:2003aw,Balasubramanian:2005zx}. The virtues  of this scenario, when it works, are:
\begin{enumerate}
 \item[+] Complex structure moduli, dilaton, D7 moduli can be stabilized classically at high mass scales by RR, NSNS and D7 worldvolume $U(1)$ fluxes.
 \item[+] Because there are always many more fluxes than moduli, there is a very high degree of discrete tunability of physical parameters, which helps in producing controlled models. In particular the cosmological constant can in principle be discretely tuned to become extremely small, easily of the order of the measured cosmological constant or less \cite{Denef:2004ze}.
 \item[+] The classical geometry of the compactification manifold remains Calabi-Yau after turning on fluxes, up to warping \cite{Giddings:2001yu}. This means in particular that many of the powerful techniques from algebraic geometry, which were invaluable to get a handle on Calabi-Yau compactifications without fluxes, can still be used to describe these vacua.
 \item[+] Strongly warped throats of Klebanov-Strassler type \cite{Klebanov:2000hb} can be achieved through the warping of the Calabi-Yau geometry. This can generate large scale hierarchies, useful for e.g.\ controlled supersymmetry breaking by adding anti-D3-branes at the bottom of the throat, or for embedding Randall-Sundrum \cite{Randall:1999ee} type scenarios in string theory.
 \item[+] These vacua can (with fine tuning of initial conditions) accommodate slow roll inflation, at least in local models \cite{Kachru:2003sx,Baumann:2007np,Baumann:2007ah}.
 \item[+] There is a rich set of explicit D-brane constructions possible in these models, useful for particle physics model building; for a review see \cite{Blumenhagen:2006ci}, and for a nice introductory overview see \cite{Marchesano:2007de}. More general F-theory model building is also possible and provides probably the most extensive class of particle physics models in string theory, allowing in particular unification to arise naturally \cite{Donagi:2008ca,beasley}.
 \item[+] The F-theory description provides $g_s$ corrections to the geometry which smooth out the O7 singularities \cite{Sen:1996vd}. This is needed if one wants a large radius geometrical description of the background because O-plane singularities tend to be of a very bad kind, ripping up space at finite distance due to the negative tension of O-planes.
\end{enumerate}
The main drawbacks are
\begin{enumerate}
 \item[--] One needs quantum corrections to stabilize the K\"ahler (size) moduli, making the Dine-Seiberg problem something to worry about.
 \item[--] Generic F-theory compactifications, away from special (orientifold) limits, do not have a globally well defined weakly coupled worldsheet description, even in the infinite volume limit, because the string coupling undergoes $S$-duality transformations when circling around generic $(p,q)$ 7-branes. Hence for generic compactifications, it is unclear how to systematically compute e.g.\ $\alpha'$ corrections even in principle.
 \item[--] Similarly, it is not known how to systematically compute such corrections in the presence of RR flux. This is a universal problem of any flux compactification involving RR fluxes.
\end{enumerate}

\subsubsection{IIA orientifolds}

This is the IIA flux model mentioned at the end of the previous subsection. A key reference for the basic setup is \cite{DeWolfe:2005uu}. The virtues of this scenario are
\begin{enumerate}
 \item[+] Classical RR and NSNS fluxes are sufficient to stabilize all geometrical moduli, in what from the low energy effective action point of view at least appears to be a parametrically controlled regime. Axions are not lifted, but these can get masses by quantum effects without triggering control issues.
 \item[+] Intersecting D-brane models can be embedded (although they are special Lagrangian, and very few explicit constructions of special Lagrangians in compact manifolds are known).
\end{enumerate}
The drawbacks are:
\begin{enumerate}
 \item[--] Because there are about as many moduli as fluxes, there is only limited discrete tunability. Warped throats cannot be generated classically, so controlled supersymmetry breaking by adding anti-D-branes is not possible in this way.
 \item[--] The presence of the localized O6 makes the solutions geometrically incomplete, since the metric and string coupling blow up at finite radius from the O6. In flat space, this can be regularized by lifting to M-theory (the analog of lifting type IIB O7-planes to F-theory), where the O6 turns into the smooth Atiyah-Hitchin manifold \cite{Seiberg:1996nz}. Unfortunately, there is no direct M-theory lift in the case at hand, due to the presence of $F_0$ flux. The control issues this implies are further discussed in \cite{Banks:2006hg}.
 \item[--] There is a no-go theorem \cite{Hertzberg:2007wc} excluding slow roll inflation without adding more ingredients than those considered in \cite{DeWolfe:2005uu}. (Quite a bit more ingredients were considered in \cite{Silverstein:2007ac} however, showing how the no-go theorem could be evaded.)
\end{enumerate}

\subsubsection{M on $G_2$}

$G_2$ flux vacua can be viewed as M-theory uplifts of IIA orientifold flux vacua, but with $F_0 = 0$. The latter restriction must be made because, although type IIA with $F_0 \neq 0$ (also known as massive type IIA or Romans theory) can be viewed as a limit of M-theory via reduction on a 2-torus and a twisted version of T-duality \cite{Hull:1998vy}, it cannot be directly be obtained by compactification of eleven dimensional supergravity.

The good news is that everything is geometrical in this setup; in particular there are no orientifold planes to worry about. The bad news is that, in accordance with our general scaling arguments, the flux potential does not have local minima. (This agrees with the fact that in IIA, when $F_0=0$, there are no vacua.)

Thus, a new ingredient is needed. In \cite{Acharya:2002kv}, a proposal was made for such a new ingredient, making use of constraints from supersymmetry, leading to a negative contribution to the potential from certain nonabelian excitations around a locus of singularities in the $G_2$ manifold. Although unfortunately no explicit example is known, and its physical origin remains to be elucidated, the virtue of such a model would be that it freezes all geometric moduli at once. The drawback, besides the fact that it is not known conclusively if such compactifications actually exist, are similar to those of type IIA models.

\subsubsection{Pure flux}

This includes all flux compactifications which exist without the addition of ``extra'' elements such as orientifold planes. The simplest class of examples are the supersymmetric Freund-Rubin vacua \cite{Freund:1980xh} of M-theory of the form AdS$_4 \times X_7$ where $X_7$ is a Sasaki-Einstein 7-manifold (for reviews see \cite{Gibbons:2002th, Boyer:2004fc, Martelli:2006yb}), supported by $N$ units of flux $F_7$ on $X$. Sasaki-Einstein manifold are obtained as the base of Calabi-Yau fourfold cones. The simplest example is $X=S^7$, obtained by considering eight dimensional flat space as a Calabi-Yau cone over a sphere.

Such vacua are very well controlled in the large $N$ limit, and have 2+1 dimensional superconformal field theory duals obtained from placing $N$ M2 branes at the tip of the cone, so in principle they are even defined nonperturbatively as quantum gravity theory. Their disadvantage as far as realistic model building is concerned is that known examples do not have a large hierarchy between KK and AdS scales, as mentioned earlier. Quite a few examples also have residual moduli, descending from the moduli of the Calabi-Yau cone.

In IIA, Freund-Rubin compactifications with just one flux are not possible because of a dilaton runaway. More involved IIA pure flux compactifications on non-Calabi-Yau manifolds with multiple fluxes do exist however; for a recent examples and a nice overview, see \cite{Tomasiello:2007eq}. The examples include $\ICP^3$ with all kinds of fluxes turned on, carrying a non-K\"ahler, non-nearly-K\"ahler, non-Einstein metric. Again though, the KK and AdS scales are observed to be of the same order.

Finally, the IIA Calabi-Yau flux compactifications of \cite{Polchinski:1995sm,Kachru:2004jr} are also orientifold-free. Their low energy effective action is gauged $\CN=2$ supergravity. In principle, these could exhibit AdS, Minkowski, or dS vacua in four dimensions without violating the Maldacena-Nu\~nez no-go theorem, once quantum corrections to the scalar metric are taken into account. Because of the $\CN=2$ supersymmetry, an infinite series of such corrections is known. I am not aware however of examples of controlled dS or Minkowski vacua with all moduli stabilized in this setup. Also, after supersymmetry breaking, control over quantum corrections will become problematic again.

\subsubsection{Heterotic}

Heterotic string or heterotic M theory \cite{Horava:1996ma} have the important advantage that it naturally gives rise to grand unified models, something which is apparently not natural in weakly coupled type II intersecting D-brane models.\footnote{However, as noted in section \ref{sec:IIBFFFF}, it is natural in more general F-theory model building.} Moduli stabilization has been more challenging in this setting, due to the absence of RR fluxes, limited tunability, and technical difficulties in working with the holomorphic vector bundles which are the core of these compactificactions. Significant progress has been made in recent years however, see e.g.\ \cite{Braun:2006th}.

\subsubsection{Nongeometric}

Not all compactifications of string theory are geometric. Some recent considerations of nongeometric compactifications include \cite{Shelton:2005cf} based on ``over''-T-dualization of toroidal flux compactifications involving $H$-flux, and \cite{Becker:2006ks} based on Landau-Ginzburg models. The latter approach in particular allows to study IIB orientifolds on the mirror of rigid CY manifolds, which do not have any K\"ahler moduli and are therefore necessarily nongeometric. The main advantage is that since there are no K\"ahler moduli, we no longer need to invoke quantum corrections, which was the main issue with geometric IIB compactifications. The disadvantage, perhaps, is that one can no longer directly use geometric notions such as fluxes, warping and so on; instead CFT equivalents have to be found, which is more challenging.

\subsubsection{Noncritical}

Finally, string theories also do not need to be critical; the dimension of the target space can exceed $d=10$. There is a whole landscape of supercritical string theories, connected to the more familiar critical landscape \cite{Hellerman:2007fc}. Noncritical string theories do not have conventional time-independent vacuum solutions, but in a cosmological setting, this is not necessarily a problem. Relatively little has been explored in this arena.

\vskip3mm

\section{F-theory and type IIB orientifold compactifications} \label{sec:FtheoryandIIB}

We now turn to the details of constructing string vacua. We will stay on the more conservative and best understood end of the spectrum of possibilities outlined in the previous section, namely type IIB (F-theory) flux vacua. As outlined in section \ref{sec:IIBFFFF}, this class of models also provides a very rich and interesting phenomenology.

For many purposes, including moduli stabilization, F-theory provides the most elegant and powerful framework to analyze questions in type IIB string theory. I will therefore spend some time first to explain what F-theory is, and how exactly it relates to IIB orientifold compactifications.

\subsection{What is F-theory?} \label{sec:whatisF}

Type IIB supergravity has $\CN=2$ supersymmetry in 10 dimensions (32
supersymmetry generators).
To write down the action, it is convenient (and it makes S-duality manifest) to
define
\begin{eqnarray}
\tau &:=& C_0 + i e^{-\phi}, \\
G_3 &:=& F_3 - \tau H_3, \\
\tilde{F_5} &:=& F_5 - \frac{1}{2} C_2 \wedge H_3 + \frac{1}{2} B_2 \wedge F_3. \\
F_p &:=& dC_{p-1} \quad (p=1,3,5), \qquad H_3:=dB_2 \,.
\end{eqnarray}
We will work with the 10d Einstein frame metric, which has
canonical Einstein-Hilbert term in ten dimensions and is related to the string frame
metric by
\begin{equation}
 g^{E}_{MN} = e^{-\phi/2} g^S_{MN},
\end{equation}
where $\phi$ is the dilaton ($g_{\rm IIB}=e^{\phi}$ is the string coupling constant). Due to the
presence of the self-dual 5-form field strength, there is
no standard manifestly covariant action for this theory,\footnote{There is a non-manifestly covariant action for selfdual $p$-forms \cite{Henneaux:1988gg}. The quantum field theory framework for such fields was developed in \cite{Belov:2006jd}.} but the following gives formally the correct equations of motion:
\begin{eqnarray}
S_{\rm IIB} &=& \frac{2 \pi}{\ell_s^8} \biggl[ \int d^{10} x \sqrt{-g} \, R \nonumber  \\
&& - \frac{1}{2} \int
\frac{1}{(\Im \tau)^2} \, d \tau \wedge * d \bar{\tau}
\,+\, \frac{1}{\Im \tau} G_3 \wedge * \overline{G_3}
\,+\, \frac{1}{2} \tilde{F_5} \wedge * \tilde{F_5} \nonumber \\
&& \qquad \,\,\,\,\,\, + \, C_4 \wedge H_3 \wedge F_3 \biggr].
\label{IIBaction}
\end{eqnarray}
This has to be supplemented with the selfduality constraint $*\tilde{F_5} = \tilde{F_5}$, {\em after}
varying the action, to get the complete equations of motion. The string length $\ell_s$ is related to $\alpha'$ by $\ell_s = 2\pi \sqrt{\alpha'}$. In this notation D$p$-brane tensions are $T_{Dp} = \frac{2\pi}{\ell_s^{p+1}}$.

The action (\ref{IIBaction}) is manifestly invariant under
$SL(2,\IZ)$ S-duality:
\begin{eqnarray}
\tau &\to& \frac{a \tau + b}{c \tau + d}, \label{tauSL2Z} \\
{\left( \!\! \begin{array}{c} H \\ F \end{array} \!\! \right) } &\to&
{\left( \!\!
\begin{array}{cc}
d & c \\ b & a
\end{array} \!\!
\right) \left(\!\! \begin{array}{c} H \\ F  \end{array} \!\!\right)}, \label{FHSL2Z}
\\
\tilde{F_5} &\to& \tilde{F_5} \\
g_{MN} &\to& g_{MN}.
\end{eqnarray}
The action (\ref{IIBaction}) looks uncannily like something obtained by compactification of a \emph{twelve} dimensional theory on a torus with modulus $\tau$, with $F_3$ and $H_3$ the components of some twelve dimensional $\widehat{F}_4$ reduced along the two 1-cycles of the torus. Moreover, the $SL(2,\IZ)$ gauge symmetry then simply becomes the geometrical $SL(2,\IZ)$ reparametrization gauge symmetry of the torus. This and other uncanniness has led to the proposal of F-theory \cite{Vafa:1996xn}, a putative twelve dimensional Father of all theories. However, this does not work as straightforwardly as one might wish. To begin with, there is no twelve dimensional supergravity with metric signature $(1,11)$. Also, if there were actually a twelve dimensional theory with some $\widehat{F}_4$, then we would have to explain why reducting $\widehat{F}_4$ along the full $T^2$ and a point do not show up as 2- resp.\ 4-form field strengths in type IIB. Even more directly, why would the complex structure modulus $\tau$ of the torus appear in (\ref{IIBaction}), but not the overall size modulus?

Proposals have been made to circumvent these problems, but there is actually an alternative geometrical interpretation in M-theory, which works perfectly in the most conservative way. The rough idea is as follows. We start with M-theory
on a small $T^2$ with modulus $\tau$. Taking one of the small $T^2$ circles to be the
M-theory circle gives weakly coupled IIA on the other small circle.
T-dualizing along this circle gives IIB on a large circle. In the
limit of vanishing M-theory $T^2$, this becomes uncompactified IIB. This can be extended to $T^2$ fibrations by performing this procedure fiberwise, resulting in type IIB compactifications with varying dilaton-axion given by the geometric $\tau$-modulus of the $T^2$, effectively realizing the F-theory idea through this chain of dualities.

\begin{figure}
\centering
\includegraphics[height=0.25 \textheight]{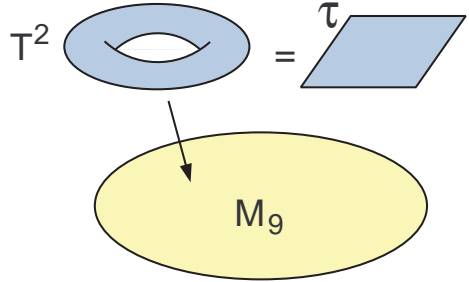}
\caption{F-theory from M-theory. Starting point: $T^2$ fibration over $M_9$.}\label{Ftheory}
\end{figure}

One might worry though that this fiberwise duality procedure might not give rise to a four dimensional Lorentz invariant solution, given the very different origin of one of the spatial directions in the IIB theory. But in fact, somewhat miraculously, it turns out that the result is fully Lorentz invariant in the limit.

Let us make this more precise. We start with M-theory, whose eleven dimensional low energy effective action is
\begin{eqnarray}
S_{\rm M} &=& \frac{2 \pi}{\ell_M^9} \biggl[ \int d^{11} x \sqrt{-g} \, R  - \frac{1}{2} \int G_4 \wedge * G_4 - \frac{1}{6} C_3 \wedge G_4 \wedge G_4 \nonumber \\
&& \quad \quad + \, \ell_M^6 \int C_3 \wedge I_8(R) \, + \cdots \biggr] \, ,
\label{Maction}
\end{eqnarray}
where $G_4 := dC_3$, $I_8(R)$ is a polynomial of degree 4 in the curvature \cite{Duff:1995wd}, and $\ell_M$ is the eleven dimensional Planck length. In this notation the tension of an M$p$ brane is $T_{Mp} = \frac{2\pi}{\ell_M^{p+1}}$, in analogy to the string case. Although the $I_8$ correction is higher derivative, we include it here because it plays a crucial role in anomaly/tadpole cancelation. Moreover further terms related to it by supersymmetry allow to evade the Maldacena-Nu\~nez no-go argument against flux compactifications to Minkowski space, providing negative energy balancing the decompactification pressure of flux.

Now, as illustrated in fig.\ \ref{Ftheory}, we compactify this theory on $T^2 \times M_9$, or more generally a $T^2$ fibration over $M_9$, with metric
\begin{equation} \label{fibrationmetric}
 ds_{M}^2 = \frac{v}{\tau_2} \biggl( (dx + \tau_1 dy)^2 + \tau_2^2 dy^2 \biggr)  + ds_9^2 \, ,
\end{equation}
where $x$ and $y$ are periodic coordinates with periodicity 1. This corresponds to a $T^2$ with complex structure modulus $\tau = \tau_1 + i \tau_2$ and total area $v$. We can allow $v$ and $\tau$ to depend on the coordinates of $M_9$; in this case we get a $T^2$ fibration rather than a direct product. We call the 1-cycle along the $x$-direction the $A$-cycle, and the one along the $y$-direction the $B$-cycle. We will reduce from M to IIA along the A-cycle, and then T-dualize to IIB along the B-cycle.

The relation between the circle compactified M-theory and type IIA metrics is in general given by
\begin{equation} \label{MtoIIA}
 ds^2_M = L^2 e^{4 \chi/3} (dx + C_1)^2 + e^{-2 \chi/3} ds_{\rm IIA}^2
\end{equation}
where $x$ is a coordinate on the M-theory circle with periodicity 1 and $L$ is a conventional length which sets the scale of the M-theory circle and which we can choose at our convenience (since rescaling $L$ can be absorbed in shifting $\chi$ by a constant). The circle bundle connection $C_1$ is the type IIA RR 1-form potential. This immediately gives
\begin{equation}
 C_1 = \tau_1 dy \, , \qquad e^{4\chi/3}=\frac{v}{L^2 \tau_2} \, , \qquad ds^2_{\rm IIA} = \frac{\sqrt{v}}{L \sqrt{\tau_2}}(v \tau_2 dy^2 + ds_9^2) \, .
\end{equation}
Now we want to T-dualize this geometry along the $y$-circle. T-duality maps IIA to IIB, the circle length $L_A$ to $L_B=\ell_s/L_A$, the RR axion becomes $C_0 = (C_1)_y$ and the string coupling $g_{\rm IIB}=\frac{\ell_s}{L_A} g_{\rm IIA}$. To compute $L_B$ and $g_{\rm IIB}$, we thus need to know $\ell_s$ and $g_{\rm IIA}$. Reducing the M2 probe action to F1 resp.\ D2 probe actions on the metric (\ref{MtoIIA}), we get the relations
\begin{equation} \label{MIIArelations}
 \frac{1}{\ell_s^2} = \frac{L}{\ell_M^3}, \qquad \frac{1}{g_{\rm IIA} \ell_s^3} = \frac{1}{e^{\chi} \, \ell_M^3} \, .
\end{equation}
This and the above allows us to express $\ell_s$ and $g_{\rm IIA}$ as a function of $v$, $\tau_2$, $L$ and $\ell_M$, and hence to compute the IIB metric and coupling in terms of these quantities. The final result is
\begin{equation}
 C_0 + \frac{i}{g_{\rm IIB}} = \tau \, , \qquad ds^2_{\rm IIB,S} = \frac{\sqrt{v\, g_{IIB}}}{L} \biggl(
 \frac{\ell_M^6}{v^2} \, dy^2 + ds_9^2 \biggr) \, .
\end{equation}
This is the metric in string frame. In Einstein frame, and trading $\ell_M$ for $\ell_s$ using (\ref{MIIArelations}), this becomes
\begin{equation} \label{genIIBmetricE}
 \qquad ds^2_{\rm IIB,E} = \frac{\sqrt{v}}{L} \biggl(
 \frac{L^2 \ell_s^4}{v^2} \, dy^2 + ds_9^2 \biggr) \, .
\end{equation}
Let us specialize now to the case $M_9 = \IR^{1,2} \times B_6$, with $B_6$ some K\"ahler manifold such as the projective space $\ICP^3$. Assume moreover that the $T^2$ depends holomorphically on the coordinates of $B_6$, that is, that we have an \emph{elliptic fibration} (with a section\footnote{This means we can globally choose a zero point for the $T^2$ fiber in a smooth way, as is the case for (\ref{fibrationmetric}).}). If we want this to be a supersymmetric solution, the resulting total space $Z_8$ must be Calabi-Yau, of complex dimension four. There are many elliptically fibered Calabi-Yau fourfolds known \cite{Klemm:1996ts}. In elliptic fibrations, $\tau$ varies holomorphically over the base $B_6$ of the fibration, but $v$ remains constant. (This is because the area of a holomorphic 2-cycle (curve) equals the integral of the K\"ahler form over the curve, and since the K\"ahler form is closed, this does not change when we slide the curve over the base.) Then we can simply take our conventional scale
\begin{equation} \label{Lvrelation}
 L \equiv \sqrt{v} \, ,
\end{equation}
and the metric becomes
\begin{equation}
 \qquad ds^2_{\rm IIB,E} = -(dx^0)^2 + (dx^1)^2 + (dx^2)^2 + \frac{\ell_s^4}{v} \, dy^2 \, + \, ds_{B_6}^2  \, .
\end{equation}
If we send now $v \to 0$ keeping $\ell_s$ finite, we see that this decompactifies to flat 3+1 dimensional Minkowski space times $B_6$, with a nontrivial dilaton profile $\tau(u)$, $u \in B_6$. Since we started with a supersymmetric solution in M-theory, this dual IIB configuration will also be a supersymmetric solution. This gives an elegant recipe to construct many nontrivial type IIB vacua with varying $\tau$, in which we can use the full power of algebraic geometry applicable to Calabi-Yau manifolds.\footnote{We have hidden an issue here. For general $v$, the actual Calabi-Yau metric will actually depend on the torus coordinates $x$ and $y$ as well (because generic Calabi-Yau manifolds do not have any isometries), so the metric is not quite of the form (\ref{fibrationmetric}). In the IIA theory this will manifest itself as nonzero values for the massive fields whose quanta are D0-branes (thus breaking the $U(1)$ gauge symmetry associated to the $U(1)$ circle isometry). However, in the limit $v \to 0$ of interest here, these fields will become infinitely massive, so one expects them to vanish. Thus, in this limit, the metric ansatz (\ref{fibrationmetric}) is correct.}

Note that remarkably, what was part of the Calabi-Yau fourfold fiber in M-theory, becomes part of noncompact, visible space in type IIB, with full Lorentz invariance in the $v \to 0$ limit! We will later see that this remains true even in the presence of fluxes, when the geometry gets warped and $v$ is no longer constant.

To conclude, ``F-theory compactified on an elliptic fibration'' should be understood as meaning the type IIB geometry obtained by compactifying M-theory on this elliptic fibration and following the procedure outlined above, in the limit of vanishing elliptic fiber size $v$. We will discuss a completely explicit example in section \ref{exK3}, but first turn to some further general considerations.

\subsection{$p$-form potentials}  \label{sec:Ffields}

\begin{figure}
\centering
\includegraphics[height=0.25 \textheight]{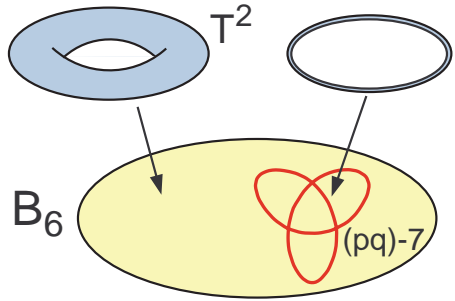}
\caption{F-theory realization of $(p,q)$ 7-branes.}
\end{figure}

Along the same lines, we can deduce F-theory equivalents of other type IIB fields. For example the M-theory 3-form potential $C_3$ can locally be decomposed as follows in the geometry (\ref{fibrationmetric}):
\begin{equation} \label{KKreduction}
 C_3 = C_3' + B_2 \wedge L \, dx  + C_2 \wedge L \, dy + B_1 \, \wedge L\, dx \wedge L\, dy \, ,
\end{equation}
where the forms $C_3'$, $B_2$, $C_2$ and $B_1$ live on $M_9$ and $L$ was defined in (\ref{MtoIIA}) and (\ref{Lvrelation}). After reduction, T-duality and taking the $L^2 = v \to 0$ limit, $B_2$ becomes the NSNS 2-form potential in type IIB, $C_2$ becomes the RR 2-form potential, $C_3'$ turns into $C_4^{(y)} = C_3' \wedge dy$, i.e.\ half the components of the self-dual 4-form potential, and $B_1$ gives rise to  off-diagonal metric components mixing the $y$-direction with the $M_9$ directions, $g_{iy} = (B_1)_i$.

Note that geometric $SL(2,\IZ)$ transformations of the $T^2$ will exactly act as (\ref{FHSL2Z}) on $(B_2,C_2)$, and as (\ref{tauSL2Z}) on $\tau$.

\subsection{Branes} \label{sec:Fbranes}

Let us see how various branes get mapped between the M-theory and IIB pictures. Of particular interest for our purposes in the case of M-theory on $\IR^{1,2} \times Z$ with $Z$ an elliptic fibration over a base $B_6$ are:
\begin{enumerate}
 \item The $\IR^{1,2}$ space-filling M2 gets mapped to a $\IR^{1,3}$ space-filling D3.
 \item At special complex codimension 1 loci of the base $B_6$, the elliptic fiber can degenerate, with generically some 1-cycle of the $T^2$ collapsing to zero size. If this is the 1-cycle along the $x$-direction (the A-cycle), this maps to a space-filling D7-brane localized at the degeneration locus in $B_6$. Note that this is a purely solitonic, geometric object from this point of view. We will see further on how the usual worldvolume degrees of freedom familiar from the perturbative string theory D-brane picture arise in a particular limit identified with the weak coupling limit in type IIB. More generally, if it is the 1-cycle $p A + q B$ which collapses, we get a $(p,q)$ 7-brane. As we will see in a detailed explicit example in section \ref{exK3}, there is an $SL(2,\IZ)$ monodromy acting on the $T^2$ fiber --- and therefore on the fields $\tau$, $B_2$ and $C_2$ --- when circling around such a degeneration point in the base. In particular around a D7-brane we have $\tau \to \tau+1$, $B_2 \to B_2$, $C_2 \to C_2 + B_2$.
 \item An M5 wrapped on a 4-cycle $\Sigma_4$ in $Z$ looks like a $1+1$ dimensional domain wall in $\IR^{1,2}$, say extended along the $(x^0,x^1)$ directions. The following cases should be distinguished, depending on the nature of $\Sigma_4$:
     \begin{enumerate}
      \item $\Sigma_4$ is an $A$-cycle fibration over $\Sigma_3 \subset B_6$. Here $\Sigma_3$ can be either a closed 3-cycle, or a 3-chain terminating on a locus where the $A$-cycle vanishes. The latter type of 3-chain in $B_6$ still produces a closed 4-cycle in $Z$, since the circle fibers collapse at the boundary of the chain. (This is analogous to the construction of a 2-sphere as a circle fibration over a line segment.) Such an M5 maps in IIB to a D5-brane wrapped on $\Sigma_3$, producing a 2+1 dimensional domain wall in $\IR^{1,3}$. If $\Sigma_3$ is a 3-chain, it maps to a D5 ending on D7 branes.

      A $B$-cycle fibration over $\Sigma_3$ similarly maps to an NS 5-brane on $\Sigma_3$. Again, $\Sigma_3$ can be a 3-chain, but now with boundary on a vanishing locus of the $B$-cycle. This gives an NS5-brane on $\Sigma_3$, which may be stretched between $(0,1)$ 7-branes.

      A $(p A + q B)$ circle fibration will map to a $(p,q)$ 5-brane possibly terminating on $(p,q)$ 7-branes.
      \item If $\Sigma_4$ is wrapping both cycles of the $T^2$, i.e.\ a $T^2$ fibration over some 2-cycle $\Sigma_2$ in the base $B_6$, we get a D3 wrapping $\Sigma_2$ and extending in the $(x^0,x^1)$-direction. In other words this is a string in four dimensions.
      \item Finally, for $\Sigma_4$ completely transversal to the $T^2$, one gets a KK-monopole extended along $\Sigma_4$ and as a string along $(x^0,x^1)$.
     \end{enumerate}
 \item M5 instantons wrap 6-cycles in $Z$. The only M5 instantons which retain finite action in the limit $v \to 0$ are those wrapped on the entire elliptic fiber. To see this, note that the M5 instanton action wrapped on $n$ directions in the $T^2$ fiber ($n=0,1,2$), has an action
     \begin{equation}
      S \sim \frac{v^{n/2}}{\ell_M^6} = \frac{v^{n/2}}{L^2 \ell_s^4} = \frac{v^{(n-2)/2}}{\ell_s^4} \, ,
     \end{equation}
     where we used (\ref{fibrationmetric}), (\ref{MIIArelations}), and (\ref{Lvrelation}). So finite action requires $n=2$. Such M5 instantons map to D3 instantons wrapped on a 4-cycle in $B_6$.
\end{enumerate}

\subsection{Fluxes} \label{sec:Ffluxes}

We can also turn on magnetic 4-form fluxes $G_4=dC_3$ on $Z$. As we will detail in section \ref{sec:turningonflux}, this will deform the geometry by warping it, but the fourfold metric remains conformal Calabi-Yau. The equations of motion give rise to the selfduality condition $G_4 = * G_4$ where $*$ is the Hodge star in the CY metric without the warp factor. In particular this implies $G_4$ is harmonic, and is uniquely determined by its (integrally quantized\footnote{More precisely, $[G_4-\frac{c_2(Z)}{2}]$ is integrally quantized, where $c_2(Z)$ is the second Chern class of $Z$ \cite{Witten:1996md}. For simplicity of exposition, we will suppress this subtlety for now. We will also always work in de Rham cohomology, to avoid torsion complications.}) cohomology class $[G_4]$.

From the discussion below (\ref{KKreduction}) we take that the only magnetic fluxes $G_4$ on $Z$ which will not violate Lorentz invariance of our eventual 4d noncompact space in type IIB are of the form
\begin{equation} \label{fluxredFIIB}
 G_4 = H_3 \wedge L \, dx + F_3 \wedge L \, dy \, ,
\end{equation}
where $H_3 = d B_2$ and $F_3 = d C_2$. But one should not make the mistake to conclude from this that all F-theory fluxes suitable for constructing flux vacua are characterized by 3-form cohomology classes $[H_3]$ and $[F_3]$ on the base. In fact, in many cases the base does not have 3-cohomology at all, while $Z$ has 4-cohomology dimension of the order of ten thousands! The mistake is that as we noted above and will detail in section \ref{exK3}, in the presence of 7-branes, the fields $(H_3,F_3)$ are \emph{not} single valued but undergo $SL(2,\IZ)$ monodromies around the 7-brane loci. This allows many more topologically nontrivial excitations, matching the large number of 4-form flux cohomology classes we have on $Z$. Typically only a small fraction of those correspond to ``bulk'' flux in IIB.
As we will see in detail in section \ref{sec:weakflux}, the twisting of $(H_3,F_3)$ around the 7-brane loci can produce topologically nontrivial excitations of $(H_3,F_3)$ whose energy and charge densities are localized very close to the 7-brane loci. In the IIB weak coupling limit, these localized excitations can be identified with D7-brane worldvolume fluxes.

Thus, the proper way to think about $(H_3,F_3)$ fluxes topologically is as $SL(2,\IZ)$-twisted cohomology, but to deal with this the proper way requires a level of formalism which involves more sequences of arrows than these lecture notes can accommodate.

There is however an alternative, more physical and intuitive way to think about these fluxes, topologically at least, and that is to consider the M5 branes (or their IIB duals) which source them, as we will now explain.

As we just noted, fluxes are characterized by their cohomology class $[G_4] \in H^4(Z,\IZ)$. In general, Poincar\'e duality canonically relates $p$-form cohomology classes and $(d-p)$-cycle homology classes, with $d$ the dimension of the space $M$ considered. A concrete way to think about this is as follows. Start with some $(d-p)$-cycle $\Sigma$ representing the homology class $[\Sigma]$, locally described by equations $f^i=0$, $i=1,\ldots,p$. Then the Poincar\'e dual $p$-form cohomology class is ${\rm PD}_M([\Sigma]) = [\delta_{\Sigma \subset M}]$, where $\delta_{\Sigma \subset M}$ is the $p$-form current
\begin{equation} \label{current}
 \delta_{\Sigma \subset M} := \delta(f^1) \, df^1 \wedge \cdots \wedge \delta(f^p) \, df^p \, .
\end{equation}
Conversely, a representative $(d-p)$-cycle of the Poincar\'e dual to a $p$-form flux can be thought of as being obtained by squeezing all flux lines maximally together. This interpretation is made clear by the observation that for any $p$-cycle $\Sigma'$, the total flux of $G={\rm PD}_M(\Sigma)$ through this cycle is
\begin{equation}
 \int_{\Sigma'} G = \int_{\Sigma'} \delta_{\Sigma \subset M} = \#(\Sigma' \cap \Sigma) \, ,
\end{equation}
where $\#$ denotes the number of intersection points counted with signs (the signs being determined by the signs of the oriented delta-function at each intersection point). Hence $[\Sigma] = PD_M([G])$ can indeed be thought of as representing the flux lines of $G$.

\begin{figure}
\centering
\includegraphics[width=\textwidth]{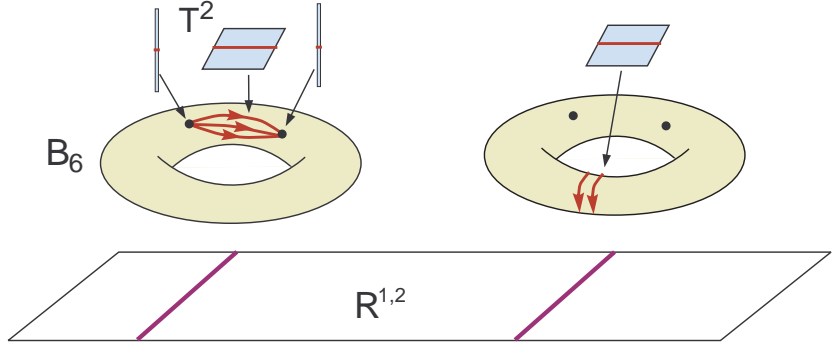}
\caption{Domain wall as a source of flux; the domain wall can be thought of as wrapping the fluxlines (or more formally the Poincar\'e dual) of the flux jump it sources. On the left we have a domain wall producing three units of a brane-type flux; on the right two units of a bulk-type flux. \label{fluxlines}}
\end{figure}

Now, magnetic $G_4$-fluxes on $Z$ are sourced by M5 domain walls wrapping a 4-cycle $\Sigma_4$ in $Z$. Across the domain wall, $[G_4]$ jumps by exactly the Poincar\'e dual of $[\Sigma_4]$. This can be seen as follows. Assuming the domain wall lies at $x^1=0$, the Bianchi identity for $G_4$ acquires a source term
\begin{equation} \label{G4jump}
 d G_4 = \ell_M^3 \, \delta_{\rm M5} = \ell_M^3 \, \delta_{\Sigma \subset Z} \wedge \delta(x^1) \, dx^1 \, ,
\end{equation}
Integrating this over $x_1$ across the wall gives the jump
\begin{equation}
 [G_4]_+ - [G_4]_- = \ell_M^3 \, {\rm PD}_Z([\Sigma_4]) \, .
\end{equation}
This is illustrated in fig.\ \ref{fluxlines}.

Thus, to classify magnetic fluxes, we only need to classify possible domain wall wrappings; the wrapped cycle can be thought of as representing the flux lines of the corresponding flux. Now, if we want the flux to preserve Poincar\'e invariance in four dimensions --- which we do if we want to construct vacua --- the only allowable sources are M5 domain walls which remain domain walls in IIB.\footnote{Strictly speaking, domain walls break Poincar\'e invariance too, of course. But away from a domain wall, all 4d fields attain their vacuum values. Other sources, such as strings, will source non-constant fluxes with legs (i.e.\ field gradients) in the noncompact directions; the 4d fields around these objects will be excited away from their vacuum values.} Luckily, we have already analyzed this: These are the M5 domain walls wrapped on 4-cycles which are well-defined $pA + qB$ 1-cycle fibrations over 3-cycles $\Sigma_3$, or over 3-chains $\Gamma_3$ which terminate on $(p,q)$ 7-branes. So the corresponding fluxes in type IIB are $p$ units of RR 3-form flux and $q$ units of NSNS 3-form flux, with flux lines closing upon themselves in the case of a 3-cycle, and terminating on the 7-branes in the case of a 3-chain.

If the 7-branes are D7-branes ($q=0$) at weak string coupling, the RR 3-form flux emanating from the branes along a 3-chain can be understood in the perturbative string picture as being sourced by worldvolume gauge flux $F_2$ on the D7-brane (through the coupling $\int_{\rm D7} C_6 \wedge F_2 $), where $F_2$ is Poincar\'e dual to the boundary of the 3-chain on the 4-cycle wrapped by the D7 in $B_6$. Keep in mind though that the perturbative D-brane picture is a \emph{different} picture than the F-theory picture we are working in: in F-theory there are no D7-branes added to the geometry, hence no D7 worldvolume fluxes sourcing bulk fields --- the 7-branes and all associated flux degrees of freedom emerge purely as solitonic excitations of the fields $\tau$, $H_3$ and $F_3$. Nevertheless we should be able to reproduce the perturbative string theory picture from the F-theory picture in a suitable weak coupling limit. This will be explained in detail further on.

\subsection{M2/D3 tadpole}

The M-theory action (\ref{Maction}) supplemented with M2-brane sources gives the following equation of motion for $G_4$:
\begin{equation}
 d* G_4 = \frac{1}{2} G_4 \wedge G_4 - \ell_M^6 I_8(R) + \ell_M^6 \sum_i \delta_{{\rm M2}_i} \, .
\end{equation}
Integrating this over the fourfold $Z$ gives
\begin{equation} \label{tadpolecancelation}
 \frac{1}{2 \ell_M^6} \int_Z G_4 \wedge G_4 \, + \, N_{\rm M2} = \frac{\chi(Z)}{24} =: Q_c \, .
\end{equation}
Here $N_{M2}$ is the number of $\IR^{1,2}$-filling M2-branes and we used the fact that $24 \, I_8(R)$ integrates to the Euler characteristic $\chi$ on a Calabi-Yau fourfold \cite{Duff:1995wd}. What this equation says is that the total M2 charge transversal to $Z$ must vanish --- indeed there is nowhere for the flux lines sourced by the M2 charge to go in a compact space. As we will see in chapter \ref{sec:statistics}, this tadpole cancelation condition turns out to be what renders the number of metastable F-theory flux vacua within any compact region of low energy parameter space finite. If it hadn't been for this constraint, string theory would have been infinitely finely tunable.

The type IIB equivalent of this is, using (\ref{fluxredFIIB}),
\begin{equation} \label{D3tadpole}
  \frac{1}{\ell_s^4} \int_{B_6} F_3 \wedge H_3 \, + \, N_{\rm D3} = Q_c \, .
\end{equation}
Again, as mentioned in the previous subsection, the contributions to the D3 tadpole attributed to D7-brane worldvolume fluxes in the perturbative IIB string picture are in fact already contained in (\ref{D3tadpole}). We will come back to this in section \ref{sec:weakflux}, after determining how to take the weak coupling limit in F-theory.

Before we do this, it is probably useful to consider a simple example.

\subsection{Example: K3} \label{exK3}

\begin{figure}
\centering
\includegraphics[height=0.25 \textheight]{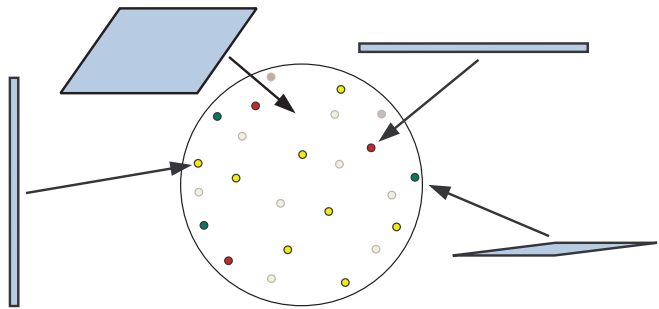}
\caption{F-theory on K3 ellipticaly fibered over a sphere. The dots indicate the 24 degeneration loci of the elliptic fiber.}\label{K3example}
\end{figure}

We now turn to an explicit example \cite{Sen:1996vd}, F-theory on an elliptically fibered K3, times (in IIB) some eight dimensional manifold which for definiteness we take to be $\IR^{1,7}$, but any other manifold solving the equations of motion would be fine too. (In particular if we wanted to compactify on a fourfold $Z$, we could take $Z=T^4 \times K3$ or $Z=K3 \times K3$.) From the previous discussion, we know that this means we should consider M-theory on $\IR^{1,6} \times {\rm K3}$. An elliptically fibered K3 can be described by the equation
\begin{equation} \label{K3eqproj}
 y^2 = x^3 + f(u,v) \, x  \, z^4 + g(u,v) \, z^6 \, ,
\end{equation}
where $x,y,z,u,v \in \IC$, modulo the projective equivalences
\begin{eqnarray}
 (u,v,x,y,z) &\simeq& (\lambda u, \lambda u, \lambda^4 x,\lambda^6 y,z) \\
 &\simeq& (u,v, \mu^2 x, \mu^3 y, \mu z) \, , \label{equiv2}
\end{eqnarray}
where $\mu, \lambda \in \IC^* = \IC \backslash \{0\}$, and $(u,v) \neq (0,0)$, $(x,y,z) \neq (0,0,0)$. The functions $f$ and $g$ are homogeneous polynomials of degree eight and twelve in $u$, $v$. Notice that all of these weight assignments are consistent with the embedding equation (\ref{K3eqproj}), and that it indeed describes a two complex dimensional surface, five coordinates minus one equation minus two equivalence relations. As we will explain systematically in section \ref{geometrictools}, the rule to determine whether such a hypersurface is Calabi-Yau is simply that the (weighted) degree of the defining polynomial equals the sum of the weights, for each equivalence. For the first equivalence, the degree is 12, and the sum of the weights is $1+1+4+6+0=12$. For the second equivalence, the degree is 6, and the sum of the weights is $0+0+2+3+1=6$. So this is indeed a Calabi-Yau twofold, hence a K3.

In a coordinate patch in which we fix the projective equivalences by putting $z \equiv 1$, $v \equiv 1$, the equation simplifies to
\begin{equation} \label{K3eqsimpl}
 y^2 = x^3 + f(u) \, x + g(u) \, ,
\end{equation}
with $f$ and $g$ ordinary polynomials of degree eight and twelve.

To see that we indeed have an elliptic fibration, it is sufficient to note that at fixed $(u,v)$, (\ref{K3eqproj}) describes a Calabi-Yau onefold, that is, an elliptic curve, that is, a $T^2$, embedded in $(x,y,z)$ space. Again this follows because the sum of the weights of the remaining equivalence (\ref{equiv2}) equals the degree (six here).

More formally, the projection of the fibration is $\pi:K3 \to \ICP^1:(x,y,z,u,v) \to (u,v)$, where $\ICP^1 = \{ (u,v) \neq (0,0) | (u,v) \simeq (\lambda u,\lambda v) \}$. You can easily check that this is a well-defined map, in the sense that equivalent points get mapped to equivalent points and no point maps to $(0,0)$. So, we have an elliptic fibration over base $\ICP^1 = S^2$.

How can we relate this algebraic description of a $T^2$ to the standard representation $T^2 = \IC/(\IZ \oplus \tau \IZ)$? This is done by relating holomorphic coordinates. On $T^2 = \IC/(\IZ \oplus \tau \IZ)$, the holomorphic coordinate is $z=x+\tau y$, which for any point $P$ can be written as
\begin{equation} \label{zcoorddef}
 z(P) = \int_0^P \Omega_1 \, , \qquad \Omega_1 = dz \, .
\end{equation}
Up to normalization, $\Omega_1$ is the unique holomorphic 1-form on the torus. In section \ref{geometrictools}, we will see how to construct in general the up to normalization unique holomorphic $n$-form on any algebraic Calabi-Yau $n$-fold. The result for the $T^2$ described by (\ref{K3eqsimpl}) at fixed $u$ is
\begin{equation} \label{hol1form}
 \Omega_1 = \frac{c \, dx}{y} \, ,
\end{equation}
where $c$ is some normalization constant. Choosing a basis of 1-cycles $(A,B)$ on the algebraic $T^2$, the modulus $\tau$ is then given by
\begin{equation}
 \tau = \frac{\oint_B \Omega_1}{\oint_A \Omega_1} \, .
\end{equation}
The ambiguity in choosing a basis is just the $SL(2,\IZ)$ S-duality frame ambiguity we expect.

In principle we could now figure out the relation between $\tau$ and $f$ and $g$ ourselves by computing period integrals, but it turns out that industrious mathematicians figured this out already more than a century ago, and fortunately left notes. The result can be expressed as
\begin{equation} \label{jfunction}
 j(\tau) = \frac{4 \cdot (24 \, f)^3}{\Delta}, \qquad \Delta = 27 \, g^2 + 4 \, f^3 \, ,
\end{equation}
where $j(\tau)$ is the $SL(2,\IZ)$ modular invariant $j$-function, $j(\tau) = e^{-2 \pi i \tau} + 744 + \CO(e^{2 \pi i \tau})$.

Although we will not need it, let me mention that once we know $\tau(u)$, we know the exact metric on the base too \cite{Greene:1989ya}:
\begin{equation} \label{basemetric}
 ds^2 = \frac{a \, \tau_2(u) \, |\eta(\tau(u))|^4}{|\Delta(u)|^{1/6}} \, du \, d\bar{u} \,  ,
\end{equation}
where $\eta(\tau)$ is the Dedekind eta function and $a$ an arbitrary constant setting the size of the $\ICP^1$.

The function $\Delta$ in (\ref{jfunction}) is called the discriminant of the elliptic curve; when it vanishes, the elliptic curve becomes singular, generically with a 1-cycle collapsing to zero size. It is a homogeneous polynomial of degree 24 on the base $\ICP^1$, so $\Delta$ has 24 zeros. Generically they will all be distinct, distinct from the zeros of $f$ and $g$, and distinct from the point $(1,0)$. Assuming the latter we fix the $\ICP^1$ scaling by putting $v \equiv 1$. Near a generic zero $u=u_i$ ($i =1, \ldots,24$), (\ref{jfunction}) becomes
\begin{equation}
 j(\tau(u)) \sim \frac{1}{u-u_i} \, ,
\end{equation}
which is solved as
\begin{equation} \label{taulogsol}
 \tau(u) \approx \frac{1}{2\pi i} \, \ln(u-u_i) \, ,
\end{equation}
up to $SL(2,\IZ)$ transformation.

Note that when $u \to u_i$, $\tau \to i \infty$. Geometrically, this means the ratio of A-cycle and B-cycle lengths of the $T^2$ vanishes. Recalling the physical meaning of $\tau$ in type IIB, namely $\tau = C_0 + \frac{i}{g_{\rm IIB}}$, we see that this corresponds to weak coupling, $g_{\rm IIB} \to 0$.

Moreover, when circling once around $u=u_i$, following a path $u(\theta) = u_1 + \epsilon \, e^{2 \pi i \theta}$, we see that $\tau$ undergoes \emph{monodromy} when $\theta:0 \to 2\pi$:
\begin{equation} \label{simplemonodromy}
 T:\tau \to \tau+1 \, .
\end{equation}
Equivalently, $C_0 \to C_0 + 1$, or $\oint_{u_i} F_1 = \oint_{u_i} dC_0 = 1$, which means \emph{there is a D7-brane at $u=u_i$}.

This immediately leads to a paradox: now it looks like we have 24 D7-branes in a compact transversal space, $\ICP^1$. There can be no net charge in a compact space, since the flux lines have nowhere to go. More directly, the sum of all contour integrals $\sum_{i=1}^{24} \oint_{u_i} F_1$ must vanish, since the total contour is contractible on the sphere. How can this be?

The resolution lies in the innocent looking ``up to $SL(2,\IZ)$ transformation'' under (\ref{taulogsol}). While it is true that we can always \emph{locally} go to an $SL(2,\IZ)$ frame where $\tau(u)$ lies in the fundamental domain, we cannot do this \emph{globally}. We can pick one point $u_*$ where we choose $\tau(u_*)$ to lie in the fundamental domain, but once we start walking around on the $\ICP^1$, $\tau(u)$ might move off to some other region of the upper half plane. Of course, near any other zero of $\Delta$ that we might encounter on our trip, the value of $\tau(u)$ will still be related to (\ref{taulogsol}) by an $SL(2,\IZ)$ transformation $M$, but then the monodromy (\ref{simplemonodromy}) will be related by conjugation, $MTM^{-1}$ instead of $T$. As a result, we will in general no longer have a D7-brane there, but a more general $(p,q)$ 7-brane, related to the D7 (i.e.\ the $(1,0)$ 7-brane) by the S-duality transformation $M$. As mentioned earlier already in section \ref{sec:Fbranes}, a general $(p,q)$ 7-brane can be characterized in F-theory by the vanishing of the $pA+qB$ 1-cycle of the $T^2$, the image under $M$ of the $A$-cycle which vanishes for a D7.

In general, it is not an easy task to figure out exactly what kind of $(p,q)$-branes we have at various points; worse even, this in fact depends on the path we take through the base! More important than the confusion this is bound to instill in anyone who sets out to explore these IIB solutions, is the fact that this makes it entirely impossible to do conventional string perturbation theory on such backgrounds. Even if we make our base $\IP^1$ as large as the solar system, and we go to an $SL(2,\IZ)$ duality frame where the monodromy closest to home is of the D7 form (\ref{simplemonodromy}), and the string coupling is very weak near home, there will always be $(p,q)$ 7-brane monodromies somewhere else with $q \neq 0$, which send $g_{\rm IIB}$ from weak to strong coupling. For example if we have a $(0,1)$ 7-brane somewhere, this will map $\tau \to -1/\tau$. Equivalently, we can say that if we send off a fundamental string and let it loop around the $(0,1)$ 7-brane, it will come back to us as a D1 string. So it is not possible to set up conventional perturbation theory for fundamental strings in a globally well defined way.

This is why people say F-theory is intrinsically strongly coupled. Of course, if we are only interested in getting nontrivial solutions of type IIB supergravity, we do not need to care about this; it is only when we need to compute string scattering amplitudes that we get into trouble.

All of this leaves us with a new puzzle: We definitely know there \emph{are} regimes in which type IIB theory in principle has a perturbative string expansion. How can we see this in F-theory?

\subsection{The weak coupling limit: Orientifolds from F-theory, K3 example}

The answer to the puzzle just raised is given by taking a clever limit of the F-theory description, pointed out by Sen \cite{Sen:1996vd,Sen:1997gv}. For our K3 example, the simplest such limit accomplishing this is as follows. We want to go to a point in the K3 moduli space where $\tau(u)$ is constant and has large imaginary part. We see from (\ref{jfunction}) that constancy requires $f^3/g^2$ = constant, which is solved by
\begin{equation} \label{fgpdef}
 g=p^3, \qquad f = \alpha p^2 \, ,
\end{equation}
with $\alpha$ a constant and $p$ a homogeneous polynomial of degree four. Let us go again to a coordinate patch $v \equiv 1$. By a rescaling of $y$ and $x$ we can set the coefficient of $u^4$ equal to one, so $p(u)$ has the form
\begin{equation}
 p = \prod_{i=1}^4 (u-u_i) \, ,
\end{equation}
where the $u_i$ are constants. Plugging this in (\ref{jfunction}), we get
\begin{equation} \label{Deltaorlimit}
 \Delta = (4 \alpha^3 + 27) \prod_{i=1}^4 (u-u_i)^6 \, , \qquad
 j(\tau) = \frac{4 \cdot (24 \alpha)^3}{27+4 \alpha^3} \, .
\end{equation}
Thus, if we tune
\begin{equation}
 \alpha \approx -3/4^{1/3} \, ,
\end{equation}
we get weak IIB string coupling everywhere on the base!

Although $\tau$ is now constant everywhere, this does not necessarily mean there is no $SL(2,\IZ)$ monodromy at all, because there is one nontrivial $SL(2,\IZ)$ element which acts trivially on $\tau$, namely
\begin{equation} \label{orientifoldmonodromy}
 M = \left( \begin{array}{cc} -1 & 0 \\ 0 & -1 \end{array} \right) \, .
\end{equation}
This may seem overly paranoid, but actually it turns out that we do get this monodromy around each of the $u_i$. To see this, note that after a change of coordinates $x=p \tilde{x}$, $y = p^{3/2} \tilde{y}$ and using (\ref{fgpdef}), we can rewrite (\ref{K3eqsimpl}) as
\begin{equation}
 \tilde{y}^2 = \tilde{x}^3 + \alpha \tilde{x} + 1 \, .
\end{equation}
This makes it completely manifest that the modulus of the torus does not vary with $u$. However, note that in the new coordinates, $\Omega_1$ defined in (\ref{hol1form}) becomes $\Omega_1 = p^{-1/2} \frac{d \tilde{x}}{\tilde{y}}$. Therefore, when we circle around a zero of $p(u)$ in the $u$-plane, we map $\Omega_1 \to - \Omega_1$. This implies is particular $\oint_A \Omega_1 \to - \oint_A \Omega_1$, $\oint_B \Omega_1 \to - \oint_B \Omega_1$, and from (\ref{zcoorddef}), $z \to -z$. In a representation of the torus where we think of $\Omega_1$ as being fixed, such as the standard $T^2 = \IC/(\IZ \oplus \tau \IZ)$, this monodromy boils down to
\begin{equation}
 (A,B) \to (-A,-B) \, ,
\end{equation}
that is, (\ref{orientifoldmonodromy}). Note that in the type IIB picture, this monodromy implies that the fields $(B_2,C_2)$ are double valued on the $\ICP^1$, flipping sign when circling around the zeros of $p$.

We can conveniently think of this situation in the following way. First we construct a double cover of the base  $\ICP^1$, which we call $X$, defined by adding a coordinate $\xi \in \IC$ and the equation
\begin{equation}
 X:\xi^2 = p(u,v) \, , \qquad (u,v,\xi) \simeq (\lambda u,\lambda v,\lambda^2 \xi) \, ,
\end{equation}
with $(u,v,\xi) \neq (0,0,0)$. The extension of the projective equivalence of $\ICP^1$ is imposed by compatibility with the equation. Notice that this equation again satisfies the Calabi-Yau condition that its degree equals the sum of the weights (four). So this describes again a $T^2$, but a different one than the fiber $T^2$ we had before.\footnote{Again this $T^2$ can be mapped to the standard representation; now the holomorphic 1-form in a patch $v \equiv 1$ is $\Omega_1'=du/\xi$. You can check that (\ref{basemetric}) reduces to the flat metric in standard coordinates $z'=\int \Omega_1'$.} The original base $\ICP^1$ is recovered from $X$ as the quotient $\ICP^1 = X/\sigma$, where
\begin{equation}
 \sigma: \xi \to - \xi \, .
\end{equation}
When circling around the zeros of $p$ on the base, we go from one sheet of the double cover to the other. At the same time, we have the $\IZ_2$ transformation (\ref{orientifoldmonodromy}) acting on the $T^2$ fiber. Thus, on the covering space, everything is single valued; the double valuedness appears in this picture by taking the simultaneous $\IZ_2$ quotient of $X$ and the $T^2$ fiber. Correspondingly, in this limit, we can represent our F-theory K3 as
\begin{equation}
 {\rm K3} = (T^2 \times T^2)/\IZ_2 \, .
\end{equation}

By now, a bell should be ringing: What we have here in the type IIB setting is exactly the same as what one would get from orientifolding $X$ by $\sigma \cdot (-1)^{F_L} \cdot \omega$, where $\omega$ denotes worldsheet orientation reversal, i.e. exchange of left- and rightmoving modes on the worldsheet, and $(-1)^{F_L}$ changes the sign of the Ramond sector states of the leftmoving sector. You can see this from the way our $\IZ_2$ action acts on the different RR and NSNS potentials obtained from F-theory as explained in section \ref{sec:Ffields} and comparing this with the perturbative worldsheet result.

Thus, the fixed loci of the $\IZ_2$ involution $\sigma$, that is, the four zeros $u_i$ of $p$, are identified with O7-planes, which have D7-charge $-4$ as measured in the base space $\ICP^1$. Since there is no monodromy of $C_0$, there is no net charge at the fixed points, so there must be four D7-branes located on top of the O7-planes.

We expect to be able to move away the D7-branes from the O7-planes. Zooming in on an O7 located say at $u=0$, and assuming we have moved the four D7-branes which were on top to nearby positions $u=u^{(a)}$, there should now be a D7 $T$-monodromy (\ref{simplemonodromy}) around each $u^{(a)}$, and a compensating $T^{-4}$ monodromy around the naked O7 which stays behind. So naively, we might expect $\tau(u)$ to be of the form
\begin{equation}
 \tau(u) = \tau_0 +\frac{1}{2\pi i} \biggl( \sum_{a=1}^4 (u-u^{(a)}) - 4 \ln u \biggr) \, .
\end{equation}
However, upon further reflection, this does not make sense, since too close to $u=0$, this gives a large \emph{negative} value for $\Im \tau$! More precisely for $\CO(1)$ values of the $u^{(a)}$ and large $\Im \tau_0$ (which we can identify with the large $u$ asymptotic inverse string coupling), this occurs when $|u|<e^{-\frac{\pi}{2} \Im \tau_0}$. Note that this is exponentially small at weak coupling, but nevertheless, since nothing fixes the overall size of the base $\ICP^1$ at this point, this breakdown could still occur at a distance much larger than the string length.

This is a typical phenomenon occurring for naive supergravity solutions in the presence of orientifold planes: one finds nasty singularities at finite distance from the O-plane. This can be traced back to the fact that these objects have negative tension.

\begin{figure}
\centering
\includegraphics[width=0.9 \textwidth]{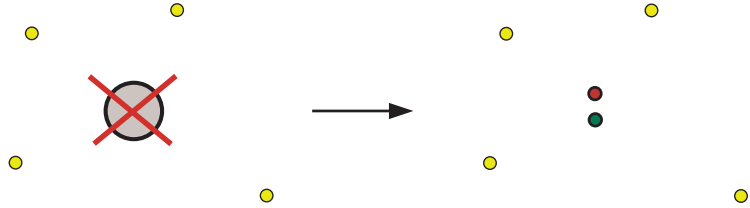}
\caption{The naive finite distance orientifold singularity gets resolved in F-theory by splitting the O-plane into two 7-branes.}\label{Osplit}
\end{figure}

The correct solution obtained from F-theory does not have this pathology, and is completely well-behaved. What actually happens is that at nonzero string coupling $1/\Im \tau_0$, the orientifold plane \emph{splits} in two $(p,q)$ 7-branes, with separation of the order of the distance $e^{-\frac{\pi}{2} \Im \tau_0}$ where our naive solution breaks down. This follows from (\ref{Deltaorlimit}): the limit in which the four D7-branes are coincident with the O7 corresponds to a zero of $\Delta$ of multiplicity \emph{six}, that is, it corresponds to a limit in which the 24 generic $(p,q)$ 7-branes coincide in four groups of six. Out of the six, four get identified with D7-branes, and the remaining two must correspond to the orientifold plane. More details can be found \cite{Sen:1996vd}.
Notice that this splitting cannot be seen in perturbation theory, as it is nonperturbatively small in the string coupling.

\subsection{Orientifolds from F-theory: general story} \label{sec:orlimitgen}

I will now give a refinement and generalization of the orientifold limit we discussed for $K3$, again due to Sen \cite{Sen:1997gv}. The refinement consists of allowing the D7-branes to move away from the O7-planes, while retaining weak coupling. The generalization consists of allowing general elliptically fibered Calabi-Yau $n$-folds.

For definiteness, we will again work with an example, but it will be clear how to generalize it (if not, see \cite{Sen:1997gv}). The example is an elliptically fibered Calabi-Yau fourfold, fibered over $\ICP^3$, described by the equation, analogous to (\ref{K3eqproj}),
\begin{equation} \label{CY4eq}
 Z: y^2 = x^3 + f(\vec u) \, x \, z^4 + g(\vec u) \, z^6
\end{equation}
where $\vec u := (u_1,u_2,u_3,u_4)$. We also impose the projective $\IC^*$ equivalences
\begin{eqnarray}
 (u_1,u_2,u_3,u_4,x,y,z) &\simeq& (\lambda u_1,\lambda u_2,\lambda
u_3,\lambda u_4,\lambda^8 x,\lambda^{12} y,z) \label{Cstargen1} \\
 &\simeq& (u_1,u_2,u_3,u_4,\mu^2 x,\mu^3 y,\mu z) \, , \label{Cstargen2}
\end{eqnarray}
where $\vec u \neq \vec 0$ and $(x,y,z) \neq (0,0,0)$.
In the case at hand, $f(\vec u)$, $g(\vec u)$ are homogeneous polynomials of degrees
16 and 24.  At fixed $u$, (\ref{CY4eq}) describes an elliptic curve, hence this equation indeed defines an elliptic
fibration over $\ICP^3$. The sum of the weights equals the degree, so we do have a Calabi-Yau fourfold.
The number of complex structure moduli is $h^{3,1}(Z)=3878$, which
can be computed directly by counting the number of coefficients
of $f$ and $g$ modulo $GL(4,\IC)$ coordinate transformations: ${16+3
\choose 3} + {24+3 \choose 3} - 16 = 3878$.

To define the orientifold limit, we first parametrize, without loss of
generality, following Sen:
\begin{eqnarray}
 f &=& -3h^2 + \epsilon \eta,\nonumber\\
 g &=& -2h^3 + \epsilon h \eta - \epsilon^2 \chi/12,
\end{eqnarray}
where $h$, $\eta$ and $\chi$ are a homogeneous polynomials of
degrees 8, 16 and 24 in the $u_i$, and $\epsilon$ is a constant. (Notice that for $\epsilon=0$, this is essentially the limit discussed in the previous subsection with $\alpha=-3/4^{1/3}$.) When $\epsilon
\to 0$ keeping everything else fixed, one finds for the discriminant and $j(\tau)$
\begin{equation} \label{jtaurelation}
 \Delta \approx - 9 \, \epsilon^2 h^2 (\eta^2 - h \chi),
 \qquad j(\tau) \approx \frac{(24)^4}{2} \, \frac{h^4}{\epsilon^2(\eta^2 - h
 \chi)}.
\end{equation}
Thus, in this limit,
\begin{equation} \label{couplepsrel}
 g_{\rm IIB} \sim -\frac{1}{\log |\epsilon|} \to 0
\end{equation}
everywhere except near $h=0$, and the $\epsilon \to 0$ limit can therefore be
interpreted as the IIB weak coupling limit. A monodromy analysis similar to what we did in the previous subsection
\cite{Sen:1997gv} shows that in this limit the two components of
$\Delta=0$ should be identified with an O7-plane and a D7-brane as
follows:
\begin{equation} \label{O7D7eq}
 {\rm O7}: h(\vec u)=0, \qquad {\rm D7}: \eta(\vec u)^2 = h(\vec u) \, \chi(\vec u),
\end{equation}
where the orientifolded Calabi-Yau 3-fold is given by the equation
\begin{equation} \label{CYeq}
 X: \xi^2 = h(\vec u)
\end{equation}
with $\IC^*$ equivalence $(\vec u,\xi) \simeq (\lambda \vec u,\lambda^4 \xi)$, and
orientifold involution
\begin{equation}
 \sigma: \xi \to - \xi.
\end{equation}
The CY threefold $X$ is a double cover of $\ICP^3$ branched over
$h(u)=0$; quotienting by $\sigma$ gives back $\ICP^3$. In the case at hand it has $149$ complex structure deformations, given by the coefficients of $h(u)$ modulo $GL(4,\IC)$ coordinate transformations $\vec u \to A \vec u$,
and one K\"ahler deformation, its volume. In addition to this, there are
D7-brane moduli, counted by the number of inequivalent deformations
of the D7 equation in (\ref{O7D7eq}), i.e.\ ${16+3 \choose 3} + {24+4 \choose 3} - {8+3
\choose 3} - 1 = 3728$, where the first subtraction comes from the
fact that we can shift $\eta \to \eta + h \psi$ with $\psi$ an
arbitrary degree 8 polynomial and shift $\chi$ accordingly, without
changing the form of the D7 equation (\ref{O7D7eq}), and the last
subtraction corresponds to overall rescaling of the coefficients. As
a check note that indeed the number of D7 moduli plus the number of
3-fold complex structure moduli plus one for the dilaton-axion modulus $\epsilon$
equals 3878, the number of fourfold complex structure moduli.

Observe that the number of D7-brane moduli is vastly larger than the number of bulk moduli.

Finally, for future reference, we relate the holomorphic 4-form living on the Calabi-Yau fourfold $Z$ to the holomorphic 3-form living on the Calabi-Yau threefold $X$. For an elliptic fibration of the form $(\ref{CY4eq})$, the holomorphic 4-form is, say in a patch $z \equiv 1 \equiv u_4$, $y \neq 0$:
\begin{equation}
 \Omega_4 = c \, \frac{dx \wedge du_1 \wedge du_2 \wedge du_3}{y} \, ,
\end{equation}
where $c$ is some normalization constant. (We will see in detail how this expression is obtained in section \ref{geometrictools}.) Define a 3-form on the base of the elliptic fibration (here $\ICP^3$) by ``integrating'' out the $A$-cycle of the $T^2$:
\begin{equation} \label{Om3def}
 \Omega_3 := \oint_A \Omega_4 \, .
\end{equation}
In general, this would not give a single-valued 3-form on the base, because the $A$-cycle undergoes various $SL(2,\IZ)$ monodromies when circling around $(p,q)$ 7-branes. However, in the weak coupling limit, the only monodromy acting on $A$ is $A \to -A$, when circling around the O7 locus $h=0$, and this disappears altogether when going to the double cover $X$. This can be seen explicitly by performing the integral in (\ref{Om3def}). To do this, first note that when $\epsilon=0$, we have
\begin{equation}
 y^2=(x+h)^2 (x-2h) + \CO(\epsilon) \, .
\end{equation}
The $A$-cycle is the loop in the $x$-plane collapsing in the limit $\epsilon=0$, i.e.\ the loop around the zeros of $y$ which collapse to the double zero $x = -h$ when $\epsilon=0$. Performing the contour integral, we get
\begin{eqnarray}
 \Omega_3 &=& c' \, \frac{du_1 \wedge du_2 \wedge du_3}{\sqrt{h}} + \CO(\epsilon) \\
 &=& c' \, \frac{du_1 \wedge du_2 \wedge du_3}{\xi} +\CO(\epsilon) \, , \label{Om3expr}
\end{eqnarray}
where $c' = 2 \pi c / \sqrt{3}$. In the last step we used (\ref{CYeq}) and consider $\Omega_3$ to live on $X$.  To leading order in $\epsilon$ this is indeed exactly the holomorphic 3-form on the Calabi-Yau 3-fold $X$. Note that (\ref{couplepsrel}) implies that the size of the corrections is about $e^{-1/g_{\rm IIB}}$, that is, nonperturbatively small.

If we integrate out the $B$-cycle instead, we get, by definition of the modular parameter $\tau$:
\begin{equation}  \label{Bcycleint}
 \oint_B \Omega_4 = \tau \, \Omega_3 \, \approx \left( \tau_0 + \frac{i}{2 \pi} \ln \frac{P_{O7}}{P_{D7}} \right) \Omega_{3,CY} \, ,
\end{equation}
where in the last step we used (\ref{jtaurelation}) and $j(\tau) \approx e^{-2 \pi i \tau}$, putting
\begin{equation} \label{tau0val}
 \tau_0 := \frac{i}{2\pi} \ln \frac{288}{\epsilon^2} \, , \qquad  P_{O7}:=h^4, \qquad P_{D7}:=\eta^2-h\chi \, .
\end{equation}
All approximations made here have errors at most of order $\epsilon \sim e^{-\pi/g_s}$, i.e.\ nonperturbatively small at weak coupling.

A recent explicit study of the weak coupling limit of F-theory can be found in \cite{Braun:2008ua}, with in particular the example of K3 worked out in detail.

\subsection{Localization, fluxes and tadpoles at weak coupling} \label{sec:weakflux}

At weak string coupling $g_{\rm IIB} \to 0$, we expect it to be possible to separate charges, energies and other physical quantities in ``bulk background'' and ``D-brane'' contributions. It is instructive to see explicitly how this happens for fluxes.

Let us consider first a local model, F-theory on an elliptic fibration $Z$ over $B_6 = S \times D$, with $S$ an arbitrary K\"ahler manifold of complex dimension two and $D$ the unit disk, parametrized by a complex coordinate $u$, with elliptic fiber modulus
\begin{equation} \label{tauuuu}
 \tau_1 + i \tau_2 := \tau(u) = \frac{\ln u}{2\pi i} \, .
\end{equation}
This is a local model for what we have earlier identified as a D7 brane wrapped on a 4-cycle $S$. The metric on $D$ can be anything conformal to the flat metric. Let $g_s$ be the type IIB string coupling at the boundary of the disk, so we can write in polar coordinates $u=:r \, e^{i \theta}$
\begin{equation} \label{taurtheta}
 \tau_1 = \frac{\theta}{2 \pi}, \qquad \tau_2 = \frac{1}{g_s} + \frac{\ln(r^{-1})}{2 \pi} \, .
\end{equation}
Using the metric (\ref{fibrationmetric}), you can check that there is a particular normalizable harmonic 2-form on the elliptic fibration over the disk, given by
\begin{equation}
 \omega = \frac{1}{g_s} \, d \left(  \frac{dx + \tau_1 \, dy}{\tau_2}  \right) \, ,
\end{equation}
where the normalization is chosen such that $\oint_{\partial D} \int_{y=0}^1 \omega \equiv 1$. It is anti-self-dual:
\begin{equation}
 * \omega = - \omega \, .
\end{equation}
In fact, this is true even if (\ref{tauuuu}) is replaced by any other holomorphic function $\tau(u)$, as you can check by noting that the metric on $D$ is conformal to $d \tau_1^2 + d\tau_2^2$.\footnote{The most general anti-self-dual form on the elliptic fibration over the disk is of the form $d \, \Re[f \, (dx + \bar{\tau} dy)/\tau_2 ]$, with $f(u) = f_0 + f_1 \, u + f_2 \, \frac{u^2}{2} + \cdots$ a holomorphic function on the disk. While the constant term (i.e.\ $\omega$), as we will see, leads to strongly localized energy and charge densities, the $\CO(u)$ corrections do not, and should be considered as part of the background in which the D7 is placed.}

\begin{figure}
\centering
\includegraphics[height=0.25 \textheight]{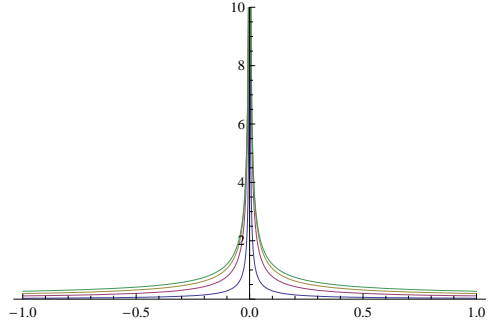}
\caption{D7-brane localization of flux energy in the weak coupling limit. The x-axis is the position along a diagonal in the unit disk surrounding the D7, and the y-axis is the charge and energy density normalized to a total of 1. The four curves correspond to four different values of the string coupling constant starting at $g_s = 0.1$ at the bottom and going up in steps of $0.25$. \label{braneloc}}
\end{figure}

Define now the following 4-form flux $G_4$ on $Z$:
\begin{equation} \label{fluxlocal}
 G_4 = F_2 \wedge L \, \omega
\end{equation}
where $F_2$ is some closed 2-form on $S$. Note that if we take $F_2$ anti-selfdual too, $G_4$ will be self-dual, as is required by the classical equations of motion (see section \ref{sec:turningonflux}). For now we will leave $F_2$ arbitrary though. Following our usual reduction, in IIB language, this $G_4$ corresponds to
\begin{eqnarray}
 H_3 &=& \frac{1}{g_s} \, F_2 \wedge d \left( \frac{1}{\tau_2} \right) \, , \label{H3eqeq} \\
 F_3 &=& \frac{1}{g_s} \, F_2 \wedge d \left( \frac{\tau_1}{\tau_2} \right) \, , \label{F3eqeq} \\
 G_3 \equiv F_3 - \tau H_3 &=& \frac{1}{g_s} \, F_2 \wedge \frac{d\tau}{\tau_2} \, .
\end{eqnarray}
Plugging this in (\ref{IIBaction}) and using (\ref{taurtheta}), we note that the $|G_3|^2$ part of the Lagrangian density is
\begin{equation} \label{Endensity}
 \frac{2 \pi}{\ell_s^8} \, \frac{1}{2} (F_2 \wedge * F_2) \wedge
 d \left( \frac{1}{[1 + \frac{g_s}{2 \pi} \ln(r^{-1})]^{2}} \right) \wedge  \frac{d \theta}{2 \pi}  \, .
\end{equation}
Integrating the last two factors over the disk gives 1, and what remains is exactly the Yang-Mills Lagrangian density for a D7-brane wrapped on $S$. Note that the radial energy distribution diverges at $r=0$ as $dr/r \log^3 r$, but in an integrable way. Moreover, in the weak coupling limit $g_s \to 0$, almost all energy is localized exponentially close to $r=0$, within a radius
\begin{equation}
 r_* \sim e^{-2\pi/g_s} \, .
\end{equation}
This is illustrated in fig. \ref{braneloc}. Similarly, the D3-charge density from the $F_3 \wedge H_3$ term is
\begin{equation} \label{Qdensity}
 -\frac{2 \pi}{\ell_s^8} \, \frac{1}{2} (F_2 \wedge F_2) \wedge
 d \left( \frac{1}{[1 + \frac{g_s}{2 \pi} \ln(r^{-1})]^{2}} \right) \wedge  \frac{d \theta}{2 \pi}  \, .
\end{equation}
Comparing to the D3 action, we thus see that the total D3-charge is
\begin{equation}
 Q_3(D7) = - \frac{1}{\ell_s^4} \int_S \frac{1}{2} \, F_2 \wedge F_2 \, .
\end{equation}
(This is in conventions in which $\ell_s^{-2} F_2$ is integrally quantized.) This is indeed as expected from the standard D7-brane action.

In the language of section \ref{sec:Ffluxes}, a representative of the Poincar\'e dual to (i.e.\ the maximally squeezed together flux lines of) the flux $G_4$ we just constructed is the 4-cycle constructed as a fibration of the $A$-cycle over the 3-chain consisting of a ray emanating from the origin of the disk times $\Sigma_2$, where $\Sigma_2$ is the Poincar\'e dual to $F_2$ on $S$. Note that although in section \ref{sec:Ffluxes} we identified such a cycle topologically with RR flux, $H_3$ in (\ref{H3eqeq}) is not zero identically, although it is an exact form on $D \backslash \{0 \}$. This is how we can still get a nonzero charge density $F_3 \wedge H_3$. From the point of view of the Poincar\'e dual cycle to $G_4$, the charge $\frac{1}{2} \int G_4 \wedge G_4$ is half the self-intersection product of this cycle, which can be seen directly to be equal to half the self-intersection product of $\Sigma_2$ on $S$, in agreement with (\ref{Qdensity}).

In contrast, fluxes like (\ref{H3eqeq})-(\ref{F3eqeq}) cannot exist localized on O7-planes. This is because $H_3$ and $F_3$ transform with a minus sign under the orientifold involution (equivalently, in the base, they change sign when looping around the orientifold point), which is not satisfied for (\ref{H3eqeq})-(\ref{F3eqeq}). This agrees with the absence of gauge fields on orientifold planes in perturbation theory.

\begin{figure}
\centering
\includegraphics[height=0.32 \textheight]{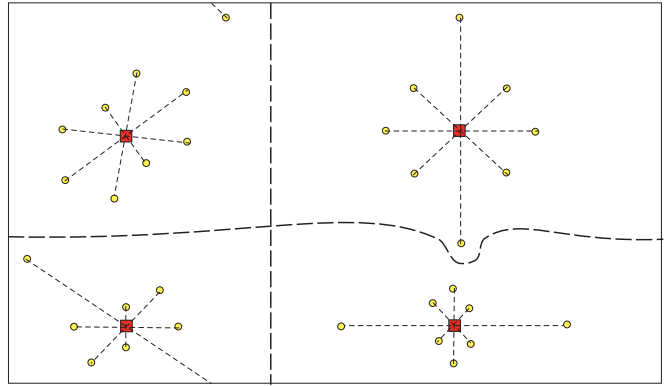}
\caption{Example of 3-chains / 3-cycles used to construct a basis of brane /bulk fluxes for the case of
$Z = K3 \times S$, with $S=T^4$ or K3. The $S$ part is suppressed in the drawing. What is shown are the corresponding 1-chains and 1-cycles in $T^2$ which is the CY orientifold double cover of the $\ICP^1 = T^2/\IZ_2$ base of the elliptically fibered K3. The upper and lower and the left and right boundaries of the rectangle are identified to form the $T^2$. The 4 red squares are the O7 planes, and the $2 \times 16$ yellow circles are the D7-branes, in brane-image-brane pairs. The dotted lines between the D7-image-D7 pairs represent the choice of 1-chains. When combined with a basis of 2-cycles in $S$, this give the 3-chain basis $\Gamma_\alpha$, which in turn determine the 4-cycles Poincar\'e dual to the brane type fluxes. The wider dashed horizontal and vertical line are the 1-cycles which when combined with 2-cycles in $S$ give 3-cycles determining the bulk type (RR and NSNS) fluxes, by fibering the $A$- resp.\ $B$-cycle of the elliptic fiber. Note that (forgetting about $S$), this construction gives 16+2+2=20 independent 20-cycles of K3. The ``missing'' two are the base $\ICP^1$ and the elliptic fiber. Since these do not wrap a single 1-cycle of the elliptic fiber, they do not give rise to suitable F-theory fluxes, as discussed in section \ref{sec:Ffluxes}.
  \label{bulkbranecycles}}
\end{figure}

We now extend these local considerations to global constructions. The basic idea is to just patch together these brane localized fluxes. Potential obstructions to this are topological in nature. To think about topological issues, the Poincar\'e dual picture is particularly useful. Consider an elliptically fibered Calabi-Yau fourfold $Z$ in the weak coupling IIB orientifold limit, and let $X$ be the associated Calabi-Yau threefold (\ref{CYeq}) doubly covering the base $B$. Let $F_2$ be a 2-form worldvolume flux class on a D7-brane wrapping a 4-cycle $S$ in $X$. Then we can associate to this a globally well defined 4-form flux on $Z$ as follows.

First, it is convenient to introduce the auxiliary space $\tilde{Z}$, which we formally construct as the elliptic fibration over $X$ instead of over $B = X/\IZ_2$, with fiber at a given point in $X$ given by the fiber at the corresponding point in $B$.

Let $\Sigma_2$ be the Poincar\'e dual 2-cycle of $F_2$ in $S$. The orientifold projects out net D5-brane charge. Therefore $\Sigma_2$ although nontrivial in homology on $S$, must be trivial in homology on $X$, that is, it must be the boundary of a 3-chain $\Gamma_3$. Now let $\tilde{\Sigma}_4$ be the 4-cycle in $\tilde{Z}$ obtained by fibering the $A$-cycle of the $T^2$ over it. Because on $X$, $A$ does not suffer monodromies anywhere, this fibration is guaranteed to be well defined, and produces a closed 4-cycle on $\tilde{Z}$. This projects to a closed 4-cycle $\Sigma_4$ in $Z = \tilde{Z}/\IZ_2$. The cohomology class of $G_4$ is defined to be the Poincar\'e dual to this 4-cycle $\Sigma_4$. See fig.\ \ref{bulkbranecycles}.

Locally near the D7, this 4-cycle looks like the local one constructed above. The 4-form flux we constructed will therefore have a part localized on the D7, given by $F_2$.

Choosing a basis $\{ \Sigma_{2,\alpha} \}_\alpha$ of the 2-form flux lattice of $S$, and corresponding 3-chains $\Gamma_\alpha$, and corresponding 4-cycles $\Sigma_{4,\alpha}$, and calling linear combinations of these the (Poincar\'e duals of) ``brane'' fluxes, we declare the lattice of ``bulk'' fluxes to be the fluxes orthogonal to all of the brane fluxes, i.e.\ $\{ G_4 | \int_{\Sigma_{4,\alpha}} G_4 = 0 \}$. So we can think of the bulk fluxes as those who have Poincar\'e duals (flux lines) ``away'' from the D7-branes. These 3-cycles are classified by ordinary 3-homology on $X$, so the bulk flux space can be thought of as being isomorphic\footnote{There is one potential subtlety, and that is that $H_3$ must vanish on the D7; otherwise $F_3$ would be ill-defined at the D7 location due to the monodromy $F_3 \to F_3 + H_3$ around it. The vanishing of $H_3$ is readily seen to be the case for the localized flux (\ref{H3eqeq}). In particular therefore the bulk $H_3$ cohomology class must vanish on the D7. In many cases, the 4-cycles wrapped by the D7 have vanishing 3-cohomology because of the Lefshetz hyperplane theorem, so this is automatically satisifed. We will assume this is the case in what follows.} to $H^3(X,\IZ)$. The construction is illustrated in fig.\ \ref{bulkbranecycles}.

Note however that the precise distinction between bulk and brane fluxes is \emph{not} canonical; it depends on the choice of 3-chains associated to the 2-form fluxes, and there is in general no canonical choice; it may be possible to loop around singularities in the deformation moduli space of $S$ such that a 3-chain does not come back to itself, but to itself plus a closed 3-cycle. For example if we turn on a brane type flux with flux lines stretching between a particular pair of D7-branes in fig.\ \ref{bulkbranecycles}, and we move the pair around a 1-cycle of the torus, then the brane flux lines will transform to the original ones, plus bulk flux lines looping around that 1-cycle.

Bearing this in mind, we can denote the bulk flux cohomology classes as $[H_3]_{\rm b}$ and $[F_3]_{\rm b}$, and write the tadpole cancelation condition (\ref{D3tadpole}) as
\begin{equation}
 - \frac{1}{\ell_s^4} \int_S \frac{1}{2} \, F_2 \wedge F_2
 + \frac{1}{\ell_s^4} \int_{X} [F_3]_{\rm b} \wedge [H_3]_{\rm b} \, + \, 2\, N_{\rm D3} = 2 \, Q_c \, ,
\end{equation}
where $Q_c$ is the curvature induced D3-charge measured on $B_6$, $Q_c = \chi(Z)/24$, and the factor of 2 appears in front of $N_{\rm D3}$ and on the right hand side because we are integrating over the double cover $X$ of $B_6$ on the left hand side.

When some D7-branes coincide, nonabelian configurations are possible, and then the first term gets replaced by the second Chern character of the holomorphic vector bundle.

It is possible to write $Q_c$ in terms of curvature induced charges on the orientifold plane and the D7-branes. The naive formula for this is
\begin{equation}
 2 \, Q_c = \frac{\chi(D7) + 4 \, \chi(O7)}{24} \, ,
\end{equation}
where $\chi(D7)$ and $\chi(O7)$ are the Euler characteristics of the 4-cycles wrapped by the D7 and the O7.
But, recalling from (\ref{O7D7eq}) and (\ref{CYeq}) that at weak coupling, the 4-cycle wrapped by the D7 is described by the equation $\eta^2 = \xi^2 \chi$ in $X$, one sees that this complex surface has double point singularities on the complex curve $\xi=\eta=0$, and additional pinch point singularities on this curve at the points where also $\chi=0$. This makes the usual notions of Euler characteristic and other topological quantities ambiguous for the D7, and more care has to be taken to define and compute these numbers. This has been analyzed in \cite{Aluffi:2007sx,CDE}.

\subsection{Enhanced gauge symmetries and charged matter} \label{sec:enhancedGS}

When D-branes coincide, one gets enhanced nonabelian gauge symmetries. For example $n$ D7-branes on $X$ coincident with an O7$^-$\footnote{The O7-planes we have encountered so far are $O7^-$-planes. O7$^+$-planes also exist, arising from a slightly different representation of the $\IZ_2$ on the string worldsheet degrees of freedom. They have positive D7-charge and $2n$ D7-branes coincident with them give rise to a ${\it USp}(2n)$ gauge group, while $n$ D3-branes give an $SO(n)$.} give rise to an $SO(n)$ gauge group, $2n$ D3-branes on an O7$^-$ give rise to ${\it USp}(2n)$, and a stack of $n$ coincident branes away from the O7 together with its orientifold image generically gives $SU(n)$. In M/F-theory, coincident D7 or more generally $(p,q)$ 7-branes correspond to a singular elliptic fibration; the massless gauge bosons are M2 branes wrapping collapsed 2-cycles. These 2-cycles can be blown up in M-theory, and the way the resulting blown up 2-cycles intersect each other can be encoded in a Dynkin diagram, which is exactly the Dynkin diagram of the enhanced gauge group. We will briefly revisit this beautiful picture in section \ref{sec:gauginoM5}.

The different possible singularities are classified according to the vanishing order of the polynomials $f$, $g$ and $\Delta=27 \, g^2 + 4 \, f^3$ with $f$ and $g$ as in (\ref{CY4eq}). The corresponding gauge groups are given in the following table \cite{Morrison:1996pp}:
\begin{center}
\begin{tabular}{|c|c|c|c|} \hline
${\rm ord}(f)$&${\rm ord}(g)$&${\rm ord}(\Delta)$
&group\\ \hline
$\ge0$&$\ge0$&$0$&none\\
$0$&$0$&$n$&$SU(n)$\\
$\ge1 $&$   1  $&$    2 $&  none\\
$  1 $&$   \ge2 $&$   3 $&$SU(2)$\\
$ \ge2 $&$  2  $&$    4 $&$SU(3)$\\
$2$&$\ge3$&$n+6$&$SO(2n+8)$\\
$\ge2$&$3$&$n+6$&$SO(2n+8)$\\
$\ge3$&$  4$  &$  8$&$   E_6$\\
$  3 $&$   \ge5 $&$   9 $&$  E_7$\\
$ \ge4$&$   5   $&$   10 $&$  E_8$ \\ \hline
\end{tabular}
\end{center}
More precisely, the above table holds under the assumption that no monodromies act on the collapsing 2-cycles; if such monodromies do occur, the classifications is more complicated and also includes $SO(2n+1)$, ${\it USp}$, $F_4$ and $G_2$ gauge groups.

Note in particular that more gauge groups are possible in the general F-theory setup than in type IIB at weak coupling. (For some obscenely huge gauge groups, with many exceptional group factors and ranks up to 121328, see \cite{Candelas:1997eh}.)

Massless charged matter on the other hand arises when two stacks of D-branes intersect. More generally, in F-theory, it is associated to singularity enhancement along the singular locus of the elliptic fibration.

The approach we are following here is a little too crude to properly analyze enhanced gauge symmetries and charged matter content, and we will therefore usually assume in what follows that we are at some point in the complex structure moduli space without enhanced gauge symmetry. More information can be found for example in \cite{Morrison:1996pp,Morrison:1996na,Bershadsky:1996nh,Donagi:2008ca,beasley}.

\section{Type IIB / F-theory flux vacua} \label{sec:IIB}

\subsection{Moduli}

We are now in the position to determine the four dimensional low energy effective field theory corresponding to F-theory compactified on a Calabi-Yau fourfold $Z$ ellipticaly fibered over
a three complex dimensional base manifold $B$, or equivalently type IIB on $B$ containing 7-branes, which in the weak coupling orientifold limit can be thought of as the $\IZ_2$ orientifold quotient of type IIB on a Calabi-Yau threefold $X$ with O7 and D7 branes. More generally, we could also have O3-planes --- these correspond to codimension eight $\IZ_2$ singularities in F-theory. We also consider space-filling D3-branes.

The following table shows the massless moduli we have before turning on fluxes, their M/F-theory and weakly coupled type IIB orientifold interpretations and the Hodge numbers counting them:
\vskip5mm
\begin{tabular}[h]{llll}
 M/F-theory & \# real moduli & IIB orientifold & \# real moduli \\
 \hline
 K\"ahler & $h^{1,1}(Z)-1$ & K\"ahler & $h^{1,1}_+(X)$ \\ \\
 && Complex structure & $2 \, h^{2,1}_-(X)$ \\
 Complex structure & $2 \, h^{3,1}(Z)$ & D7 deformations & $2 \, \hat{h}^{2,0}_-(S)$ \\
 && Dilaton-axion & 1 \\ \\
 $C_6$ axions & $h^{1,1}(Z)-1$ & $C_4$ axions & $h^{1,1}_+(X)$ \\ \\
 $C_3$ axions & $2 \, h^{2,1}(Z)$ & $B_2$, $C_2$ axions & $h^{1,1}_-(X) + h^{1,1}_-(X)$ \\ \\
 M2 positions & $6 \, N_{\rm D3}$ & D3 positions & $6 \, N_{\rm D3}$
\end{tabular}
\vskip5mm
The subscripts $\pm$ denote the Hodge numbers counting the even resp.\ odd parts of the relevant cohomology under the geometrical orientifold involution \cite{Brunner:2003zm}.\footnote{The hat on $\hat{h}^{2,0}_-(S)$ is there to indicate subtleties in the definition of this number due to the singularities of $S$ \cite{CDE}.} For simplicity we will assume that $h^{2,1}(Z)=0$. This implies in particular that there are no $C_3$ axions in the M-theory picture, and no $B_2$/$C_2$ axions in the IIB picture, nor $U(1)$ vectors from the reduction of $C_4$.
Many such $h^{2,1}(Z)=0$ examples are known \cite{Klemm:1996ts}. In any case, axions are never really a problem, since they are not control parameters, cannot destabilize compactifications and will generically get lifted as soon as supersymmetry is broken.

\subsection{Low energy effective action in F-theory framework} \label{sec:Fleeee}

The four dimensional low energy effective theory is $\CN=1$ supergravity, with zero potential classically and to all orders in perturbation theory. The $\CN=1$ supersymmetry constraints imply that the moduli parametrize a K\"ahler manifold. The complex coordinates in the M-theory representation of F-theory are the complex structure moduli $z^a$, $a=1,\ldots,h^{3,1}(Z)$ (which can be thought of as the coefficients of the defining equation, modulo coordinate redefinitions, if the Calabi-Yau is algebraic), the D3 moduli $y_i^m$, $m=1,2,3$, $i=1,\ldots,N_{\rm D3}$, and the complexified K\"ahler moduli
\begin{equation}
 T_A = \frac{1}{\ell_M^6} \int_{D_{6,A}} C_6 + i \, dV \, .
\end{equation}
Here $\{ D_{6,A} \}$ is a basis of 6-cycles (divisors) in $Z$ wrapping the $T^2$ fiber, and $dV$ is the volume element of the 6-cycle. As always, we take the F-theory limit of vanishing fiber area $L^2=v=\int_{T^2} dV$. According to the general F-theory - type IIB reduction scheme of section \ref{sec:whatisF}, using in particular the relation $L/\ell_M^3 = 1/\ell_s^2$, we can also write this in IIB language as
\begin{equation} \label{Kahlermoduli}
 T_A = \frac{1}{\ell_s^4} \int_{D_{4,A}} C_4 + i \, dV \, ,
\end{equation}
where now $\{ D_{4,A} \}$ is the corresponding basis of 4-cycles (divisors) in the base $B$ (or, if we add a prefactor $\frac{1}{2}$, in $X$).

In general the D3 moduli and K\"ahler moduli mix in a rather intricate way in the K\"ahler potential. To avoid this complication, we will assume there are no D3-branes present for now. See \cite{DeWolfe:2002nn,Grana:2003ek,Grimm:2004uq,Giddings:2005ff} for the effective K\"ahler potential and action for D3-branes, and \cite{Baumann:2006th} for concretely applied examples.

The classical K\"ahler potential then splits in a K\"ahler part and a complex structure part:
\begin{equation}
 \CK = \CK_K(T,\bar T) + \CK_c(z,\bar z) \, .
\end{equation}

The K\"ahler part is determined by the volume:
\begin{eqnarray}
 \CK_K &=& -2 \ln \biggl( \frac{1}{\ell_M^8} \int_Z dV \biggr)
 = -2 \ln \biggl( \frac{1}{\ell_s^6} \int_B dV \biggr)
 = -2 \ln V(B) \\ &=& -2 \ln \biggl( \frac{1}{6} \int_B J^3 \biggr) = -2 \ln \biggl( \frac{1}{6} D_{ABC} J^A J^B J^C \biggr) \, . \label{KKpot}
\end{eqnarray}
Here $V(B)$ is the volume of $B$ in string units, and the $J^A$ are the components of the K\"ahler form: $J = J^A \, D_{4,A}$, where, slightly abusively, we used the same notation for the 4-cycle $D_{4,A}$ (above) and its Poincar\'e dual (here). The coefficients $D_{ABC}$ are the triple intersection numbers of these divisors:
\begin{equation}
 D_{ABC} := \#(D_A \cap D_B \cap D_C) = \int_B D_A \wedge D_B \wedge D_C \, .
\end{equation}
The $J^A$ are related to the $T_A$ by
\begin{equation}
 \Im T_A = \partial_{J^A} V(B) =  \frac{1}{2} D_{ABC} J^B J^C \, .
\end{equation}
Inverting $J^A(T,\bar T)$ may or may not be possible explicitly, depending on the model. Note that $V(B) = V(X)/2$.

The complex structure part is
\begin{eqnarray}
 K_c &=&  - \ln \int_Z \Omega_4 \wedge \overline{\Omega_4} \\
 &=&  - \ln \biggl( \Pi_I(z) \, Q^{IJ} \, \overline{\Pi_J(z)} \biggr) \, .
\end{eqnarray}
Here $\Omega_4$ is the unique holomorphic 4-form on $Z$, the $\Pi_I$ are its periods:
\begin{equation}
 \Pi_I(z) := \int_{\Sigma_{4,I}} \Omega_4(z) \, ,
\end{equation}
with $\{ \Sigma_{4,I} \}_I$, $I=1,\ldots,b'_4(Z)$, a basis of 4-cycles wrapping a 1-cycle in the elliptic fiber. Finally, $Q^{IJ}$ is the inverse of $Q_{IJ}$, and $Q_{IJ}$ is the intersection form of the basis:
\begin{equation}
 Q_{IJ} := \Sigma_I \cdot \Sigma_J = \# (\Sigma_{4,I} \cap \Sigma_{4,J}) = \int_Z \Sigma_{4,I} \wedge \Sigma_{4,J} \, ,
\end{equation}
where again we used the same notation for cycle and Poincar\'e dual form.

Since we have assumed $h^{2,1}(Z)=0$, there are no further massless fields in four dimensions besides the metric tensor, so our specification of the low energy effective action is complete.

Let us consider a very simple toy model as illustration for the complex structure moduli sector. The model can morally be thought of as $Z= X \times T^2$ with $X$ a rigid Calabi-Yau threefold (i.e.\ $X$ has no complex structure moduli).
Thus, $Z$ has a single complex structure modulus $\tau$, the modular parameter of the $T^2$. The moduli space of the model is the fundamental domain in the upper half $\tau$-plane. A rigid Calabi-Yau has two independent 3-cycles, so we can make four 4-cycles by combining these with the $A$ and $B$ 1-cycles in the $T^2$. The nonvanishing periods of $\Omega_4$ are then
\begin{equation} \label{fourPi}
 \Pi_I = (1,\omega,\tau,\omega \tau) \, ,
\end{equation}
where $\omega$ is some complex number depending on $X$ and our choice of 3-cycles. For simplicity we just put $\omega \equiv i$. If the 3-cycles have intersection product 1, the intersection form for the 4-cycles is
\begin{equation} \label{toyQ}
 Q_{IJ} = \left( \!\! \begin{array}{rrrr} 0 & 0 & 0 & -1 \\
                                     0 & 0 & 1 & 0 \\
                                     0 & 1 & 0 & 0 \\
                                     -1 & 0 & 0 & 0 \\ \end{array} \!\! \right) \, .
\end{equation}
The K\"ahler potential on the complex structure moduli space is thus
\begin{equation} \label{toyKc}
 \CK_c = - \ln \, \Pi_I(z) \, Q^{IJ} \, \overline{\Pi_J(z)} = - \ln(4 \, \Im \tau) \, ,
\end{equation}
and its complex structure moduli space metric
\begin{equation} \label{toymetric}
 g_{\tau \bar\tau} = \partial_\tau \bar{\partial}_{\bar \tau} \CK_c = \frac{|d \tau|^2}{4(\Im \tau)^2} \, ,
\end{equation}
the standard Poincar\'e metric on the upper half plane.

\subsection{Low energy effective action in IIB weak coupling limit} \label{sec:leeaweak}

In the type IIB weak coupling orientifold limit, we can reproduce the structure of the low energy effective action expected from the perturbative string picture as follows. First recall that at the end of section \ref{sec:weakflux}, we introduced a (formal) elliptic fibration $\tilde{Z}$ over the CY 3-fold $X$, the varying field $\tau$ on $X$ being the modulus of the elliptic fiber. The space $\tilde{Z}$ can be thought of as a double cover of $Z$ in the orientifold limit. In particular in this limit (ignoring $e^{-\pi/g_s}$ corrections) we have
\begin{equation} \label{ZZtilde}
 \CK_c = -\ln \int_Z \Omega \wedge \bar{\Omega}  = -\ln \, \frac{1}{2} \int_{\tilde{Z}} \Omega \wedge \bar{\Omega} \, .
\end{equation}
We also saw there that we can define bulk and brane 4-cycles on $\tilde{Z}$ (or $Z$), the brane 4-cycles being $A$-cycle fibrations over 3-chains ending on the D7 locus $S$ in $X$, and the bulk cycles being those with zero intersection product with those, which are $A$- or $B$-cycle fibrations over 3-cycles in $X$.  Let us denote the chosen basis for the 3-chains by $\{ \Gamma_\alpha \}_\alpha$, $\alpha = 1,\ldots,\hat{b}^2_-(S)$, and for the bulk 3-cycles by $\{ \Sigma_i \}_i$, $i=1,\ldots,b^3(X)$. Denote the 4-cycles in $\tilde{Z}$ obtained by fibering the $A$-cycle over $\Sigma_i$ by $\Sigma_i \times A$, those obtained by fibering the $B$-cycle over $\Sigma_i$ by $\Sigma_i \times B$, and those obtained by fibering the $A$-cycle over $\Gamma_\alpha$ by $\Gamma_\alpha \times A$. Then the corresponding periods are, using (\ref{Om3def}) and (\ref{Bcycleint}), and denoting 3-fold complex structure moduli by $\psi$ and D7 moduli by $\phi$, up to $e^{-\pi/g_s}$ corrections:
\begin{eqnarray}
 \int_{\Sigma_i \times A} \Omega_4 &=& \int_{\Sigma_i} \Omega_3(\psi)  =: \Pi_i(\psi) \, , \label{Pidefinition}\\
 \int_{\Sigma_i \times B} \Omega_4 &=& \int_{\Sigma_i} \tau \, \Omega_3(\psi) = \int_{\Sigma_i}
 \left( \tau_0 + \frac{i}{2 \pi} \ln \frac{P_{O7}(\psi)}{P_{D7}(\psi,\phi)} \right) \, \Omega_3(\psi) \nonumber \\
 & =: & \tau_0 \, \Pi_i(\psi) \, + \chi_i(\psi,\phi) \, , \label{logint} \\
 \int_{\Gamma_\alpha \times A} \Omega_4 &=& \int_{\Gamma_\alpha(\phi)} \Omega_3(\psi) =: \Pi_\alpha(\psi,\phi) \, , \label{Pidefinition2}
\end{eqnarray}
where $\Omega_3$ is the holomorphic 3-form on $X$ and $\tau_0 = i/g_s$ is the ``bulk'' value of $\tau$ as defined in (\ref{tau0val}). The dependence of the various terms on the threefold complex structure moduli $\psi$ and the D7 moduli $\phi$ (up to $e^{-\pi/g_s}$ corrections) is indicated.

Furthermore, the intersection form $Q_{IJ}$ splits in bulk and brane blocks. The nonzero entries are
\begin{eqnarray}
 Q_{\alpha \beta} &:=& (\Gamma_\alpha \times A) \cdot (\Gamma_\beta \times A)
 \, = \, - (\partial \Gamma_\alpha) \cdot (\partial \Gamma_\beta) \, |_S \, , \\
 Q_{ij} &:=& (\Sigma_i \times A) \cdot (\Sigma_j \times B) \, = \, - (\Sigma_i \cdot \Sigma_j) \, |_X \, .
\end{eqnarray}
Thus using (\ref{ZZtilde}) we can write, up to nonperturbative $e^{- \pi/g_s}$ corrections:
\begin{equation}
 \CK_c = - \ln \frac{1}{2}\biggl( \, (\tau_0 -\bar{\tau}_0) \, \Pi_i \, Q^{ij} \, \overline{\Pi}_j
 + \, \chi_i \, Q^{ij} \, \overline{\Pi}_j + \overline{\chi}_i \, Q^{ij} \, \Pi_j
  \, - \, \Pi_\alpha \, Q^{\alpha\beta} \, \overline{\Pi}_{\beta} \, \biggr)
\end{equation}
In a perturbative $g_s=1/\Im \tau_0$ expansion, this becomes
\begin{eqnarray}
 \CK_c = \CK_{\tau_0} + \CK_{X}(\psi) + g_s \, \CK_{\rm D7}(\psi,\phi) + \cdots
\end{eqnarray}
where
\begin{eqnarray}
 \CK_\tau(\tau_0)(\tau_0) &=& - \ln(\Im \tau_0) \\
 \CK_{X}(\psi) &=& - \ln \left( i \, \Pi_i \, Q^{ij} \, \overline{\Pi}_j \right) = - \ln i \int_X \Omega_3 \wedge \overline{\Omega_3} \\
 \CK_{\rm D7}(\psi,\phi) &=& \mbox{$\frac{1}{2}$} e^{\CK_{X}} \left(
  \Pi_\alpha \, Q^{\alpha\beta} \, \overline{\Pi}_{\beta}
  + \, \chi_i \, Q^{ij} \, \overline{\Pi}_j + \overline{\chi}_i \, Q^{ij} \, \Pi_j
  \right)
  \, .
 \end{eqnarray}
The first two parts of the K\"ahler potential are the standard dilaton and complex structure K\"ahler potentials one gets by direct reduction of type IIB on $X$. The third part governs the D7-brane moduli. Note that it enters at order $g_s = 1/\Im \tau_0$ compared to the bulk moduli part; it is nevertheless the first nontrivial order at which the D7 degrees of freedom $\phi$ appear. This means the backreaction of the D7-branes on the bulk geometry is suppressed by a power of $g_s$, as it should. At fixed $\psi$, the $\chi_i$-dependent part of $\CK_{\rm D7}$ is merely a K\"ahler gauge transformation, and therefore does not contribute to the D7 moduli space metric to leading order:
\begin{equation}
 g_{r\bar{s}} = \partial_{\phi^r} \bar{\partial}_{{\bar \phi}^s} \CK_{\rm D7} =
 \mbox{$\frac{1}{2}$} e^{\CK_{X}}
  \partial_r \Pi_\alpha \, Q^{\alpha\beta} \, \bar{\partial}_{\bar s} \overline{\Pi}_{\beta} =
 \mbox{$\frac{1}{2}$} e^{\CK_{X}} \int_S \omega_r \wedge \overline{\omega}_{\bar{s}} \, .
\end{equation}
Here $\omega_r := (\Omega_3 \cdot \delta_r n)|_S$, with $\delta_r n$ the holomorphic deformation vector field normal to $S$ corresponding to a variation $\delta \phi^r$ of the moduli of $S$, and ``$\cdot$'' denotes index contraction. The forms $\omega_r$ are holomorphic $(2,0)$ forms on $S$. Notice that the D7 moduli metric does not depend on the choice of 3-chains; only the K\"ahler potential does.

All of this fits well with what we expect from the perturbative string point of view.

The geometrical structures underlying open-closed string moduli spaces were explored in \cite{Lerche:2002yw}.

\subsection{The effect of turning on fluxes} \label{sec:turningonflux}

We now consider the effect of turning on F-theory $G_4$ flux.

\subsubsection{Effective potential}

We will first work in the Kaluza-Klein approximation, i.e.\ the M-theory metric remains $ds^2 = -(dx^0)^2 + (dx^1)^2 + (dx^2)^2 + ds_Z^2$, where $ds_Z^2$ is an unwarped Ricci flat Calabi-Yau metric, and the flux is harmonic. The flux is quantized as
\begin{equation}
 \ell_M^{-3} \, [G_4] = N^I \, \Sigma_{4,I} \, ,
\end{equation}
where $\{\Sigma_I\}$ is a basis of integral 4-form cohomology classes, and $N^I$ is integral modulo a possible half-integral shift equal to $\frac{c_2^I(Z)}{2}$ \cite{Witten:1996md}. The energy density in $\IR^{1,2}$ due to the flux is
\begin{equation}
 V_M(G) = \frac{2\pi}{\ell_M^9} \, \frac{1}{2} \, \int_Z G_4 \wedge *G_4 \, ,
\end{equation}
where $*$ is the Hodge star on $Z$. If there were no negative energy contributions to the potential, we would be squarely in the no-go situation described in section \ref{sec:nogo}. We already know from (\ref{Maction}) and more explicitly from (\ref{tadpolecancelation}) that the curvature of $Z$ provides negative M2 \emph{charge} $-Q_c = - \chi(Z)/24$. Since consistent Minkowski solutions exist with $Q_c$ space-filling M2-branes canceling this charge, and $Q_c$ space-filling M2-branes have an energy density equal to $\frac{2\pi Q_c}{\ell_M^3}$, this implies there must be additional higher order curvature terms in the action providing an energy density exactly equal to minus this. This is indeed the case \cite{Haack:2001jz}. Thus, assuming (\ref{tadpolecancelation}) is satisfied, we find for the total potential including contributions from M2-branes, curvature and flux:
\begin{equation}
 V_M = \frac{2\pi}{\ell_M^9} \, \frac{1}{2} \, \int_Z (G_4 \wedge *G_4 - G_4 \wedge G_4) \, ,
\end{equation}
Splitting $G_4$ in its self-dual and anti-self-dual part, $G_4=G_{4,+} + G_{4,-}$, this becomes
\begin{equation}
 V_M = \frac{2\pi}{\ell_M^9} \, \int_Z G_{4,-} \wedge * G_{4,-} \, .
\end{equation}
From the general scheme of section \ref{sec:whatisF}, it follows that the corresponding energy density in type IIB is
\begin{equation} \label{IIBpot}
 V_{\rm IIB} = \frac{2 \pi}{\ell_s^4} \, \frac{1}{\ell_M^6} \int_Z G_{4,-} \wedge *_Z G_{4,-} =
 \frac{2 \pi}{\ell_s^8} \int_B \frac{1}{\Im \tau} \, G_{3,-} \wedge *_B \overline{G_{3,-}} \,  ,
\end{equation}
where $G_{3,-}$ is the imaginary anti-self-dual part of $G_3 = F_3 - \tau H_3$, i.e.\ $*G_{3,-} = - i G_{3,-}$. (The $i$ must be there because $*^2=-1$ on 3-forms in $B$.)

Whether in IIB or in M-theory, after a Weyl rescaling to bring the 3d/4d Einstein-Hilbert term in canonical form, the potential gets an additional prefactor proportional to an inverse power of the volume, as in (\ref{fluxpot}). Therefore to avoid a runaway, we must have $G_{4,-} = 0$ (equivalently $G_{3,-}=0$) identically, i.e.\
\begin{equation} \label{G4selfdual}
 G_4 = *_Z G_4  \qquad (\mbox{equivalently } G_3 = i *_B G_3)\, .
\end{equation}
This puts constraints on the complex structure and K\"ahler moduli of $Z$. To make this explicit, we need some results about Hodge decompositions, which we develop in the following intermezzo. (This can be skipped by the reader not interested in the general framework.)

\subsubsection{Intermezzo: Lefshetz $SU(2)$ and diagonalizing the Hodge star operator}

The vector space of harmonic forms on a K\"ahler manifold of complex
dimension $n$ can be organized according to representations of the
\emph{Lefshetz $SU(2)$ algebra}. Up to normalization, the raising
operator $L_+=L_1+i L_2$ is given by wedging with the K\"ahler form
$J$, the lowering operator $L_-$ by contraction with $J$, and the
$L_3$ operator is the form degree up to a constant shift. For a
harmonic $p$-form $\omega$:
\begin{equation}
L_3 \, \omega = \frac{p - n}{2} \, \omega, \qquad L_+ \, \omega
\sim J \wedge \omega, \qquad L_- = L_+^\dagger,
\end{equation}
where the adjoint is taken with respect to the inner product on
forms defined by the Hodge *-operator. Although in general it is not true that wedging two harmonic forms together produces a new harmonic form, it is true that wedging a harmonic form with the K\"ahler form gives again a harmonic form, so the above operations are well defined on the space of harmonic forms.
One checks that
$L_+ \, * = * \, L_-$, $L_- \, * = * \, L_+$, and $L_3 \, * = - * \,
L_3$, so in particular $[{\bf L}^2,*] = 0$, and we can
simultaneously diagonalize the Lefshetz spin $\ell$ and the Hodge *.
Explicitly, on a K\"ahler manifold of complex dimension $n$, one has for
a spin $\ell$ harmonic $(n-k,k)$-form $\omega$:
\begin{equation} \label{staraction}
 * \, \omega = (-1)^{k+\ell} \, \omega
 \mbox{ ~~($n$ even)}, \qquad * \, \omega = (-1)^{k+\ell} (-i) \,
 \omega \mbox{ ~~ ($n$ odd)}.
\end{equation}
For example, for $n$ even, the $(n/2,n/2)$ form $J^{n/2}$ comes in a
spin $\ell=n/2$ multiplet $(1,J,J^2,\ldots,J^{n})$ and is self-dual.

A $p$-form has at most spin $\ell=(n-p)/2$. A \emph{primitive}
$p$-form is a form with spin exactly equal to this. For the middle
cohomology this means that it has spin zero, or $\omega \wedge J=0$.
Thus for a 2-fold for example, primitive (1,1)-forms are anti-self-dual, while
for a 4-fold, primitive (2,2)-forms are self-dual.

For an $SU(4)$ holonomy Calabi-Yau 4-fold, we thus get the decompositions
\begin{eqnarray}
 H^4_{+} &=& H^{0,4}_{\ell=0} \, \oplus \, H^{2,2}_{\ell=0} \, \oplus \, H^{2,2}_{\ell=2}
 \, \oplus \, H^{4,0}_{\ell=0} \, , \\
 H^4_{-} &=& H^{1,3}_{\ell=0} \, \oplus H^{2,2}_{\ell=1} \, \oplus H^{3,1}_{\ell=0} \,.
\end{eqnarray}
There is a unique $\ell=2$ multiplet given by
$(1,J,J^2,J^3,J^4)$, and there are $h^{1,1}-1$ independent $\ell=1$
multiplets $(\omega_k,\omega_k J,\omega_k J^2)$,
where $\{ \omega_k \}_k$ is a set of $h^{1,1}-1$ independent
(1,1)-forms such that $\omega_k J^3=0$.

It is worth pointing out that for harmonic forms, equations in cohomology are equivalent to pointwise equations for the differential forms. In particular if say $J \wedge \omega = 0$ in cohomology, it is zero pointwise. This is because there is always a unique harmonic representative of a cohomology class.

\subsubsection{Superpotential formulation} \label{sec:superpotform}

%

From the intermezzo we take that the self-duality condition $G_4 = * G_4$ is equivalent to
\begin{equation} \label{hodgeconditions}
 G_4^{1,3} = G_4^{3,1} = 0 \, , \qquad G_4^{2,2}|_{\ell=1} = 0 \, .
\end{equation}
A basis of $H^{3,1}(Z)$ is provided\footnote{A variation of the complex structure can be thought of as a variation of complex coordinates $\delta y^m = \epsilon f^m(y,\bar{y})$ on the CY fourfold $Z$, where $f$ is a nonholomorphic function. The resulting variation $\delta \Omega_4$ of the holomorphic $(4,0)$-form will thus be of type $(4,0) + (3,1)$, with nonzero $(3,1)$ part. Hence $\partial_a \Omega$ is of type $(4,0) + (3,1)$. It is easily checked that going to the K\"ahler covariant derivative $D_a \Omega$ subtracts off precisely the $(4,0)$ part. So the $D_a \Omega$ form a set of linearly independent $(3,1)$ forms. The number of these equals the number of complex structure deformations, which equals $h^{3,1}$ (see e.g.\ \cite{Greene:1996cy} for the analogous case of a CY 3-fold). Thus, the $D_a \Omega$ form a basis of $H^{3,1}$.} by the covariant derivatives with respect to the complex structure moduli of $Z$:
\begin{equation}
 D_a \Omega_4 := (\partial_a + \partial_a \CK_c) \Omega_4 \, .
\end{equation}
Therefore, the first condition is (\ref{hodgeconditions}) is equivalent to $G_4$ being orthogonal to all $D_a \Omega_4$, i.e.:
\begin{equation} \label{GVWcond}
 D_a W(z) = 0 \, , \qquad W(z) := \frac{1}{\ell_M^3} \int_Z G_4 \wedge \Omega \, .
\end{equation}
We recognize this as a superpotential condition. The superpotential $W(z)$ appearing here lives on the complex structure moduli space of $Z$, and was first derived by Gukov, Vafa and Witten, in \cite{Gukov:1999ya}. We included the factor $\ell_M^{-3}$ to make $W(z)$ dimensionless.

This formulation makes it manifest that turning on flux constrains the complex structure moduli. In fact, for sufficiently generic $W(z)$, one expects isolated critical points, suggesting that \emph{all} complex structure moduli can be stabilized in this way. (This is plausible but not completely obvious in the case at hand, because the fluxes are quantized and constrained by the tadpole cancelation condition (\ref{tadpolecancelation}).)

Since the $\ell=1$ part of $H^{2,2}(Z)$, by definition, consists of the $(2,2)$ forms which are not annihilated by $J$, minus the $(2,2)$ forms proportional to $J \wedge J$, the second condition in (\ref{hodgeconditions}) can be written as
\begin{equation} \label{Dtermconstr}
 G_4 \wedge J = c \, J \wedge J \wedge J \, ,
\end{equation}
for some constant $c$, which can be computed as $c=(\int_Z G_4 \wedge J^2)/(\int_Z J^4)$.
This can again be written in a superpotential-like form:
\begin{equation} \label{tildeWeq}
 D_J \tilde{W}(J) = 0 \, \qquad \tilde{W}(J) := \frac{1}{\ell_M^3} \int_Z G_4 \wedge J \wedge J \, ,
\end{equation}
where we introduced another covariant derivative
\begin{equation}
 D_J \tilde{W} := \partial_J \tilde W + (\partial_J \CK_J) \tilde W \, \qquad
 \CK_J:=-\ln \int_Z \frac{J^4}{4!} \, .
\end{equation}
However, being a real function, this does not have an actual superpotential interpretation in four dimensions. Instead, (\ref{Dtermconstr}) should be interpreted in four dimensions as a D-term constraint.

Actually, given the form (\ref{fluxredFIIB}) of the fluxes $G_4$ which preserve four dimensional Poincar\'e invariance in the IIB description, (\ref{Dtermconstr}) is automatically satisfied for harmonic $G_4$ on smooth, full $SU(4)$ holonomy Calabi-Yau fourfolds $Z$. To see this, first note that if $Z$ has $SU(4)$ holonomy,\footnote{This in contrast to for example $Z=K3 \times K3$ and indeed in this case (\ref{Dtermconstr}) is not automatically satisfied.} $H^2(Z)=H^{1,1}(Z)$, so all 2-cohomology classes have Poincar\'e dual representatives which are divisors (linear combinations of holomorphic 6-cycles). Let $\{D_M\}_M$ be a basis of $H^{1,1}(Z)$ or equivalently of divisors in $Z$. Then we claim that $G_4 \wedge D_A$ is zero in cohomology for all $D_M$, i.e.\
\begin{equation} \label{restrictedflux}
 \int_Z G_4 \wedge D_M \wedge D_N = 0  \qquad \forall M,N \, ,
\end{equation}
or equivalently by going to the Poincar\'e dual representation $\int_{D_M \cap D_N} G_4 = 0$. To prove this, note that for smooth $Z$ at least, all divisors in $Z$ with the exception of the base $B$ itself (more precisely the section of the elliptic fibration) can be thought of as elliptic fibrations over divisors in the base. Hence intersections of divisors $D_M \cap D_N$ are linear combinations of divisors in $B$ and elliptic fibrations over holomorphic curves in $B$; in particular they wrap the elliptic fiber either completely, or not at all. Fluxes of the form (\ref{fluxredFIIB}) integrate to zero on such surfaces, since there are no components with two legs on the elliptic fiber or with no legs on the elliptic fiber at all. This shows that $G_4 \wedge D_M$ is zero in cohomology, so in particular $G_4 \wedge J$ is zero in cohomology. Since we take $G_4$ to be the harmonic representative in its cohomology class, $G_4 \wedge J$ is harmonic too, and must therefore be zero pointwise. Thus, as claimed,
\begin{equation} \label{autoprimitive}
 G_4 \wedge J = 0
\end{equation}
and (\ref{Dtermconstr}) is automatically satisfied.

This should not surprise us, since for a smooth, full $SU(4)$ holonomy CY $Z$, there are no suitable massless $U(1)$ vectors in the four dimensional effective theory which could generate a D-term.\footnote{Again, this is not true for reduced holonomy CY manifolds such as $K3 \times K3$, where $U(1)$s generating D-terms do survive; in the weak coupling limit, they correspond to the relative $U(1)$s of (disjoint) D7-image-D7 pairs. For smooth, genuine $SU(4)$ holonomy CYs on the other hand, D7-image-D7 pairs get recombined into single branes, and the $U(1)$ is broken.} Note that this need not be the case when $Z$ is singular, so in those cases we might still have a D-term condition to take into account. In particular this will be the case when there is enhanced gauge symmetry or when there are intersecting 7-branes. Our approach has been somewhat too crude to properly deal with singularities however, so we will continue to operate under the assumption of smoothness for now.

This leaves us with (\ref{GVWcond}), and when this is satisfied, the effective potential for the remaining moduli, in particular all the K\"ahler moduli, is flat.

Finally, note that using $G^{2,2}|_{\ell=1} = 0$ and expanding $G^{3,1}$ in the basis $D_a \Omega$, (\ref{IIBpot}) can also be written as
\begin{equation} \label{IIBpot2}
 V_{\rm IIB} = \frac{m_p^4}{4\pi} \, e^{\CK_c+\CK_K} g^{a \bar{b}} D_a W \overline{D_b W} \, ,
\end{equation}
where $m_p$ is the 4d Planck mass defined such that the coefficient of the Einstein-Hilbert term in the 4d action is $\frac{m_p^2}{2}$. This is reminiscent of the standard $\CN=1$ formula for the potential in terms of a superpotential, but seems to be missing a term proportional to $-3 |W|^2$. This term is indeed there when \emph{all} moduli are taken into account in the standard formula, but happens to cancel out exactly against the $|DW|^2$ part generated by the moduli different from the complex structure moduli, leaving (\ref{IIBpot2}) behind.

Because nothing sets the scale of the internal manifold, these compactifications are called ``no scale'' compactifications. The good thing about them is that this allows us to go to the large radius regime where all of our approximations are justified. The bad thing is that we will need to invoke quantum corrections again to lift this degeneracy.

As an illustration, we return to the toy model introduced at the end of section \ref{sec:Fleeee}. If we turn on flux with flux quanta $N^I=(A_1,A_2,B_1,B_2) \in \IZ^4$, the superpotential becomes
\begin{equation}
 W(\tau) = A + B \tau \, , \qquad A:=A_1 + i A_2, \quad B:=B_1+iB_2 \, .
\end{equation}
The tadpole cancelation conditon (\ref{tadpolecancelation}) becomes, using \ref{toyQ},
\begin{equation}
 \Im(\bar B A) + N_{\rm D3} = Q_c \, .
\end{equation}
Strictly speaking $Q_c = 0$ if $Z = T^2 \times X$, but for the sake of the toy model we will take it to be some arbitrary given number. Using (\ref{toyKc}), we find
\begin{equation}
 D_{\tau} W := (\partial_{\tau} + \partial_{\tau} \CK_c) W = \frac{A+B \bar{\tau}}{\bar{\tau}-\tau} \, .
\end{equation}
Hence $V_{\rm IIB} \sim |A+B \bar \tau|^2$. Solving $DW=0$ for $\tau$ gives
\begin{equation}
 \tau = - \frac{\bar{A}}{\bar{B}} \, .
\end{equation}
Note that although this naively looks like an infinite number of vacua, we should not count vacua related by $SL(2,\IZ)$ transformations separately. To avoid overcounting, we could for example restrict the solutions $\tau$ to the fundamental domain. Fig.\ \ref{T2vacua} in section \ref{sec:toystat} shows the set of vacua for $Q_c = 150$.

You can find more simple, explicit examples of (bulk) flux vacua in \cite{Kachru:2002he,Giryavets:2003vd}. In particular, in the latter reference some flux vacua for the example (\ref{CY4eq}) are constructed. Needless to say though, systematically finding fully explicit examples --- let alone enumerating them --- for compactifications with many moduli and fluxes becomes effectively intractable. Fortunately it is not necessary to construct things explicitly to find approximate distributions of flux vacua over parameter space. We will get to this in section \ref{sec:statistics}.

\subsubsection{Weak coupling limit} \label{weakcouplinglimitflux}

In the IIB weak coupling limit, we can expand the flux in the basis introduced in section \ref{sec:leeaweak}:
\begin{equation}
 \frac{1}{\ell_M^3} [G_4] =  \sum_i N^i \, [\Sigma_i \times A]
 - \sum_i M^i \, [\Sigma_i \times B] + \sum_\alpha N^\alpha \, [\Gamma_{\alpha} \times A] \, .
\end{equation}
In terms of the weak coupling type IIB bulk and brane fluxes introduced in section \ref{sec:weakflux}, this is
\begin{equation}
 \frac{1}{\ell_s^2} [F_3]_{\rm b} = \sum_i N^i \, [\Sigma_i] \, , \quad
 \frac{1}{\ell_s^2} [H_3]_{\rm b} = \sum_i M^i \, [\Sigma_i] \, , \quad
 \frac{1}{\ell_s^2} [F_2] = \sum_\alpha N^\alpha \, [\partial \Gamma_{\alpha}] \, .
\end{equation}
The corresponding Gukov-Vafa-Witten superpotential is then, up to $e^{-\pi/g_s}$ corrections:
\begin{eqnarray}
 W(\tau_0,\psi,\phi) &=& \sum_i (N^i - \tau_0 M^i) \Pi_i(\psi)  \\
 && - \sum_i M^i \chi_i(\psi,\phi) + \sum_\alpha N^\alpha \Pi_\alpha(\psi,\phi) \, \\
 &=:& W_{\rm b}(\tau_0,\psi) + W_{\rm D7}(\psi,\phi) \, ,
\end{eqnarray}
with $\Pi_i$, $\chi_i$ and $\Pi_\alpha$ defined in (\ref{Pidefinition})-(\ref{Pidefinition2}).

For suitable\footnote{The choice of representative matters for $H_3$, since the logarithmic branch cut in the integrand of the first term of $W_{\rm D7}$ makes the integral not invariant under $H_3 \to H_3 + d \beta_2$.} closed 3-forms $F_3$ and $H_3$ representing $[F_3]_{\rm b}$ resp.\ $[H_3]_{\rm b}$, this can also be written as
\begin{eqnarray}
  W_{\rm b} &=& \int_X \left(F_3-\tau_0 H_3 \right) \wedge \Omega_3 \\
  W_{\rm D7} &=& - \frac{i}{2 \pi} \int_X H_3 \wedge \ln \frac{P_{O7}}{P_{D7}} \, \Omega_3 \, \,  + \, \, \int_{\Gamma(F_2)} \Omega_3 \, ,
\end{eqnarray}
where $\partial \Gamma(F_2) = [F_2]$.
The bulk superpotential is the same as in the absence of 7-branes. To formally make contact with the general D7 superpotential of \cite{Martucci:2006ij} and the work of \cite{Gomis:2005wc}, put $H_3 = d B_2$. Note that the integrand in the first term has a branch cut 5-chain $\Gamma_5$ going between the D7-brane $S: P_{D7}=0$ and the O7-plane $P_{O7}=0$, on which it jumps by an amount $H_3$.\footnote{The presence of the cut is due to the fact that there is a $SL(2,\IZ)$ $T$-monodromy $F_3 \to F_3 + H_3$ around the D7 branes for the original $F_3$ and $H_3$ defined in (\ref{fluxredFIIB}).} Now extend $F_2$ as a closed form onto this 5-chain $\Gamma_5$ (by taking it to be the Poincar\'e dual of $\Gamma(F_3)$, which we can take to lie in $\Gamma_5$), and perform an integration by parts on the first term in $W_{\rm D7}$. This gives:
\begin{equation}
  W_{\rm D7}(\phi) \, \mbox{``}=\mbox{''} \, \int_{\Gamma_5} (F_2-B_2) \wedge \Omega_3 \, ,
\end{equation}
reproducing the D7 superpotential of \cite{Martucci:2006ij}. I have put the equality sign between quotation marks because it is not quite justified to simply put $H_3 = dB_2$, since $H_3$ is not globally exact. I will not try to make this correspondence more precise; in practice, if one wishes to explicitly compute the superpotential in specific models, the expression involving the periods is computationally superior anyway.

To leading order in $g_s$, the critical point condition $DW=0$ splits up as
\begin{eqnarray}
 &&\partial_{\tau_0} W+ (\partial_{\tau_0} \CK_\tau) W = 0 \\
 &&\partial_\psi W + (\partial_{\psi} \CK_X) W = 0 \\
 &&\partial_\phi W_{\rm D7} = 0 \, .
\end{eqnarray}
When the D7-branes coincide with the O7, or more precisely when $\eta^2-h\chi=h^4$, $W_{\rm D7}=0$ and the first two equations are equivalent to
\begin{equation}
 [G_3]_{\rm b}^{3,0} = 0, \qquad [G_3]_{\rm b}^{1,2} = 0, \qquad [G_3]_{\rm b} := [F_3]_{\rm b} - \tau_0 [H_3]_{\rm b} \, ,
\end{equation}
that is, $[G_3]_{\rm b}$ is of type $(2,1)$ + $(0,3)$; a harmonic representative would be imaginary self-dual.
This is no longer true when the branes move off the O7. This was to be expected, since the distinction between bulk and brane flux is not canonical: looping around in D7 configuration space can create bulk flux out of brane flux. And certainly, if the bulk flux changes, we expect the bulk flux equations to be changed too.

Taking $[H_3]_{\rm b}=0$, the third equation is equivalent to
\begin{equation} \label{F02cond}
 [F_2]^{0,2} = 0 = [F_2]^{2,0}  \, ,
\end{equation}
that is, $F_2$ is of type $(1,1)$. In addition, note that automatically $[F_2] \wedge [J] = 0$ if the D7 is generic. This is because $[F_2]$ is odd under the orientifold involution, while $[J]$ is even, so $\int F_2 \wedge J = 0$. Genericity of the D7 implies in particular it has a single component, in which case the latter equation implies $[F_2] \wedge [J] = 0$. This fits with our earlier observation that (\ref{Dtermconstr}) is automatically satisfied in our setup. Thus, a harmonic representative $F_2$ would be anti-self-dual.
This is the condition for a D7 configuration with flux to preserve the supersymmetries of the orientifold background, as obtained from the perturbative string description \cite{Marino:1999af}.

Explicitly finding flux vacua by computing the periods $\Pi_\alpha$, $\Pi_i$, $\chi_i$ and finding critical points is prohibitively difficult in almost any example. However, the condition (\ref{F02cond}) has a simple geometrical interpretation:
\begin{equation}
 [F_2]^{0,2} = 0 \quad \Leftrightarrow \quad [F_2] \mbox{ is a divisor in S} \, .
\end{equation}
In other words, in the absence of $[H_3]_{\rm b}$ and to lowest order in $g_s$, the D7 embedding must be such that $N^\alpha \partial \Gamma_\alpha$ can be represented in homology as a linear combination of holomorphic curves. It is infinitely much simpler to explicitly construct holomorphic curves as surfaces containing them. Thus, explicitly constructing D7 flux vacua at weak string coupling is in fact more tractable than one might have naively feared.

It is not known if a similar geometrization of flux vacua exists for nonzero bulk fluxes or away from the weak coupling limit.

\subsubsection{Supersymmetry}

Generically, the flux vacua obtained by solving $D_a W=0$ break supersymmetry. This is simply because generically $W \neq 0$ at the critical point, and therefore the covariant derivatives with respect to the K\"ahler moduli, $D_{T_A} W = (\partial_{T_A} \CK)W$, do not vanish.

For the flux vacuum to preserve supersymmetry, we need in addition $W=0$, or equivalently $G_4^{4,0}=G_4^{0,4}=0$; for the fluxes we are considering, this means $G_4$ is of type (2,2) and primitive.

Since $W=0$ is one more constraint than there are variables, we generically do not expect solutions. A notable exception to this is when $D_a W=0$ alone does not fix all complex structure moduli; in this case we can conceivably move along the flat direction till we hit a zero of $W$. Of course what we really are after are vacua without any remaining flat directions at all, so this would not be a desirable situation from that point of view.

However it may happen that $W=0$ ``accidentally'' even for isolated vacua \cite{DeWolfe:2004ns}.

\subsubsection{Warping} \label{sec:warping}

So far we have done our analysis in the effective field theory approximation. It is possible to do better. In the absence of flux and M2-branes, our metric was flat space times the Calabi-Yau metric on $Z$. In the presence of $G_4$ flux and/or M2-branes, this metric no longer solves the Einstein equations, but a warped version of it still does \cite{Becker:1996gj,Gukov:1999ya,Dasgupta:1999ss}. The metric is of the form
\begin{equation} \label{warpedmetric}
 ds^2 = e^{-w(y)} [-(dx^0)^2+(dx^1)^2+(dx^2)^2] + e^{w(y)/2} ds^2_Z \, ,
\end{equation}
where $w(y)$ is the warp factor, which depends only on the coordinates $y^m$ of the internal space $Z$. The metric $ds^2_Z$ is our original Ricci-flat Calabi-Yau metric on $Z$. The warp factor satisfies the following Poisson equation on $Z$:
\begin{equation}
 d*d(e^{3 w/2}) = \frac{1}{2 \, \ell_M^6} G_4 \wedge G_4 - I_8(R) + \sum_i \delta_{{\rm M2}_i} \, ,
\end{equation}
where the Hodge * is with respect to the $ds^2_Z$ metric. The 4-form $G_4$ satisfies
\begin{equation}
 G = *_Z G \, ,
\end{equation}
again with respect to $ds^2_Z$. Furthermore $G_{\mu\nu\rho m} = \epsilon_{\mu\nu\rho} \partial_m e^{-3w/2}$.

The metric (\ref{warpedmetric}) still fits in our original $T^2$ fibered metric ansatz (\ref{fibrationmetric}), namely
\begin{equation}
 ds_{M}^2 = \frac{v}{\tau_2} \biggl( (dx + \tau_1 dy)^2 + \tau_2^2 dy^2 \biggr)  + ds_9^2 \, ,
\end{equation}
provided we now allow $v$ to vary over $M_9$:
\begin{equation}
 v = v_0 e^{w/2}, \qquad ds_9^2 = e^{-w}[-(dx^0)^2+(dx^1)^2+(dx^2)^2] + e^{w/2} ds_B^2 \, .
\end{equation}
Plugging this back into (\ref{genIIBmetricE}), taking $L \equiv \sqrt{v_0}$ and defining $x^3:=\frac{\ell_s^2}{\sqrt{v_0}} \, y$, we find for the Einstein frame IIB metric
\begin{equation}
 ds^2_{\rm IIB} = e^{-3w/4}[-(dx^0)^2+(dx^1)^2+(dx^2)^2+(dx^3)^2] + e^{3w/4} \, ds_B^2 \, .
\end{equation}
In the F-theory limit $v_0 \to 0$, the periodicity $\ell_S^2/\sqrt{v_0}$ of $x^3$ goes to infinity, and we retrieve a fully four dimensional Poincar\'e invariant solution. Note in particular the remarkable fact that the warp factors have combined in just the right way to make full Poincar\'e invariance possible, despite the different origin of the $x^3$ direction in M-theory.

The IIB metric obtained here is indeed of the warped form considered in \cite{Giddings:2001yu} in the weak coupling limit of type IIB.

It was pointed out in \cite{Giddings:2001yu} that the warping can become very substantial for some flux vacua. A simple local, non-orientifolded, D-brane free model for this is the following. Consider the (noncompact) deformed conifold Calabi-Yau threefold embedded in $\IC^4$:
\begin{equation}
 X:u_1^2 + u_2^2 + u_3^2 + u_4^2 = \epsilon^2 \, .
\end{equation}
The holomorphic 3-form is $\Omega_3 = \frac{du_1 du_2 du_3}{u_4}$, and its $(\alpha,\beta)$ 3-cycle periods are
\begin{equation}
 \int_\alpha \Omega_3 = z, \qquad \int_\beta \Omega_3 = z \frac{\log z}{2 \pi i} + g(z) =: \CG(z) \, .
\end{equation}
where $z \sim \epsilon^2$ and $g(z)$ denotes an order 1 part regular analytic in $z$ which depends on how this local model is embedded into a compact model. Choosing fluxes $F_3 = M \, \beta$, $H_3 = K \, \alpha$ (where we use again the same notation for cycles and Poincar\'e dual fluxes), the superpotential takes the form
\begin{equation}
 W(z) = - K \, \tau \, z + M \, \CG(z).
\end{equation}
We take $\tau:=i/g_s$ fixed (in more complete models it would be fixed by other fluxes). For large $K/g_s$ and small $z$, we can consistently solve $0 = D_z W \approx \partial_z W$ as
\begin{equation}
 z \sim \exp(-2 \pi K/g_s M) \, .
\end{equation}
Due to the high concentration of D3-charge $F \wedge H$ near the exponentially small 3-cycle $A$, such a solution will be strongly warped, the cone of the conifold times $\IR^{1,3}$ being deformed into an AdS$_5$ throat capping off at a redshift factor $e^A = e^{-3u/4}$ of order
\begin{equation} \label{warpzrel}
 e^{A_{\rm min}} \sim |z|^{1/3} \sim e^{-2 \pi K/3 M g_s} \, .
\end{equation}
This is the Klebanov-Strassler solution \cite{Klebanov:2000hb}.

Computing the four dimensional effective action including warping effects requires more work, see e.g.\ \cite{Giddings:2005ff,Baumann:2006th,Douglas:2007tu,torroba}.

\subsection{Quantum corrections}

We have seen that all fourfold complex structure moduli can be stabilized in principle by turning on $G_4$ flux, but that this leaves the K\"ahler moduli directions exactly flat at tree level.

To stabilize those, we therefore need to consider quantum corrections to the superpotential and K\"ahler potential. There can be no K\"ahler moduli dependent perturbative corrections to the superpotential. This is because the K\"ahler moduli appear in chiral multiplets whose complex scalar components are given by (\ref{Kahlermoduli}):
\begin{equation}
 T_A = \frac{1}{\ell_M^6} \int_{D_{6,A}} C_6 + i \, dV \,  = \frac{1}{\ell_s^4} \int_{D_{4,A}} C_4 + i \, dV \, .
\end{equation}
Shifting the axionic modes by a constant is an exact symmetry of the classical action, so all perturbative corrections will also be invariant under such shifts. Combining this with the holomorphicity of the superpotential then shows that there can be no $T$-dependent perturbative corrections to $W$, as such corrections would come as powers of the $T_A$, which are not invariant under axion shifts.

\subsubsection{Instantons}

However, supersymmetric instantons can give corrections to the superpotential. They are of the form
\begin{equation}
 W_{\rm inst} = \Lambda^3 \, e^{2 \pi i n^A T_A} \, ,
\end{equation}
where $\Lambda^3$ is some holomorphic function of the other (non-K\"ahler) scalars in the theory, including the complex structure moduli $z$. In the type IIB picture, these can be thought of as arising from D3 instantons wrapping divisors $D_4 = n^A D_{4,A}$. In the M-theory picture, these correspond to M5 instantons wrapping the elliptic fiber and $D_4$, i.e.\ wrapping $D_6 = n^A D_{6,A}$. As we saw in \ref{sec:Fbranes}, these are indeed the only M5 instantons which have finite action in the F-theory limit of vanishing fiber size.

Some classic references for instantons effects in string theory are \cite{Dine:1986zy,Dine:1987bq,Becker:1995kb,Witten:1996bn,Harvey:1999as}. The general calculus of instantons was reviewed in \cite{Dorey:2002ik}, and the lecture notes \cite{Vandoren:2008xg} give an introduction to instanton effects in quantum mechanics and field theory. D-brane instanton effects in string theory are an active area of current research and a relatively large literature exists by now. A short recent overview with the relevant references, in particular for IIB appications, can be found in \cite{Blumenhagen:2007sm}.

Every holomorphic divisor $D_4$ gives rise to an instanton, and there are infinitely many holomorphic divisors. However, the existence of the instanton does not imply there will be an actual nonzero contribution to $W$: if there are too many fermionic zeromodes, the instanton will not contribute. Guaranteed to contribute are instantons with the absolute minimal amount of fermionic zeromodes, i.e.\ two of them, corresponding to the two broken supersymmetries of the original four preserved by the vacuum. In particular, such instantons must wrap rigid cycles (cycles without infinitesimal holomorphic deformations), because the superpartners of the deformation moduli provide additional fermionic zeromodes.

A rough sketch of why instantons with exactly two zeromodes contribute to the superpotential goes as follows. Let $\psi_A$ be the fermion in the chiral multiplet with complex scalar $T_A$. The instanton corrections to the superpotential can in principle be extracted by computing the $(\partial_T^2 W) \psi \psi$ term in the effective action generated by the instanton. If we denote the fermionic zeromodes (or collective coordinates) of the instanton by $\Theta_i$, $i=1,\cdots,N$, this term is given by the zero momentum correlator
\begin{eqnarray}
 \frac{\partial^2 W}{\partial T_A \partial T_B} &\sim& \langle \psi_A \psi_B \rangle_{\rm inst} \\
 &\sim& \int_{\rm inst} \CD \Theta \, \CD \psi \, \CD(\cdots) \, e^{-S} \, \psi_A \psi_B \,  \, .
\end{eqnarray}
where the path integral is over all fields of the theory, expanded around the instanton background. Expanding the action as
\begin{equation}
 S = S^0_{\rm inst} + V^i_A \, \Theta_i \bar{\psi}^A  + S'(\Theta,\psi,\cdots) \, ,
\end{equation}
where $S^0_{\rm inst} = 2 \pi i \, n^A T_A$ is the classical instanton action and $S'$ contains all terms not bilinear in $(\Theta,\bar\psi)$, this becomes
\begin{eqnarray}
 \frac{\partial^2 W}{\partial T_A \partial T_B} \sim e^{-S_{\rm inst}^0} \int \CD \Theta \, \CD \psi
 \, \CD(\cdots) \, e^{-V^i_C \Theta_i \bar{\psi}^C - S'} \, \psi_A \psi_B  \,  \, . \nonumber
\end{eqnarray}
If there are exactly two zeromodes $\Theta_1$, $\Theta_2$, integrating over them gives
\begin{eqnarray}
  \frac{\partial^2 W}{\partial T_A \partial T_B} &\sim& e^{-S_{\rm inst}^0} \int \CD \psi \, \CD(\cdots) \,
  e^{ -S'} \, V^1_C \,
  V^2_D \, \bar{\psi}^C \bar{\psi}^D \psi_A \psi_B  \nonumber \\
  &\sim&e^{-S_{\rm inst}^0} \int \CD(\cdots) \, e^{-S'} \, V^1_{(A} \, V^2_{B)} \,,
\end{eqnarray}
 where we contracted $\bar{\psi}$s with $\psi$s. If on the other hand there are more than two zero modes $\Theta_i$, then integrating over them would bring down more than two $\bar{\psi}$s, which is more than there are $\psi$s to contract with, possibly resulting in a zero amplitude.

\subsubsection{Gaugino condensation}

Another nonperturbative effect that can cause a contribution to the effective superpotential in four dimensions is gaugino condensation. For example if $N$ D7-branes wrap a \emph{rigid} divisor $D=n^A D_A$, we get pure $SU(N)$ super Yang-Mills in four dimensions with gauge coupling constant
\begin{equation}
 T = \frac{\theta}{2 \pi} + \frac{4 \pi i}{g_{\rm YM}^2} = n^A T_A \, .
\end{equation}
Its ground states have a nonzero gaugino condensate, $\langle \lambda_\alpha \lambda^\alpha \rangle \neq 0$, which can be obtained from an effective superpotential as follows. Let $S = {\rm Tr} \, W_{\alpha} W^{\alpha}$ be the composite chiral superfield with scalar component $s = \rm Tr \lambda_\alpha \lambda^\alpha$. The effective action for $S$ involves the Veneziano-Yankielowicz superpotential \cite{Veneziano:1982ah,Taylor:1982bp}:
\begin{equation}
 W(T,S) = 2 \pi i T S - N S \left( \ln \frac{S}{\Lambda^3} \, - 1 \right) \, ,
\end{equation}
where $\Lambda$ is the UV cutoff. Solving $\partial_S W=0$ gives
\begin{equation} \label{gluinocondW}
 S = \Lambda^3 e^{2 \pi i T/N} \quad \Rightarrow \quad W(T) = N \Lambda^3 e^{2 i \pi T/N} \, .
\end{equation}
This gives the vev of the gluino condensate and the effective superpotential for the K\"ahler modulus $T$.

If on the other hand the divisor has deformation moduli, there will be massless adjoint matter coupled to the gauge theory, and no gluino condensation occurs.

\subsubsection{Relation between 4d gaugino condensation and M5 instantons} \label{sec:gauginoM5}

\begin{figure}
\centering
\includegraphics[width=\textwidth]{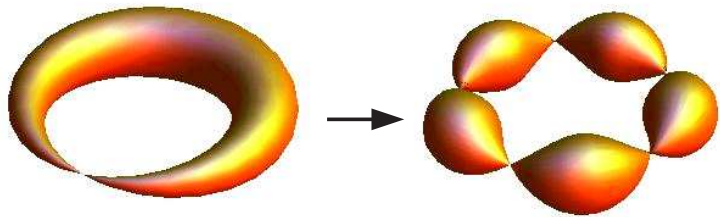}
\caption{Moving onto the Coulomb branch: $N = 5$ coincident M-theory KK monopole cores are moved apart along the circle T-dual to the 4d $\to$ 3d compactification circle. The transversal circle is the M-theory circle. }\label{fibdeg}
\end{figure}

The four dimensional gaugino condensate superpotential can be related to a three dimensional M5 instanton superpotential \cite{Katz:1996th}. The idea is as follows. Recall from section \ref{sec:whatisF} that our 4d IIB theory compactified on a circle of length $\ell=\ell_s^2/\sqrt{v}=\ell_M^3/v$ is dual to M-theory on an elliptically fibered Calabi-Yau fourfold with fiber size $v$. The four dimensional IIB limit is obtained by sending $v \to 0$, but let us not do that now, but instead take $v/\ell_M^2$ large, so we are looking at M-theory on a large Calabi-Yau fourfold. Now, imagine we had a pure $\CN=1$ $SU(N)$ Yang-Mills theory in four dimensions, engineered by letting $N$ D7-branes wrapped on a rigid 4-cycle $S = n^A D_{4,A}$ coincide, as just discussed. After T-dualizing on the circle to IIA, these D7-branes become D6-branes. Wilson lines along the circle on the D7-branes become positions of the D6-branes along the dual circle. These moduli correspond to fields in the adjoint of the gauge group. Turning them on breaks the $SU(N)$ gauge symmetry to $U(1)^{N-1}$ --- i.e.\ this puts us on the Coulomb branch. In M-theory, D6-branes lift to KK monopoles with core at the location of the D6-branes.\footnote{In the limit of vanishing elliptic fiber size we have been considering before, the localization of these KK monopoles on the elliptic fiber could consistently be neglected, but in the finite size case this is no longer true.} At the core locus, the M-theory circle shrinks to zero size. Thus we get the situation depicted in fig.\ \ref{fibdeg}: moving apart the $N$ KK centers deforms the degenerate elliptic fiber into $N$ spheres intersecting according to the extended Dynkin diagram of $SU(N)$.

The M5 instantons wrapping $S$ and any of the $N$ spheres $S^2_k$ generate a superpotential
\begin{equation}
 W = \Lambda^3 \sum_{k=1}^N e^{2 \pi i \rho_k}  \, ,
\end{equation}
where $\Lambda$ can depend on the fourfold complex structure moduli and
\begin{equation}
 \rho_k := \frac{1}{\ell_M^6} \int_{S \times S^2_k} C_6 + i \, dV_6 \, .
\end{equation}
Since the sum of the $N$ spheres is homologous to the full elliptic fiber at generic position we have
\begin{equation} \label{sumconstr}
 e^{2 \pi i \sum_{k=1}^N \rho_k} = e^{2 \pi i T} \, , \qquad T = n^A T_A \, ,
\end{equation}
$-2 \pi i T$ being the instanton action of an M5 wrapping the full elliptic fiber at generic position, or equivalently (as we saw before) the action of a D3 instanton wrapping $S=n^A D_{4,A}$.  Extremizing $W(\rho)$ subject to the constraint (\ref{sumconstr}) gives $N$ solutions:
\begin{equation}
 e^{2 \pi i \rho_k} = e^{2 \pi i m/N} \, e^{2 \pi i T /N} \, , \qquad m \in \{0,1,\ldots,N-1\} \, ,
\end{equation}
resulting in
\begin{equation}
 W = N \Lambda^3 \, e^{2 \pi i m/N} \, e^{2 \pi i T/N} \, .
\end{equation}
Note that this is completely independent of $v$, so we expect this superpotential to survive in the $v \to 0$ limit, despite the fact that the large radius geometric M5-instanton picture is no longer valid in this regime. And indeed, we see that this superpotential exactly coincides with the gluino condensate superpotential (\ref{gluinocondW})! The $N$ different solutions we find here correspond to the $N$ vacua we have in 4d, characterized by different complex phases of the gluino condensate (implicit in (\ref{gluinocondW})).

This beautiful geometrical unification in M-theory of all nonperturbative effects extends to more complicated gauge groups as well \cite{Katz:1996th}.

\subsubsection{Geometric conditions for nonvanishing contributions} \label{sec:geomcondnonv}

In the absence of flux, a necessary condition for instantons to contribute to the superpotential has been given by Witten \cite{Witten:1996bn}. Let $h^{p,0}(D_6)$ be the number of holomorphic $(p,0)$-forms on the holomorphic divisor $D_6$ wrapped by an M5 instanton. Then a \emph{necessary} condition for the instanton to contribute is that its holomorphic Euler characteristic (or arithmetic genus) equals 1:
\begin{equation} \label{holeulcond}
 1 = \chi_0(D_6) := \sum_{p=0}^3 (-1)^p \, h^{p,0}(D_6) \, .
\end{equation}
The derivation is based on a $U(1)$ charge selection rule. A sufficient condition to have exactly two fermionic zeromodes and therefore a nonzero contribution to the superpotential is that the divisor is completely rigid; more precisely:
\begin{equation} \label{suffcond}
 h^{0,0}=1 \, , \qquad h^{0,p} = 0 \quad p \geq 1 \, .
\end{equation}
This satisfies (\ref{holeulcond}), of course.

A good thing about the criterion (\ref{holeulcond}) is that it involves an index, which is relatively easily computed just from knowledge of the wrapping numbers of the divisor. We will see in detail how this works in section \ref{sec:IndexFormulae}. Computing the individual Hodge numbers is also possible, but requires a bit more work.

The necessary condition (\ref{holeulcond}) has been derived in the absence of fluxes. In the presence of fluxes, the condition appears to be no longer necessary; essentially this is because fluxes can effectively rigidify previously overly floppy M5-instantons, or, in the IIB picture, rigidify D3 instantons or D7-branes. In the latter case, the physics of this is quite intuitive: As we saw in the previous parts, fluxes can freeze D7-moduli, i.e.\ give masses to adjoint matter, thus (if all moduli are lifted and no other matter is present) reducing the theory to pure SYM below this mass scale, hence giving rise to strong coupling at low energies and gaugino condensation again.

It has not been completely clarified what replaces (\ref{holeulcond}) in the presence of flux. In \cite{Kallosh:2005gs} it was found that the presence of flux effectively modifies the value of $h^{2,0}$ in (\ref{holeulcond}) to some lower value $h^{2,0}_{\rm eff}$ given as the number of solutions to a particular flux-dependent wave equation on $D_6$. It is unfortunately not known if this number is computed by an efficiently computable index formula. But in any case, since $h^{2,0}_{\rm eff} \leq h^{2,0}$, we have the necessary condition
\begin{equation} \label{fluxnecccond}
 \chi_0 \geq \chi_{0,\rm eff} = 1 \, .
\end{equation}
More discussion of this issue and more concrete examples of flux modifications of M5 instanton effects can be found in \cite{Gorlich:2004qm,Saulina:2005ve,Tsimpis:2007sx,Lust:2005cu,Blumenhagen:2007bn}.

What is not affected by flux is the sufficiency of (\ref{suffcond}) to get a nonzero contribution.

\subsubsection{Corrections to the K\"ahler potential}

Corrections to the K\"ahler potential are much less constrained. In particular, unlike the superpotential, it can receive $T_A$-dependent perturbative corrections. In the IIB weak coupling limit, the leading correction to (\ref{KKpot}) in an expansion in inverse powers of the volume is \cite{Becker:2002nn}
\begin{equation} \label{Kahlercorr}
 \CK_K = -2 \ln \biggl( V(B) + \frac{\xi}{g_s^{3/2}} \biggr) \, , \qquad \xi = - \frac{\zeta(3)}{32 \, \pi^3} \, \chi(X) \, .
\end{equation}
where $V(B) = \int J_B^3/6 = \frac{1}{2} V(X)$ is the Einstein frame volume of the base $B$ in string units, and $X$ is the Calabi-Yau threefold which is the double cover of $B$.

More corrections have been considered. A concise recent review can be found in \cite{Berg:2007wt}, in particular in relation to the large volume scenario we will discuss in section \ref{sec:cheese}.

\subsection{The KKLT scenario} \label{sec:KKLT}

The KKLT scenario \cite{Kachru:2003aw} is a synthesis of all of the elements we introduced so far, providing a way in principle to stabilize all moduli, break supersymmetry and obtain metastable de Sitter vacua in string theory in a reasonably controlled way. The meaning of ``reasonable'' is somewhat debatable; as always when relying on quantum corrections, the Dine-Seiberg problem (explained in section \ref{sec:DineSeiberg}) makes strict parametric control impossible and remains an issue that needs to be carefully addressed in specific models.

To outline the basic idea, consider a hypothetical model with one K\"ahler modulus $T$, and assume a nonperturbative superpotential is generated depending on $T$, so the total superpotential is of the form:
\begin{equation}
 W(z,T) = W_{\rm flux}(z) + \Lambda^3 \, e^{2 \pi i a T}
\end{equation}
with $a=1$ if the correction is due to a D3 instanton, and $a=1/N$ is it is generated by $SU(N)$ gaugino condensation. The flux superpotential was defined in (\ref{GVWcond}): $W_{\rm flux} = \ell_M^{-3} \int_Z G_4 \wedge \Omega_4 = N^I \Pi_I(z)$. Recall we are working in conventions in which $W_{\rm flux}$ is dimensionless; the dimensionful flux effective potential contains a scale-setting factor $\sim e^{\CK_K} m_p^4 \sim m_s^4$. The UV scale $\Lambda^3$ should therefore be taken to be expressed in string units, and can be expected to be roughly of order 1. In general it may depend on the complex structure moduli of $Z$.

\subsubsection{Complex structure stabilization}

We will be interested in vacua for which the second term can self-consistently be considered to be a small perturbation compared to the first one as far as the fourfold complex structure moduli $z$ is concerned. Then we can first solve the classical flux vacua equations of motion for the $z^a$:
\begin{equation} \label{DWDWDW}
 D_a W_{\rm flux}(z) = 0 \, .
\end{equation}
or equivalently (given (\ref{autoprimitive})) $G_4 = * G_4$. We assume that for suitable 4-form fluxes $G_4$, all complex structure moduli $z$ get frozen. The typical mass scale of these moduli will be of order
\begin{equation} \label{mzscale}
 m_z \sim \frac{|G| \, m_p}{V} \sim \frac{|G| \, m_s}{V^{1/2}} \, ,
\end{equation}
where $|G|$ is some measure for the size of the flux (given below) and $V=V(B)$ is the volume of the IIB compactification manifold in string units. At weak string coupling there is an additional factor $g_s$ from the $e^{\CK_{\tau_0}} \sim g_s$ factor in the potential.

The size of the flux is constrained by the tadpole cancelation condition (\ref{tadpolecancelation}):
\begin{equation} \label{tpcnc}
 \frac{1}{2} Q_{IJ} N^I N^J \, + \, N_{\rm D3} = \frac{\chi(Z)}{24} = Q_c \, .
\end{equation}
The first term equals $\frac{1}{2 \ell_M^6} \int_Z G_4 \wedge G_4$. The intersection product on flux space is not positive definite, so naively it might seem there is an infinite number of possible fluxes, becoming arbitrarily large with positive and negative contributions canceling out in the first term. However, using that $G_4 = * G_4$ on solutions to the equations of motion, one sees that the first term is in fact positive definite for actual flux vacua. The second term is positive as well if we do not introduce anti-D3-branes. Therefore we can estimate
\begin{equation} \label{fluxsize}
 |G| \sim \sqrt{Q_c}
\end{equation}
in (\ref{mzscale}).

Moreover, it follows that we can roughly think of the space of possible fluxes as a ball in $b_4'$-dimensional flux space\footnote{$b_4'$ is the number of 4-form fluxes with one leg on the elliptic fiber; more formally, it is the dimension of the subspace of $H^4(Z)$ satisfying (\ref{restrictedflux}); in typical models \cite{Klemm:1996ts} $b_4'$ is close to $b_4$, both being of order $10^4$.} of radius $\sqrt{2 Q_c}$. A rough estimate for the number of flux vacua for sufficiently large $Q_c$ is therefore the volume of this ball:
\begin{equation}
 \CN_{\rm flux \, vac} \sim \frac{(2 \pi Q_c)^{b_4'/2}}{\left(b_4'/2\right)!} \, .
\end{equation}
Although this reasoning is very heuristic, it gives essentially the right result (at least for sufficiently large $Q_c$) \cite{Ashok:2003gk,Denef:2004ze}, as we will see in a much more refined counting analysis in section \ref{sec:statistics}.

To get an idea of the numbers involved, for the example of $Z$ the elliptic fibration over $\ICP^3$ described by (\ref{CY4eq}), we have $b_4'=23320$, $Q_c = 972$, so according to our estimate
\begin{equation}
 \CN_{\rm vac} \sim 10^{1787} \, .
\end{equation}
The perhaps more famous order of magnitude $\sim 10^{500}$ is obtained by restricting to the much smaller set of \emph{bulk} RR and NSNS fluxes in the IIB weak coupling limit, in which case $b_4'$ in the above formula gets replaced by $2 \, b_3(X) = 600$.

We will also (crucially) assume that $|W_{\rm flux}|$ can be made extremely small in a (small) fraction of all vacua. For generic vacua this will not be the case, as we expect typical values $|W_{\rm flux}| \sim |G| \sim \sqrt{Q_c}$. However, the different contributions to $W_{\rm flux} = N^I \Pi_I(z)$ add up with essentially random complex phases. A small fraction of random walks will happen to end up exponentially close to $W=0$. The distribution of $W_{\rm flux}$ values will have some broad Gaussian-like structure, but exponentially close to $W=0$ the density of vacua will be essentially uniform. Hence we expect roughly $\lambda \, \CN_{\rm vac}$ flux vacua within the region $|W|^2 < \lambda \ll 1$. Given the exponentially large typical values of $\CN_{\rm vac}$, vacua with exponentially small values of $|W_{\rm flux}|$ should therefore still be abundant in absolute numbers. Again this estimate can be put on a much firmer footing \cite{Denef:2004ze}.

\subsubsection{K\"ahler moduli stabilization}

The effective superpotential for the K\"ahler moduli after integrating out the complex structure moduli is
\begin{equation}
 W(T) = W_0 + \Lambda^3 \, e^{2 \pi i a T} \, ,
\end{equation}
where $W_0$ is exponentially small and $\Lambda^3$ of order 1. Solving $D_T W = 0$ makes the first term balance against the second, resulting in
\begin{equation}
 2 \pi i a T \sim \ln \frac{W_0^{-1}}{\Lambda^3} \, .
\end{equation}
Since $\Lambda \sim 1$, $W_0$ is exponentially small and $a$ is at most 1, this stabilizes the K\"ahler modulus $T$ at a moderately large value. For example taking $a=1/5$, $W_0 = 10^{-30}$, $\Lambda = 1$, we get $\Im T \approx 55$, $V \sim T^{3/2} \sim 400$, which is already more than large enough to justify neglecting for example the first correction to the K\"ahler potential (\ref{Kahlercorr}), and definitely to neglect higher instanton corrections to $W$. Even larger volumes are possible: Taking $\ln |W_0|^{-1}$ to be of the order of its estimated smallest possible nonzero value in the example given above, i.e.\ $\ln |W_0|^{-1} \sim 2000$, we get $\Im T \sim 1600$. Note however that the maximal size is bounded by the \emph{logarithm} of the number of vacua, and so the volume will never become exponentially large in this scenario.

The mass scale for the K\"ahler moduli is (dropping polynomial factors in $T$):
\begin{equation}
 m_T \sim e^{2 \pi i a T} m_s \sim |W_0| m_s \, ,
\end{equation}
which is exponentially small compared to the scale of the complex structure moduli masses (\ref{mzscale}), given that we take $|W_0|$ to be exponentially small. This shows it is self-consistent to first integrate out the complex structure moduli; the backreaction of the K\"ahler moduli on the complex structure moduli will only give rise to exponentially small corrections to the vacuum values of the $z^a$.

For realistic applications it should be kept in mind however that $m_T$ should not become too small to be in conflict with observations. (Taking $|W_0| \sim 10^{-30}$, $m_s \sim 10^{18} {\rm GeV}$ gives $m_T \sim 10^{-3} {\rm eV}$, which is the lower bound set by fifth force experiments, and well below the bound set by cosmological considerations.)

Despite the fact that the quantum corrections in this regime give merely exponentially small corrections to the complex structure flux vacua, there is one dramatic qualitative change: instead of a flat Minkowski compactification with exponentially small supersymmetry breaking (due to $D_T W_{\rm flux} \sim W_{\rm flux} \neq 0$), we now have an Anti-de Sitter vacuum with exponentially small cosmological constant
\begin{equation}
 \Lambda = -3 \, \frac{m_p^4}{4 \pi} \, e^\CK |W|^2 \sim - m_s^4 \, e^{2 \pi i a T} \sim - m_s^4 \, |W_0|^2 \, ,
\end{equation}
and supersymmetry restored! In particular, this suggests that thanks to the quantum corrections, such vacua have CFT duals. Some suggestions regarding the nature of these putative CFTs has been made in \cite{Silverstein:2003jp}, but they remain clouded in mystery. Understanding them would be a huge step forward in putting these flux vacua on a firm, uncontestable footing as genuine string theory vacua.

\subsubsection{Uplifing to dS}

The third crucial element in \cite{Kachru:2003aw} was a way to uplift these vacua with exponentially small negative cosmological constant to vacua with exponentially small positive cosmological constant. As we saw in section \ref{sec:warping}, it is possible to tune fluxes such that regions of exponentially strong warping occur, locally described by a Klebanov-Strassler throat. Using statistical methods, it can moreover be argued that such vacua, which lie exponentially close to a conifold degeneration of the compactification manifold, are relatively common (see section \ref{sec:statistics}).

Once we have such a region, supersymmetry can be broken by an exponentially small amount by placing an anti-D3 brane at the bottom of the warped throat. This will add an exponentially small positive contribution to the effective potential. There will on the other hand still be an exponentially large discretuum of flux vacua with approximately the same warped throat (we again refer to \ref{sec:statistics} for justification), so in particular if the number of such vacua is still larger than $10^{120}$, there should be at least some of them with a positive cosmological constant of the order of the observed value. The reason we observe such a small cosmological constant is then attributed to environmental selection, giving a concrete realization in string theory of the ideas of \cite{Weinberg:1987dv,Bousso:2000xa,Susskind:2003kw}.

It should be clear now that the existence of the finely spaced discretuum of flux vacua in F-theory/IIB is at the core of being able to circumvent the Dine-Seiberg problem to a certain extent. Although the problem persists for generic vacua, for an exponentially small fraction, but still an exponentially large absolute number, we ``accidentally'' achieve reasonable control, at least sufficient to argue for existence within the framework of supergravity. In this way, the existence of a landscape is a blessing.

Nevertheless, several of the arguments rely on genericity assumptions. Although explicit models have been constructed realizing AdS complex structure and K\"ahler moduli stabilization within the supergravity approximation, with satisfactory error estimates for neglected corrections, the same level of confidence has not been achieved for the uplift to de Sitter, in part due to the complications induced by the necessary strong warping. There could in general also always be subtle quantum consistency constraints we have overlooked so far. It would clearly be desirable to establish the existence of these vacua as genuine quantum string vacua more convincingly, perhaps by providing a holographic description.

\subsection{The large volume (Swiss cheese) scenario} \label{sec:cheese}

\begin{figure}
\centering
\includegraphics[totalheight=0.25\textheight]{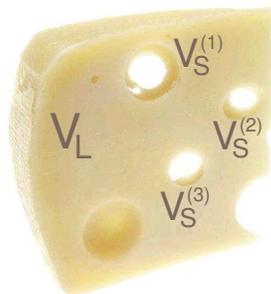}
\caption{Swiss cheese volume $V=V_L - \sum_i V_S^{(i)}$.}\label{swiss}
\end{figure}

A drawback of the KKLT scenario is that control over corrections remains relatively marginal, worsening significantly when the number of K\"ahler moduli goes up. This is because ciritical points of the superpotential balance off nonperturbative effects $\sim e^{-2 \pi V_i}$ and the tree level flux contribution $W_0$, so $2 \pi V_i \sim -\ln |W_0|$. If we want to stabilize the K\"ahler moduli at masses above $10^{-3} {\rm eV}$ to be in agreement with fifth force experiments, we need $W_0 > 10^{-30}$ and therefore $T_i < 10$ or so. If we want the K\"ahler moduli mass scale to be above the TeV scale, we need $W_0 > 10^{-15}$ and $T_i < 5$ or so. The $V_i$ are 4-cycle volumes, which can be expressed in terms of (positive) areas $J^A$ of a basis of holomorphic 2-cycles $C^A$ as $V_i = \frac{1}{2} n_i^A \, D_{ABC} J^A J^B$, where the $D_{ABC}$ are triple intersection numbers of the divisors dual to the $C^A$, which form a basis of the K\"ahler cone (see section \ref{sec:curves}). The intersection numbers for a basis of the K\"ahler cone are nonnegative integers,\footnote{This is because such a basis element by definition has only positive intersection numbers with holomorphic curves, and intersections of two such holomorphic basis divisors are holomorphic curves.} so if the number of moduli is large, the expression of $V_i$ in terms of the $J^A$ contains a large number of all positive terms. Since $V_i$ is at most of order 10, one thus expects in these cases that at least some of these terms will be small, i.e.\ that some 2-cycles will be small in string units. On top of that, these are sizes measured in Einstein frame. The string and Einstein frame K\"ahler moduli are related by $J_s = \sqrt{g_s} \, J_E$. Therefore, if we are at small string coupling, say $g_s \sim 1/10$, we find for the string frame 4-cycle volumes $V_i^s < 1$, and less even for 2-cycle volumes. Clearly, control becomes a serious issue here.

A variant of the KKLT scenario which ameliorates this problem was proposed by Balasubramanian, Berglund, Conlon and Quevedo \cite{Balasubramanian:2005zx}. The idea here is to consider models with at least two K\"ahler moduli a large one and a smaller one, and to balance a nonperturbative correction depending exponentially on the smaller one against a perturbative correction depending inversely on the larger one, thus potentially giving rise to exponentially large overall volumes.

More concretely this goes as follows \cite{Conlon:2005ki}. The dilaton, D7 and complex structure moduli stabilization proceeds as in the KKLT setup, leaving us with an effective superpotential for the large and small K\"ahler moduli $T_L$ and $T_S$ given by, say,
\begin{equation}
 W = W_0 + \Lambda^3 e^{2 \pi i a T_S} \, .
\end{equation}
We assume we have stabilized ourselves in the weak IIB coupling region of the fourfold complex structure moduli space. The K\"ahler potential for the K\"ahler moduli, including the first perturbative correction, is thus of the form (\ref{Kahlercorr}):
\begin{equation}
 \CK = - 2 \ln(V + \frac{\xi}{g_s^{3/2}}) \, ,
\end{equation}
where $\xi = - \frac{\zeta(3)}{32 \, \pi^3} \, \chi(X)$. For the large volume scenario to work, one needs
$\xi>0$, i.e.\
\begin{equation}
 \chi(X) < 0 \, .
\end{equation}
For simplicity the expression of the threefold volume in terms of the K\"ahler moduli is taken to be of ``Swiss cheese'' form:
\begin{equation}
 V \sim (\Im T_L)^{3/2} - (\Im T_S)^{3/2} \, ,
\end{equation}
as illustrated further for multiple small moduli $T_S^{(i)}$ in fig.\ \ref{swiss}. Although this seems a rather special Ansatz, several models are known to satisfy it; basically the ``hole'' contributions are due to blowup modes (see (\ref{swissvol}) for an explicit example).

Focusing on the regime $\Im T_S \ll \Im T_L \sim V^{2/3}$, putting $v_S := 2 \pi a \, \Im T_S$, taking $\Lambda \sim 1$ and setting the axions (consistently) to zero, the effective potential for the K\"ahler moduli is then of the form
\begin{equation} \label{poteq}
  U \sim \frac{\sqrt{v_S}  \, e^{-2 v_S}}{V} - \frac{|W_0| \, v_S \, e^{-v_S}}{V^2} + \frac{\xi \, |W_0|^2}{g_s^{3/2} V^3} \, .
\end{equation}
Here and it what follows we are suppressing positive numerical factors.


Minimizing (\ref{poteq}) with respect to $v_s$ results in
\begin{equation}
 e^{-v_S} \sim \frac{|W_0|}{V} \, .
\end{equation}
Plugging this back in (\ref{poteq}):
\begin{equation}
 U \sim \frac{W_0^2}{V^3} \biggl( \left( \ln \frac{V}{|W_0|} \right)^{1/2} - \,
 \ln \frac{V}{|W_0|} \, + \frac{\xi}{g_s^{3/2}}  \biggr) \, ,
\end{equation}
and minimizing with respect to $V$, we finally get
\begin{equation} \label{swissstab}
 V \sim |W_0| \, e^{\frac{\xi^{2/3}}{g_s}} \, , \qquad v_S \sim \frac{\xi^{2/3}}{g_s} \, ,
  \qquad U \sim - \frac{|W_0|^2}{V^3} \, .
\end{equation}
Thus, remarkably, even without tuning $W_0$ exponentially small, we see we can get very large, even exponentially large, volumes $V$ by tuning $g_s$ moderately small, as well as moderately large $v_S$ and very small to exponentially small negative cosmological constant. In string frame, $V$ is still essentially as large as we wish, but now $v_S \sim \xi^{2/3}$ --- whether this is satisfactorily large depends on the proportionality constant and therefore the model. It should be kept in mind however that this potentially leads us back to the Dine-Seiberg problem in specific models --- we can never parametrically escape it. But in any case the situation is significantly better than in the KKLT scenario, where sending $g_s \to 0$ causes all string frame volumes to collapse.

The minimum of the potential is nonsupersymmetric AdS. In principle it can be uplifted to dS by the same mechanism as in the KKLT scenario.

Extensive analysis of various corrections has been done in \cite{Conlon:2005ki} and  \cite{Berg:2007wt}, and it was found that these large volume compactification models are remarkably robust.

For a recent and more in-depth discussion of realizations of the large volume scenario and several of the ingredients introduced here, see \cite{Blumenhagen:2007sm}.

The large volume scenario has been the starting point for a number of phenomenological explorations, both in particle physics and cosmology. See for instance \cite{Conlon:2006gv} for a review of some of this work.

\subsection{Recap and to do list}

So far we have explained in detail the general geometry of IIB/F-theory flux vacua, we have sketched how and under which geometrical conditions quantum corrections can arise and summarized two related scenarios on how these can lead to fully moduli stabilized models with small positive cosmological constant. But, having a scenario is not the same as having an actual model that works. To construct and analyze such models, and in particular to find models that  generate the required instanton corrections to the superpotential, we need to develop more sophisticated geometrical techniques. This will be the subject of the next section, which gives a hands-on introduction to various constructions in applied algebraic and toric geometry. We will then apply these constructions to build models meeting all requirements to make one or both of the above scenarios work. In practice however, constructing fully explicit flux vacua in typical F-theory compactifications would require specifying 20,000 or so flux quanta and finding the corresponding critical points in the 3000 or so dimensional complex structure moduli space, hoping to hit the region of parameter space we are interested in (e.g.\ weak string coupling, tiny cosmological constant, \ldots). Needless to say, such a task is hopeless. Nevertheless, approximate distributions of vacua over parameter space are relatively easily derived, and from this estimates of how many vacua satisfy properties of interest, without actually having to go through the pain of constructing them explicitly. Developing these statistical techniques will be the subject of section \ref{sec:statistics}.

\section{A geometrical toolkit} \label{geometrictools}

In this section we will give a hands-on introduction to various constructions in algebraic and toric geometry useful for the construction of explicit examples of moduli stabilized type IIB/F-theory vacua. I have tried to make the exposition as concrete and accessible as possible, with emphasis on computation rather than on abstract formalism and structure. This is at the cost of some rigor and generality, and certainly there are much more sophisticated and powerful treatments, but for the purpose of constructing explicit models to play around with, the elementary approach we will follow here is more than sufficient.

We will only assume a basic familiarity with the differential geometry contained in section 2 of \cite{Greene:1996cy}.

The outline of this section is as follows.
\begin{itemize}
\item
In \ref{sec:toricvar} we introduce toric varieties as classical ground state manifolds of gauged linear sigma models. Toric varieties and algebraic subspaces of them provide a huge, fully explicit class of possible compactification manifolds for string theory, including moduli stabilized IIB flux compactifications, which is why we introduce them here.
\item
In \ref{sec:divisors} we define divisors in toric varieties and explain how one can compute their mutual intersection numbers. Once these intersection numbers are known, it is straightforward to compute various quantities of physical interest, such as charges, volumes, K\"ahler potentials, numbers of moduli, numbers of fermionic zero modes, and so on. They form the basic geometric data of everything that follows.
\item
In \ref{sec:curves} we describe the duals of divisors, namely 2-cycles, and we explain how exactly their areas are related to the Fayet-Iliopoulos terms appearing in the definition of the gauged linear sigma model. We also explain how a basis of holomorphic 2-cycles can be constructed and how this allows one to explicitly parametrize the K\"ahler moduli space.
\item
In \ref{sec:volumes} we show how volumes of toric varieties and holomorphic subspaces thereof can be explicitly computed as a function of the K\"ahler moduli. This is needed for example if one wants to compute the K\"ahler potential for a string compactification, or if one wants to compute instanton actions.
\item
In \ref{sec:charclasses}, characteristic classes are introduced. They play an important role in string theory in
extracting, from geometrical setups, various physical topological quantities such as RR charges, moduli space and flux lattice dimensions, numbers of fermionic zero modes of instantons, and so on. We show in particular how they can be computed from the divisor intersection products, for any algebraic subspace of a toric variety.
\item
In \ref{sec:residues}, the concept of Poincar\'e residue is explained. This is an elegant and useful general construction of holomorphic top forms on algebraic subspaces of toric varieties. As we have seen, periods of holomorphic forms play a crucial role in the computation of super- and K\"ahler potentials. This section will also make clear where the holomorphic forms on Calabi-Yau manifolds stated in examples came from.
\item
In \ref{sec:CYsubtor} we focus on Calabi-Yau submanifolds of toric varieties; in particular we consider some examples, one of which is the elliptically fibered CY fourfold over $\ICP^3$ introduced before in (\ref{CY4eq}). In particular we compute the divisor intersection numbers of this fourfold and its total Chern class. This allows to compute in particular the Euler characteristic of the fourfold, which in turn determines the curvature induced D3 tadpole of  F-theory compactified on it, crucial for the existence of flux vacua.
\item
In \ref{sec:IndexFormulae} we list a number of classic index theorems, relating numbers of various bosonic and fermionic zeromodes (e.g.\ those of M5 instantons) to integrals of characteristic classes, which by now we know how to compute.
\item
Finally, in \ref{sec:hodge} we show more concretely how these index theorems can be applied to compute Hodge numbers (i.e.\ numbers of moduli, zeromodes, fluxes, and so on).

\end{itemize}

By the end of this section, you should be able to construct a gigantic variety of supersymmetric compactifications of F-theory for yourself, tailor them to your liking, and compute all of their physically relevant topological characteristics.

\subsection{Toric varieties as gauged linear sigma model ground states} \label{sec:toricvar}

Toric varieties can be represented very concretely as
supersymmetric moduli spaces of gauged linear sigma models
\cite{Witten:1993yc}. This is the approach we will follow here. For more advanced introductions to
toric varieties, see \cite{Kreuzer:2006ax,Hori:2003ic,Greene:1996cy}.

Consider $n$ chiral superfields $X_i$ charged under a
$U(1)^s$ gauge group with charges $Q_i^a$, $a=1,\ldots,s$. In the
absence of a superpotential, the potential for the scalar components
$x_i$ reads
\begin{equation}
 V(x) = \sum_{a=1}^s \frac{e_a^2}{2} \left( \sum_{i=1}^n Q_i^a |x_i|^2 - \xi^a
 \right)^2.
\end{equation}
Here the $e_a$ are the $U(1)^s$ coupling constants, and $\xi^a$ are
the Fayet-Iliopoulos (FI) parameters. The space $\CM$ of classical
supersymmetric ground states is given by the zeros of $V(x)$ (i.e.\ the
D-flat configurations), modulo the $U(1)^s$ gauge symmetry:
\begin{equation} \label{CMdef}
 \CM = \{ x \in \IC^n \, | \, \sum_{i=1}^n Q_i^a |x_i|^2 = \xi^a
 \}/U(1)^s,
\end{equation}
where $U(1)^s$ acts as
\begin{equation}
 x_i \to e^{i \, Q_i^a \, \varphi_a} \, x_i.
\end{equation}
If the FI parameters $\xi_a$ are such that $d := \dim \CM = n-s$,
$\CM$ is a toric variety.

As a simple example, consider a single $U(1)$ with charges $q_i = 1$
for $i=1,\ldots,n$ and $\xi > 0$. Then $\CM = \ICP^{n-1}$. To relate
this to the usual description of $\ICP^{n-1}$ as
$(\IC^n-\{0\})/\IC^*$ where $\lambda \in \IC^*$ acts as $x_i \to
\lambda x_i$, note that the D-flatness condition $\sum_i |x_i|^2 =
\xi$ can be thought of as gauge fixing the real rescalings $x_i \to
|\lambda| x_i$ for $x \neq 0$, leaving only the $U(1)$ part to
divide out.

In general one can similarly represent a toric variety
as $\IC^n$ minus a certain set $Z$ quotiented by the complexified
gauge group $(\IC^*)^s$. The excluded set $Z$ is the set of $x \in
\IC^n$ which cannot be gauge transformed to a solution of the
D-flatness constraints. This can be shown to consist of the union of
planes obtained by putting a subset of the coordinates $x_i$ equal
to zero, such that the D-flatness constraints cannot be solved. Note
that $Z$ thus depends on the choice of FI parameters $\xi$. The
advantage of this description is that it makes holomorphic
properties manifest. The advantage of the gauge linear sigma model
description on the other hand is that it is very concrete and that
specifying a set of FI parameters is in general less complicated
than specifying $Z$. This is the approach we will follow here.

As a less trivial example, consider the space defined by five fields $x_i$ and
$U(1) \times U(1)$ gauge group, with charges
\begin{equation} \label{example2}
\left( \! \!
\begin{array}{l}
 Q^1_i \\
 Q^2_i
\end{array}
\! \! \right)
= \left(
\begin{array}{rrrrr}
 1&1&1&-n&0\\
 0&0&0&1&1
 \end{array}
\right) \, ,
\end{equation}
positive FI parameters $(\xi^1,\xi^2)$, and $n \geq 0$. Thus
\begin{equation}
 \CM_n = \left\{ x \in \IC^5 \, | \, \begin{array}{l} |x_1|^2 + |x_2|^2 + |x_3|^2 - n \, |x_4|^2 = \xi_1
 \\ |x_4|^2 + |x_5|^2 = \xi_2 \end{array} \right\} \, / \, U(1)^2 \, ,
\end{equation}
where the $U(1)^2$ act as
\begin{equation}
 (x_1,x_2,x_3,x_4,x_5) \to (e^{i \varphi_1} x_1,e^{i \varphi_1} x_2,e^{i \varphi_1} x_3,
e^{i (-n \varphi_1 + \varphi_2)} x_4, e^{i \varphi_2} x_5) \, .
\end{equation}
This is a smooth $\ICP^1$ bundle over $\ICP^2$, with the ``amount of
twisting'' determined by $n$.

Any toric variety $\CM$ is complex, with local complex coordinates given by
$U(1)^s$ invariant combinations of the $x_i$. For the $\ICP^{n-1}$
example such coordinates are e.g.\ $t_i=x_i/x_n$, $i<n$ in a patch where $x_n \neq 0$. For the second example we can take for example $t_1=x_2/x_1$, $t_2=x_3/x_1$, $t_3 = x_4 x_1^n/x_5$ in a patch where $x_1 \neq 0$, $x_5 \neq 0$.

Moreover $\CM$ inherits a K\"ahler form from the standard flat K\"ahler form
on $\IC^n$,
\begin{equation} \label{kahlerform}
 J = \frac{i}{2 \pi} \sum_i dx_i \wedge d\bar{x}_i = \frac{1}{2 \pi} \sum_i du_i \wedge
 d\phi_i,
\end{equation}
where $x_i=:\sqrt{u_i} e^{i\phi_i}$ and the normalization is chosen
for later convenience. In the case of $\ICP^{n-1}$, this gives the
K\"ahler form of the standard Fubini-Study metric, as can be seen in the coordinate patch
parametrized by $t_i=x_i/x_n$ by substituting the D-flatness solution $x_i=\sqrt{\xi} t_i
(\sum_j |t_j|^2)^{-1/2}$ in $J$, where we put $t_n \equiv 1$.

\subsection{Divisors, line bundles and intersection numbers} \label{sec:divisors}

A divisor $D=\sum_I n_I S^I$ is a formal sum of holomorphic
hypersurfaces $S^I$, with (positive or negative) integer
coefficients. Physically it can be thought of as a collection of
complex codimension one holomorphic branes and anti-branes. The holomorphic hypersurface $S^I$
is described locally on each coordinate patch $\alpha$ by a
holomorphic equation $f^I_\alpha(t)=0$, such that
$f^I_\alpha/f^I_\beta$ has no zeros or poles on the overlap between
$\alpha$ and $\beta$. To the divisor $D$ we can thus associate in
each patch $\alpha$ the meromorphic function $f_{D,\alpha}=\prod_I
(f_\alpha^I)^{n^I}$ whose zeros and poles describe the positive
resp.\ negative parts of $D$. The functions
$f_{D,\alpha}/f_{D,\beta}$ can be interpreted as transition
functions of a holomorphic line bundle on overlap regions. This
construction thus gives a one to one correspondence between the data describing holomorphic
line bundles and the data describing divisors. One denotes the line bundle corresponding
to the divisor $D$ as $\CO(D)$.

Sums and differences of divisors correspond to products and
quotients of their defining equations. A divisor given by
the zeros and poles of a \emph{globally} defined rational
function corresponds to a trivial line bundle (all transition
functions are 1) and is trivial in homology. Divisors which differ
by such a homologically trivial divisor are called linearly
equivalent.

The toric variety $\CM$ has a particularly simple set of divisors
\begin{equation}
 D_i: x_i = 0.
\end{equation}
More complicated divisors can be constructed as poles and zeros of
rational equations in the $x_i$ transforming homogeneously under the
gauge transformations. Gauge invariant rational functions of the
$x_i$ are globally defined on $\CM$ and hence correspond to
homologically trivial divisors.

For example on $\ICP^{n-1}$,
$x_i/x_j$ is gauge invariant for all $i,j$. Consequently $D_i = D_j$
for all $i,j$, where the equality should be read here as linear
equivalence. So in this case there is only one independent divisor
class. Any homogeneous polynomial equation of degree $k$ describes a
divisor linearly equivalent to $kD_1$. For instance $x_1^5 x_2^3 x_3 + x_1^7 x_4^2 =0$ describes
a divisor in class $9 D_1$ in $\ICP^3$.

In our second example (\ref{example2}), we similarly get the relations
\begin{equation} \label{example2lineq}
 D_1 = D_2 = D_3 \, , \qquad D_4 = D_5 - n \, D_1 \, .
\end{equation}

More generally, divisor classes are completely characterized by the charges
of their defining equation; there will be as many independent
divisors $D_i$ as there are $U(1)$ factors, and they generate all
divisor classes on $\CM$.

The Poincar\'e dual ${\rm PD}_\CM(D)$ of a divisor class $D$ in $\CM$ is an
element of $H^2(X,\IZ)$: If the divisor is locally described by the equation $f(x)=0$, a representative
of the Poincar\'e dual class is $\delta(f) \, df \wedge \delta(\bar{f}) \, d\bar{f}$.

The \emph{intersection product} of $d$ divisor classes plays a
fundamental role in computing just about any topological quantity.
It can be defined as
\begin{equation}
 D_A \cdots D_B = \int_{\CM} {\rm PD}(D_A) \wedge \cdots \wedge
 {\rm PD}(D_B) = \#(D_A \cap \cdots \cap D_B)
\end{equation}
where for the last equality we assumed the divisors to be transversally intersecting at
regular points. The intersection product is invariant under linear equivalence.
Intersection products of less than $d$ divisors are defined
similarly, but now represent curves, surfaces and so on rather than
numbers or points.

As a first example, say we want to compute $D_1 D_2 D_3$ on
$\ICP^3$. Setting $x_1=x_2=x_3=0$ reduces the D-term constraint to
$|x_4|^2=\xi$. The $U(1)$ gauge symmetry can be used to set
$x_4=\sqrt{\xi}$, so the triple intersection is a single (regular)
point, i.e.\ $D_1 D_2 D_3 = 1$. Now, using $D_3=D_2=D_1$, this
immediately gives $D_1^3=1$. Since $D_1$ generates all divisor
classes, this is all we need to know to compute all intersection
products.

In general, for $\ICP^d$, we have $D_1^d=1$.

As an example illustrating how to deal with intersections at orbifold singularities, consider the weighted projective space $W\ICP^2_{1,2,3}$, defined by 3 fields $x_i$ with charges $Q_i=(1,2,3)$,
and FI parameter $\xi>0$. Since $x_1^3/x_3$ and $x_1^2/x_2$ are gauge
invariant, the toric divisor classes satisfy $D_3=3
D_1$, $D_2=2 D_1$. Note that the point $x=(0,\sqrt{\xi},0)$ is fixed under a
$\IZ_2$ subgroup of the gauge group and similarly
$x=(0,0,\sqrt{\xi})$ is fixed by a $\IZ_3$ subgroup. Hence these points
are $\IZ_2$ and $\IZ_3$ orbifold singularities, and some special
care has to be taken in computing intersection products. To compute
$D_1 D_2$, note that $x_1=x_2=0$ is precisely the $\IZ_3$ orbifold
point. The correct value of $D_1 D_2$ is then $1/3$. One way to see
this is to observe that $3 D_1 D_2 = D_3 D_2 = 1$, where the latter
equality follows from the fact that $(\sqrt{\xi},0,0)$ is a regular
point. Similarly $D_1 D_3 = 1/2$, and $D_1^2 = D_1 D_3/3 = 1/6$.

In general, for $W\ICP^d_{Q}$, we have $D_j^d = Q_j^d/\prod_i Q_i = Q_j^{d-1}/\prod_{i\neq j} Q_i$.

As a last example, consider (\ref{example2}) again, for which we obtained the linear
equivalences (\ref{example2lineq}). Let us take $\{ D_1, D_5 \}$
as a basis. All independent triple intersection products can be
computed by using the linear equivalences to reduce to intersection
products of distinct divisors, which in turn can be directly
computed by solving the equations together with the D-term
constraints. It is also useful to note that the D-term constraints in (\ref{example2})
exclude $(x_1,x_2,x_3)=(0,0,0)$ and $(x_4,x_5)=(0,0)$, so $D_1 D_2
D_3 = 0$ and $D_4 D_5 = 0$. This gives the reduction relation $D_5^2
= D_5(D_4+n D_1) = n D_1 D_5$, hence
\begin{eqnarray}
 D_1^3 &=& D_1 D_2 D_3 = 0 \label{ex2int1} \\
 D_1^2 D_5 &=& D_1 D_2 D_5 = 1 \\
 D_1 D_5^2 &=& n D_1^2 D_5 = n \\
 D_5^3 &=& n D_1^2 D_5= n^2. \label{ex2int4}
\end{eqnarray}

\subsection{Curves and K\"ahler moduli} \label{sec:curves}

We will now show that, while the fields $x_i$ correspond to ($n-2$)-cycles $D_i$, the
charges $Q^a$ correspond to 2-cycles $C^a$, and that
the mutual intersection product between the divisors and these curves is
\begin{equation} \label{CDint}
 D_i \cdot C^a = Q_i^a \, ,
\end{equation}
and moreover that for the K\"ahler form $J$ on $\CM$ induced by
(\ref{kahlerform}) we have
\begin{equation} \label{Kahlervol}
 \int_{C^a} J = \xi^a,
\end{equation}
with $\xi^a$ the FI parameters from (\ref{CMdef}). We will check these claims by constructing the 2-cycle classes $C^a$ explicitly.

\begin{figure}
\centering
\includegraphics[totalheight=0.25\textheight]{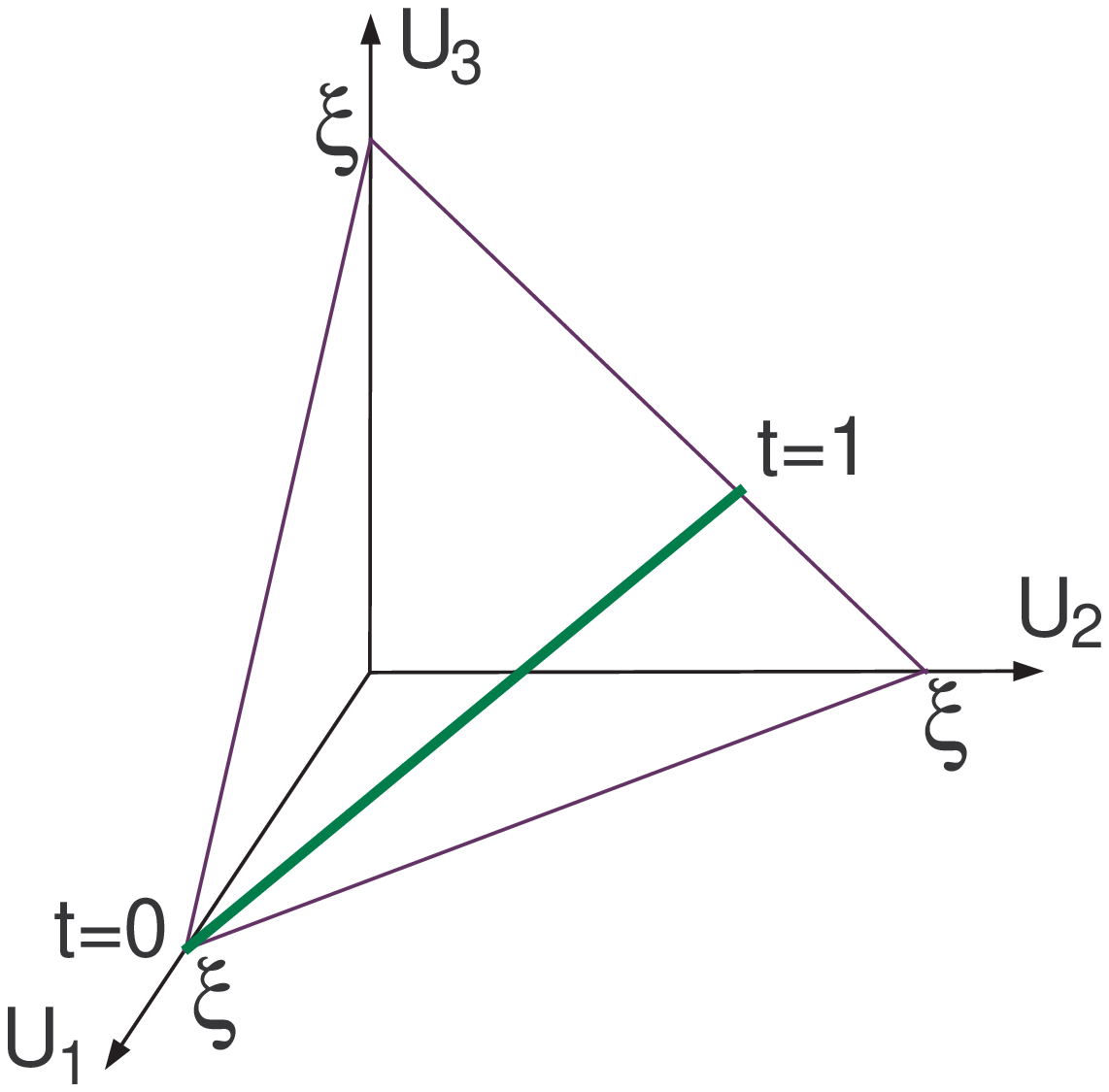}
\caption{Example of a path $U_i(\tau)$ for
$\ICP^2$.}\label{pathexample}
\end{figure}


A representative 2-cycle $C^a$ can be constructed as the
image of a map $X:[0,1] \times [0,2 \pi] \to \CM: (\tau,\sigma)
\mapsto x_i = X_i(\tau,\sigma)$ which we build as follows. First
split the coordinates $x_i$ in two suitable groups by splitting the
index set as the disjoint union $\{1,\ldots,n\}=I_1 \cup I_2$, with
the number of elements in $I_2$ equal to $s$, the number of
$U(1)$'s. Then we put
\begin{equation}
 X_i(\tau,\sigma) = \left\{
 \begin{array}{lcl}
 \sqrt{U_i(\tau)} \, \exp(i \, Q^a_i \sigma) & \mbox{ if } & i \in I_1 \\
 \sqrt{U_i(\tau)} & \mbox{ if } & i \in I_2
 \end{array}
 \right.
\end{equation}
where we choose the path $U_i(\tau)$ such that the following
conditions are met:
\begin{equation} \label{pathcond}
 \sum_{i=1}^n Q_i^a \, U_i(\tau) = \xi^a \, \, \, \forall \tau \in [0,1], \, \forall a, \, \quad U_i(0)=0 \,\, \forall i
 \in I_1, \, \quad U_i(1)=0 \,\, \forall i \in I_2.
\end{equation}
The first constraint enforces the $X_i$ to satisfy the D-flatness
constraints defining $\CM$, while the two last conditions are needed
to make the boundary circles $X_i|_{\tau=0,1}$ collapse to a point
up to gauge transformation, so the 2-cycle is closed in $\CM$. An
example for $\ICP^2$ is plotted in fig.\ \ref{pathexample}.

Then we have for $i \in I_1$:
\begin{equation}
 D_i \cdot C^a = \frac{1}{2\pi i} \oint_{\sigma=0}^{2\pi}
 \left. \frac{dX_i}{X_i} \right|_{\tau = \epsilon} = Q_i^a
\end{equation}
Similarly, after doing the proper gauge transformation $X_i \to
e^{i Q_i^a \sigma} X_i$, we find $D_i \cdot C^a = Q_i^a$ for $i \in
I_2$. This establishes (\ref{CDint}).

To prove (\ref{Kahlervol}), note that
\begin{equation}
 \int_{C^a} J = \int_{C^a} \frac{1}{2 \pi} \sum_{i \in I_1} d(Q_i^a U_i(\tau)) \wedge d\sigma \, = \, \xi^a,
\end{equation}
where in the last equality we used (\ref{pathcond}).

This gives a direct connection between the FI parameters and the
K\"ahler moduli of $\CM$.

In general the 2-cycles $C^a$ or integer linear combinations thereof
will not be holomorphic; the corresponding homology classes might
not even have any holomorphic representatives at all. To construct
holomorphic curves, one can simply take transversal intersections of
$n-1$ of the divisors $D_i$. By taking positive linear combinations,
these in fact generate the full set of all 2-cycle classes with
holomorphic representatives (this set is called the \emph{Mori
cone}). The relation with the $C^a$ can be deduced by comparing the
intersection products $D_i \cdot C^a$ and $D_i \cdot (D_{i_1} \cdots
D_{i_{n-1}})$.

For example for $\ICP^{n-1}$, we have $C^1=D_1^{n-2}$, because indeed $D_1 \cdot (D_1^{n-2}) = 1 = Q^1_1$.
For the
$n$-twisted $\ICP^1$ bundle over $\ICP^2$ defined by (\ref{example2}), we have
\begin{equation} \label{C1C2expr}
 C^1=D_1 D_4 \,, \qquad  C^2=D_1 D_2 \, ,
\end{equation}
so $C^1$ and $C^2$ are holomorphic curve classes, and moreover
since $D_1 D_5 = C^1 + n C^2$ and $n \geq 0$, they generate (with positive coefficients) the full
Mori cone of holomorphic curve classes. (Had we chosen $n \leq
0$, then the Mori cone would be generated by $C^2$ and $\tilde{C}^1
:= C^1 + n C^2 = D_1 D_5$.)

Finally, since for holomorphic curves $C$ the period $\int_C J$
equals the area of $C$, we must have $\int_C J \geq 0$ for all
generators $C$ of the Mori cone. This translates to a set of
inequalities on the $\xi^a$. The space of all $J' \in H^2(\CM,\IR)$
satisfying $\int_C J' \geq 0$ for all $C$ in the Mori cone is called
the \emph{K\"ahler cone}. This corresponds to all possible K\"ahler
classes $J$ which can be obtained by varying the $\xi^a$ without
degenerating the space $\CM$. The space of these deformations is
called the K\"ahler moduli space.

One can choose the generators of the gauge group to be such that the
$C^a$ are generators of the Mori cone. If this is a complete set of
generators, then the K\"ahler cone is given simply by $\xi^a \geq
0$, and if we then choose a basis of divisors $K_a$ dual to the
$C^a$ (i.e.\ $C^a \cdot K_b = \delta^a_b$), we can parametrize the
K\"ahler form as
\begin{equation} \label{Kahlerformpar}
 J = \xi^a K_a \, .
\end{equation}
If there are more than $s$ generators of the Mori cone, there will be additional inequality
constraints on the $\xi^a$. Intersection products of the $K_a$ are always positive.

For $\ICP^{n-1}$, we have $K_1 = D_1$ and $J = \xi^1 K_1$ with
$\xi^1 \geq 0$. For our $n$-twisted $\ICP^1$ bundle over $\ICP^2$,
we have $K_1 = D_1$, $K_2 = D_5$, and $J=\xi^1 K_1 + \xi^2 K_2$ with
$\xi^a \geq 0$.

At the boundary of the K\"ahler cone, a 2-cycle collapses to zero
area. By keeping on varying the $\xi^a$ (so formally the curve area
becomes negative), one often simply continues to a different smooth
geometry in this way, still described by (\ref{CMdef}), but with
different intersection products , and with its own K\"ahler cone.
This is called a flop transition, with the collapsing curve being
flopped. In other cases, there is no transition to a new smooth
geometry, although in the full gauged linear sigma model, the
physics remains sensible.

\subsection{Volumes} \label{sec:volumes}

Using the parametrization $J=\xi^a K_a$ of (\ref{Kahlerformpar})\footnote{One can of course also consider
more general parametrizations $J = J^A E_A$ with $\{ E_A \}_A$ some basis of divisor classes; the resulting expressions are completely analogous.} and the intersection products of the divisors, it is straightforward to compute the volume of $\CM$:
\begin{equation}
 V_{\CM} = \int_{\CM} \frac{J^d}{d!} = \frac{1}{d!} \, K_{a_1} \cdots K_{a_d} \, \xi^{a_1} \cdots \xi^{a_n} \, .
\end{equation}
Explicitly, for say $\ICP^{3}$, parametrizing $J = \xi^1 D_1$, this is, using $D_1^{3} = 1$:
\begin{equation}
 V_{\ICP^{3}} = \frac{(\xi^1)^{3}}{6} \, ,
\end{equation}
and for the $\ICP^1$ fibration over $\ICP^2$, parametrizing $J=\xi^1 K_1 + \xi^2 K_2 = \xi^1 D_1 + \xi^2 D_5$,
using (\ref{ex2int1})-(\ref{ex2int4}):
\begin{equation}
 V_{\CM_n} = \frac{1}{6} \, \xi^2 \left( 3 \, (\xi^1)^2 + 3 \, n \, \xi^1 \xi^2 + n^2 \, (\xi^2)^2 \right) \, .
\end{equation}
The volume of other holomorphic cycles can be computed similarly. In particular the volume of the divisor $K_a$ is
\begin{equation}
 V_{K_a} = \int_{K_a} \frac{J^{d-1}}{(d-1)!} = \int_{\CM} \frac{K_a J^{d-1}}{(d-1)!} = \frac{\partial}{\partial \xi^a} V_{\CM}(\xi) \, ,
\end{equation}
and all other holomorphic divisor volumes can be computed from this by linearity: $V_{\alpha K_a + \beta K_b} = \alpha V_{K_a} + \beta V_{K_b}$. Holomorphic intersections are also easy to compute:
\begin{equation}
 V_{K_a \cap \cdots \cap K_b} = \frac{\partial}{\partial \xi^a} \cdots \frac{\partial}{\partial \xi^b} V_{\CM}(\xi) \, .
\end{equation}
As an example, the volume of $D_1$ in $\ICP^{3}$ is
\begin{equation}
 V_{D_1} = \frac{(\xi^1)^2}{2} \, .
\end{equation}
In the $\ICP^1$ bundle example, we get
\begin{eqnarray}
 V_{D_1} = V_{K_1} &=& \frac{\partial}{\partial \xi_1} V_{\CM_n}(\xi) = \frac{\xi^2(2 \, \xi^1+ n\, \xi^2)}{2}  \, ,\\
 V_{D_5} = V_{K_2} &=& \frac{\partial}{\partial \xi_2} V_{\CM_n}(\xi) = \frac{(\xi_1 + n \, \xi_2)^2}{2} \, , \\
 V_{D_4} = V_{K_2} - n \, V_{K_1} &=& \frac{\xi_1^2}{2} \, .
\end{eqnarray}
More general volumes can be computed easily as well; for example the formal area of the self-intersection of $D_4$ is
\begin{equation} \label{negvol}
 \int_{D_4 \cap D_4} J = \left(\frac{\partial}{\partial \xi_2} - n \, \frac{\partial}{\partial \xi_1} \right)^2 V_{\CM}(\xi) = -n \xi_1 = - n \sqrt{2 V_{D_4}} \, .
\end{equation}
Note that this is negative. This simply indicates that the self-intersection of $D_4$ does not have a holomorphic representative. (In general only transversal intersections of holomorphic objects are again holomorphic.)
Finally note that in this example, we have
\begin{equation} \label{swissvol}
 V_{\CM_n} = \frac{\sqrt{2}}{3n} \left( V_{D_5}^{3/2} - V_{D_4}^{3/2} \right) \, .
\end{equation}
Recalling section \ref{sec:cheese}, this suggests that these manifolds (as base manifolds of elliptically fibered CY fourfolds) might provide examples of the ``Swiss cheese'' scenario of section \ref{sec:cheese}. We will see in section \ref{sec:explconstr} that this is indeed the case.

\subsection{Characteristic classes} \label{sec:charclasses}

Characteristic classes play an important role in string theory in
extracting, from geometrical setups, various physical topological
quantities such as RR charges, moduli space and flux lattice
dimensions, numbers of fermionic zero modes of instantons, and so on. In the following we will first list the
general (smooth differential geometric) definitions of the various
characteristic classes that appear in this paper, and then
specialize to computations of tangent bundle characteristic classes
of toric varieties and algebraic submanifolds thereof.

The total \emph{Chern class} $c=c_1+c_2+\cdots+c_r$ of a rank $r$
holomorphic vector bundle $V$ with $r \times r$ matrix curvature
form $F$ is defined as the cohomology class $\in H^0 + H^2 + \cdots
+ H^{2r}$ of
\begin{equation}
 c(V) = \det(1+\frac{1}{2 \pi} F) = 1 + \frac{1}{2\pi}{\rm Tr} \, F +
 \cdots.
\end{equation}
From the properties of the determinant, it immediately follows that
$c(V_1 \oplus V_2) = c(V_1) \, c(V_2)$. More generally, if we have
three vector bundles $U$, $V$, $W$ such that $U=V/W$ (so locally
$V=U \oplus W$), then we have the \emph{Whitney sum formula}:
\begin{equation}
 c(V)=c(U) \, c(W).
\end{equation}
An important example is given by the tangent and normal bundles of a
holomorphic submanifold $S$ of a manifold $X$. Because the normal bundle
$NS$ of $S$ is the quotient of the tangent bundle $TX$ of $X$
restricted to $S$ by the tangent bundle $TS$ of $S$, we have
\begin{equation} \label{adjformula}
 c(TX)|_S = c(TS) \, c(NS),
\end{equation}
which can be used to compute $c(TS)$ from knowledge of $c(TX)$ and $c(NS)$.
(If for example $S$ is given as the intersection of divisors, $S=S_1
\cap S_2 \cap \cdots \cap S_k$, then we simply have
$c(NS)=\prod_\alpha(1+[S_\alpha]|_S$). When specifically applied to the tangent and normal bundles of a manifold as just shown here, the Whitney sum formula is usually referred to as the \emph{adjunction formula}.

In terms of the eigenvalues $\lambda_m$, $m = 1, \ldots, r$, of
$\frac{1}{2 \pi} F$, we can also write
\begin{equation}
 c(V) = \prod_{m=1}^r (1+\lambda_m). \label{chernroots}
\end{equation}
Thus the $\lambda_m$ can be thought of as the formal roots of the
total Chern class, and for that reason are called the \emph{Chern
roots}. They are very useful to define and relate various
characteristic classes. The Whitney sum formula given above can be
thought of as simply splitting the Chern roots of $V$ into Chern
roots of $U$ and Chern roots of $W$.

The \emph{Euler class} of a holomorphic vector bundle is its top
Chern class:
\begin{equation} \label{eulerclassdef}
 e(V) = c_n(V) = \prod_m \lambda_m.
\end{equation}
In particular the Euler characteristic of a complex manifold $M$
equals the integrated Euler class of its holomorphic tangent bundle:
\begin{equation}
 \chi(M) = \int_{M} e(TM).
\end{equation}
Similarly one defines the \emph{Chern character} as
\begin{equation}
 {\rm ch}(V) := {\rm Tr} \, e^F = \sum_m e^{\lambda_m} = r + c_1 + \frac{1}{2}(c_1^2-2c_2) +
 \frac{1}{6}(c_1^3-3 c_1 c_2 + 3 c_3) + \cdots
\end{equation}
and this satisfies the sum and product formulas ${\rm ch}(V \oplus W) = {\rm ch}(V) + {\rm
ch}(W)$ and ${\rm ch}(V \otimes W)={\rm ch}(V)\, {\rm ch}(W)$. The
\emph{Todd class} is given by
\begin{equation} \label{Todddef}
 {\rm Td}(V) = \prod_m \frac{\lambda_m}{1-e^{-\lambda_m}} = 1 +
 \frac{1}{2} c_1 + \frac{1}{12} (c_1^2+c_2) + \frac{1}{24} c_1 c_2 +
 \cdots
\end{equation}
and is multiplicative, like the Chern class. Finally, the
\emph{Hirzebruch L-genus} is
\begin{equation} \label{Lgenus}
 L(V) = \prod_m \frac{\lambda_m}{\tanh \lambda_m} = 1 + \frac{1}{3} \sum_m
 \lambda_m^2 + \cdots = 1 + \frac{1}{3}(c_1^2-2c_2) + \cdots
\end{equation}
and the \emph{A-roof genus} is
\begin{equation}
 \widehat{A}(V) = \prod_m \frac{\lambda_m/2}{\sinh (\lambda_m/2)} = 1 - \frac{1}{24} \sum_m \lambda_m^2 + \cdots
 = 1 - \frac{1}{24}(c_1^2-2c_2) + \cdots.
\end{equation}

The Chern class of a toric variety is given by the particularly
simple expression
\begin{equation}
 c(\CM) \equiv c(T\CM) = c(\oplus_{i=1}^n \CO(D_i)) = \prod_{i=1}^n (1+D_i),
\end{equation}
where $D_i$ should be read in the last expression as the Poincar\'e
dual to the divisor $D_i:x_i=0$ in $\CM$. More generally we get
formulas for all characteristic classes defined above by the
substitutions $m \to i$, $r \to n$ and
\begin{equation}
 \lambda_i \to D_i.
\end{equation}

For example for $\CM=\ICP^{3}$ we have, after using the relations
between the toric divisors, and putting $H \equiv D_1$:
\begin{equation}
 c(\CM)=(1+H)^4=1+ 4 H + 6 H^2 + 4 H^3 \, ,
\end{equation}
$\chi(\CM)=4$, and ${\rm Td}(\CM)=1+2 H + \frac{11H^2}{6} + H^3$. For weighted projective
space $\CM=\ICP^{n-1}_Q$, defining $H$ by $D_i =: Q_i H$, we have
\begin{equation}
 c(\CM)=\prod_i (1+Q_i H) \, .
\end{equation}
For the $n$-twisted $\ICP^1$ bundle over
$\ICP^2$, we get
\begin{equation}
 c=(1+D_1)^3(1+D_5-n D_1)(1+D_5) \, ,
\end{equation}
so $c_1= (3-n) D_1 + 2 D_5$, $c_2 = 6 C^1 + 3(n+1) C^2$, and $\chi(\CM) = 6$.

It is also straightforward to compute Chern classes of algebraic
submanifolds of toric varieties, by making use of the adjunction
formula (\ref{adjformula}). For a submanifold $\CS$ of $\CM$ defined
by
\begin{equation}
 \CS = S_1 \cap S_2 \cap \cdots \cap S_k
\end{equation}
where the $S_k$ are hypersurfaces given by polynomial equations in
the $x_i$, this yields
\begin{equation} \label{chernCI}
 c(\CS) = \left. \frac{c(\CM)}{\prod_\alpha c(S_\alpha)} \right|_{\CS} =
 \left. \frac{\prod_i(1+D_i)}{\prod_{\alpha}(1+S_\alpha)}
 \right|_{\CS} = 1 + \sum_i D_i - \sum_{\alpha} S_{\alpha} + \cdots |_{\CS} \, .
\end{equation}
Similar formulas hold for the other multiplicative characteristic
classes. It is important to remember however that these formulas can only be directly applied when the
complete intersection $\CS$ is smooth.

As a classic example, consider the quintic hypersurface in $\CM = \ICP^4$:
\begin{equation} \label{quinticeq}
 \CS: \sum_{i=1}^5 x_i^5 = 0.
\end{equation}
Then, putting $H \equiv D_1$,
\begin{equation}
 c(\CS) = \left. \frac{(1+H)^5}{1+5H} \right|_{\CS} = (1 + 10 H^2 -
 40 H^3)|_{\CS}.
\end{equation}
Note that the first Chern class vanishes: the quintic is a
Calabi-Yau manifold. Furthermore
\begin{equation} \label{chiquintic}
 \chi(\CS) = \int_{\CS} c_3(\CS) = (5 H) \cdot (-40 H^3) = -200.
\end{equation}

\subsection{Holomorphic forms and Poincar\'e residues}
\label{sec:residues}

Periods of holomorphic forms play an important role in the computation of super- and K\"ahler potentials. An elegant and useful general construction of such forms is as a Poincr\'e residue, as we now explain.

Any \emph{gauge invariant} meromorphic form on $\IC^n$
\begin{equation} \label{omegadef}
 \omega = R(x) \, dx^1 \wedge \cdots \wedge dx^n,
\end{equation}
where $R(x)$ is a homogeneous rational function
of the $x^i$, descends to a well-defined meromorphic $d$-form (more precisely a $(d,0)$ form) on the
$d$-dimensional toric variety $\CM$. The reduction goes as follows. Let
\begin{equation}
 V^a := \sum_i Q^a_i x_i \frac{\partial}{\partial x_i}
\end{equation}
be the holomorphic vector fields generating the gauge symmetries
(i.e.\ $\delta_a x_i = i \epsilon V^a x_i = i \epsilon Q^a_i x_i$).
Then the contraction of $\omega$ with all vector fields,
\begin{equation}
 \Omega := \omega \cdot \prod_a V^a
\end{equation}
is a globally defined meromorphic $d$-form on $\CM$. If $R(x)$ is a
polynomial, then $\Omega$ is holomorphic.

As a first example, consider $\ICP^{2}$. To make $\omega$ gauge
invariant, $R(x)$ must have charge $-3$, so $R(x)$ cannot be
polynomial and we do not get any holomorphic 2-forms on $\ICP^2$,
but e.g.\ $R(x)=(x_1 x_2 x_3)^{-1}$ will do, leading to the
meromorphic 2-form
\begin{equation}
 \Omega = \frac{dx_2}{x_2} \wedge \frac{dx_3}{x_3} +
          \frac{dx_3}{x_3} \wedge \frac{dx_1}{x_1} +
          \frac{dx_1}{x_1} \wedge \frac{dx_2}{x_2}.
\end{equation}
For the weighted projective space $\CM=W\ICP^3_{1,1,1,-n}$, $R(x)$
must have charge $n-3$. So in particular when $n \geq 3$, there are
holomorphic 3-forms on $\CM$, and when $n=3$, there is a unique one
up to overall scale, namely $R$ a constant:
\begin{eqnarray}
 \Omega &=& x_1 \, dx_2 \wedge dx_3 \wedge dx_4 \, - \,  x_2 \, dx_1 \wedge dx_3 \wedge dx_4 \nonumber \\
 && \, + \,  x_3 \, dx_1 \wedge dx_2 \wedge dx_4 \, + \, 3 \, x_4 \, dx_1 \wedge dx_2 \wedge dx_3. \nonumber
\end{eqnarray}
Indeed when $n=3$, the first Chern class is trivial, $c_1 = D_1 +
D_2 + D_3 + D_4 = 0$, so $\CM$ is a Calabi-Yau 3-fold and has a
unique holomorphic 3-form. Note however that it is noncompact. (In
fact there are no compact toric Calabi-Yau manifolds.)

It is clear that the above construction will always give a unique
holomorphic $d$-form when $c_1=\sum_i D_i$ vanishes.

There is also a natural way to construct meromorphic $(p,0)$-forms
on $p$-dimensional algebraic subspaces of $\CM$ defined by a system
of homogenous polynomial equations. Consider first the case of a
hypersurface $\CS$ given by $P(x)=0$. Let $\omega$ again be as in
(\ref{omegadef}), but now we take the charges of $R(x)$ such that
$\omega/P(x)$ is gauge invariant instead of $\omega$. Then
\begin{equation} \label{Omegaresidue1}
 \Omega := \frac{1}{2\pi i} \oint_{P=0} \frac{\omega \cdot \prod_a
 V^a}{P},
\end{equation}
where the contour is taken to be an infinitesimal loop around $P=0$,
defines a globally well defined meromorphic top form on $\CS$. The
contour integral picks up the so-called \emph{Poincar\'e residue}.
This can be defined more precisely as follows.
Let $\eta$ be a meromorphic $d$-form in an $d$-dimensional space
with a single pole along a a smooth hypersurface $\CS$, locally
described by the equation $z=0$. Near $z=0$ write
\begin{equation}
 \eta = \frac{dz}{z} \wedge \rho + \eta_0,
\end{equation}
where $\rho$ and $\eta_0$ are locally defined holomorphic $d$-forms.
Then the Poincar\'e residue of $\eta$ is the restriction of $\rho$
to $\CS$:
\begin{equation}
 {\rm res}_{\CS} \, \eta := \frac{1}{2\pi i} \oint_{z=0} \eta :=
 \rho|_{\CS},
\end{equation}
which is unique and extends globally on $\CS$.

For a space $\CS$ given by a complete intersection of $k$ divisors
$S_\alpha$ given by the polynomial equations $P_\alpha(x)=0$ in a
toric variety, we can generalize (\ref{Omegaresidue1}) by picking
$R(x)$ to be such that $\omega/\prod_\alpha P_\alpha(x)$ is gauge
invariant, and putting
\begin{equation}
 \Omega := \frac{1}{2\pi i} \oint_{P_1=0} \, \cdots \, \frac{1}{2\pi i}
 \oint_{P_k=0} \,
  \frac{\omega \cdot \prod_a
 V^a}{\prod_\alpha P_\alpha}.
\end{equation}
Again when $c_1=\sum_i D_i - \sum_\alpha S_\alpha$ vanishes, so
$\CS$ is Calabi-Yau, this gives rise to a unique holomorphic top
form on $\CS$.

As an example consider the quintic hypersurface in $\ICP^4$, defined
by a degree 5 homogeneous polynomial equation $P(x)=0$. On it we
have the unique holomorphic 3-form
\begin{equation} \label{intformresid}
 \Omega = \frac{1}{2\pi i} \oint_{P=0} \frac{x_1 \, dx_2 \wedge dx_3 \wedge dx_4 \wedge dx_5 \, + \, {\rm
 cycl.}}{P(x)}
\end{equation}
which in a patch where we gauge fix say $x_1 \equiv 1$ can be evaluated
as
\begin{equation}
 \Omega = \frac{dx_2 \wedge dx_3 \wedge dx_4}{\partial P/\partial
 x_5}.
\end{equation}
Although explicit patch-dependent expressions like this one are often easily computed, the gauge invariant integral form of the residue, like (\ref{intformresid}), is often more useful to compute periods and differential equations satisfied by them. For techniques to explicitly compute periods, see e.g.\ \cite{Berglund:1993ax}.

Another application of the Poincar\'e residue is the one to one map between holomorphic deformations of a divisor $\CS:P(x)=0$ in a Calabi-Yau $n$-fold and holomorphic $(n-1,0)$-forms on $\CS$. For a deformation $\delta P$ of the polynomial $P$, the corresponding $(n-1,0)$-form is
\begin{equation}
 \omega_{\delta P} = \frac{1}{2 \pi i} \oint_{P=0} \Omega \, \frac{\delta P}{P} \, ,
\end{equation}
where $\Omega$ is the holomorphic $n$-form on the CY. Indeed the number of deformations of a holomorphic divisor in a Calabi-Yau is $h^{n-1,0}(\CS)$.

\subsection{Calabi-Yau submanifolds of toric varieties}
\label{sec:CYsubtor}

Complete intersections in toric varieties with vanishing first Chern class,
\begin{equation} \label{CYcond}
 c_1 = \sum_i D_i - \sum_\alpha S_\alpha = 0 \, ,
\end{equation}
provide a large, concrete set of
examples of Calabi-Yau manifolds which can be used as target
manifolds for string, M or F theory. These manifolds inherit their
K\"ahler moduli spaces from the ambient toric variety, and their
complex structure moduli spaces can be identified with the
deformations of the defining polynomials modulo coordinate redefinitions.\footnote{In some cases, there may be additional complex structure deformations which do not correspond to defining polynomial deformations.}

We consider some examples.

The most general quintic submanifold $\CS$ of $\ICP^4$ is given by
an equation of the form $\CS:P_5(x) = 0$ with $P_5$ a homogeneous
degree 5 polynomial. Such a polynomial has ${5+4 \choose 4} = 126$
coefficients. Polynomials which differ only by a $GL(5,\IC)$
coordinate transformations of the $x_i$ are isomorphic, so we have
$126-25=101$ independent complex structure moduli.

We can check our moduli counting by computing the Euler
characteristic from the Hodge numbers the counting implies and comparing to
(\ref{chiquintic}). With 1 K\"ahler modulus and 101 complex
structure moduli, the independent Hodge numbers of $\CS$ are
$h^{1,1}=1$, $h^{2,1}=101$, so the independent Betti numbers are
$b^0=1$, $b^1=0$, $b^2=1$ and $b^3=204$, and $\chi(\CS)=4-204=-200$,
in agreement with (\ref{chiquintic}).

The quintic inherits a K\"ahler class $J_\CS$ from the K\"ahler
class $J=\xi D_1$ of $\ICP^4$, by pulling $J$ back to $\CS$. It is
usually convenient to express $J_\CS$ in terms of a basis of
$H^{1,1}(\CS)$. Such a basis is obtained by intersecting the divisor
basis of the ambient variety with the hypersurface. In this case
this consists of the single element $H_{\CS}:=D_1|_{\CS}$. Then
$J_{\CS}=\xi H_{\CS}$. The intersection numbers for this basis
follow directly from the intersection numbers of the ambient toric
variety:
\begin{equation}
 H_{\CS}^3 = [\CS] D_1^3 = (5 D_1) D_1^3 = 5.
\end{equation}

A somewhat more complicated example is the Calabi-Yau fourfold elliptically fibered over $\ICP^3$ as considered in section \ref{sec:orlimitgen}. To cast this in gauged linear sigma model language, we first define a five complex dimensional toric variety by
introducing seven fields $x_i$, identified with the variables used in section \ref{sec:orlimitgen} as
\begin{equation}
 (x_1,x_2,x_3,x_3,x_5,x_6,x_7)=(u_1,u_2,u_3,u_4,x,y,z) \, .
\end{equation}
We assign these fields the following $U(1) \times U(1)$ charges:
\begin{equation}
\left( \! \!
\begin{array}{l}
 Q^1_i \\
 Q^2_i
\end{array}
\! \! \right)
= \left(
\begin{array}{rrrrrrr}
 1&1&1&1&0&0&-4 \\
 0&0&0&0&2&3&1
 \end{array}
\right) \, .
\end{equation}
And take the corresponding FI parameters $\xi^1$ and $\xi^2$ positive. The assignment of the charges is uniquely fixed (up to change of basis of $U(1) \times U(1)$ generators) by the Calabi-Yau condition (\ref{CYcond}) and the form of the equation $Z:y^2=x^3+fxz^4+gz^6$. This also fixes the charges of the polynomials $f$ and $g$.
We picked a slightly different basis of $U(1) \times U(1)$ generators compared to (\ref{Cstargen1}), to make the associated curves $C^1$ and $C^2$ to form a basis
of the Mori cone. The divisors
\begin{equation}
 K_1=D_1 \, , \qquad K_2=D_7 + 4 \, D_1
\end{equation}
dual to $C^1$, $C^2$ form a basis for the K\"ahler cone. Note the relations
\begin{equation} \label{divrelations}
 D_2=D_3=D_4=K_1 \, , \qquad D_5=2 \, K_2 \, , \qquad D_6=3 \, K_2 \, .
\end{equation}
Using the techniques described in section \ref{sec:divisors}, we find the intersection products
\begin{equation}
 K_1^5=0, \quad K_1^4 K_2 = 0, \quad K_1^3 K_2^2 = \frac{1}{6}, \quad K_1^2
 K_2^3 = \frac{2}{3}, \quad K_1 K_2^4 = \frac{8}{3}, \quad K_2^5 =
 \frac{32}{3}.
\end{equation}
There is a $\IZ_2$ quotient singularity at $y=0$, $z=0$ and a $\IZ_3$ quotient singularity at $x=0$, $z=0$, explaining the fractional intersection numbers. For example the third intersection number is obtained by noting that  $6 K_1^3 K_2^2 = D_1 D_2 D_3 D_5 D_6$, and that the intersection between those five distinct divisors is given by an up to gauge transformations unique, regular point $(0,0,0,\sqrt{\xi_1+4\xi_2},0,0,\sqrt{\xi^2})$.

The elliptically fibered Calabi-Yau $Z$ itself is given by (\ref{CY4eq}):
\begin{equation} \label{nou}
 Z: y^2 = x^3 + f(\vec u) \, x \, z^4 + g(\vec u) \, z^6 \, .
\end{equation}
This hypersurface avoids the quotient singularities of the ambient toric variety: for example if $x=z=0$, then
(\ref{nou}) implies $y=0$, but the point $(x,y,z)=(0,0,0)$ is excluded by the D-term constraints. The homology class of $Z$ is
\begin{equation}
 [Z]= 6 \, K_2 \, ,
\end{equation}
and the intersection products of the pullbacks $\tilde{K}_a \equiv
{K_a}|_{Z}$ are, by the rule $\tilde{K}_a \cdots \tilde{K}_b = [Z] K_a \cdots K_b$:
\begin{equation} \label{tildeintprods}
 \tilde{K}_1^4 = 0, \quad \tilde{K}_1^3 \tilde{K}_2 = 1, \quad \tilde{K}_1^2
 \tilde{K}_2^2 = 4, \quad \tilde{K}_1 \tilde{K}_2^3 = 16, \quad \tilde{K}_2^4 =
 64.
\end{equation}
Note in particular that
\begin{equation} \label{Krel}
 \tilde{K}_2^2=4 \, \tilde{K}_1 \tilde{K}_2 \, .
\end{equation}
The K\"ahler class on $Z$ is $J_Z=\xi^a \tilde{K}_a$ with $\xi^a >
0$, and hence the volume of $Z$ is
\begin{equation}
 V_Z = \frac{1}{24} \int_Z J_Z^4 =  \frac{\xi^2}{6} \left(
 (\xi^1)^3 + 6 \, (\xi^1)^2 \xi^2 + 16 \, \xi^1 (\xi^2)^2
 + 16 \, (\xi^2)^3 \right) \, .
\end{equation}
A representative elliptic fiber of $Z$ is given by $u_1=u_2=u_3=0$, so
its homology class in $Z$ is $E=\tilde{K}_1^3$. Hence the area of
the elliptic fiber is
\begin{equation}
 v=\int_E J = \xi^2 \, .
\end{equation}
This is exactly the parameter $v$ introduced in section \ref{sec:whatisF}.
A section of the base of the fibration is given by $z=0$, so its homology class is
$B=D_7|_Z =\tilde{K}_2-4 \tilde{K}_1$, and its volume
\begin{equation}
 V_B=\int_B \frac{J^3}{6} = \frac{(\xi^1)^3}{6} \, .
\end{equation}
In the F-theory limit on M-theory, we send $v \to 0$. In this case
\begin{equation}
 V_Z \to v \, V_B \, ,
\end{equation}
as expected from the explicit form of the metric (\ref{fibrationmetric}) in this limit.

The total Chern class of the fourfold is, using (\ref{chernCI}) and (\ref{divrelations}) and the intersection numbers:
\begin{eqnarray} \label{cfourfoldex}
 c(Z) &=& \left. \frac{(1+K_1)^4(1+2K_2)(1+3K_2)(1+K_2-4K_1)}{(1+6 K_2)} \right|_Z  \\
 &=& 1 + (48 \tilde{K}_1 \tilde{K}_2 - 10 \tilde{K}_1^2) - 20 (48 \tilde{K}_1^2 \tilde{K}_2 +
 \tilde{K}_1^3) + 23328 \, \omega_Z \, . \nonumber \\
\end{eqnarray}
We repeatedly used $\tilde{K}_2^2 = 4 \tilde{K}_1
\tilde{K}_2$, and $\omega_Z$ denotes the unit volume element on $Z$. In particular we thus read off the
Euler characteristic of the fourfold:
\begin{equation} \label{eulercharfourfoldex}
 \chi(Z) = 23328 \, .
\end{equation}
Hence for this example, the number $Q_c$ appearing in the tadpole cancelation condition (\ref{tpcnc}), i.e.\ minus the curvature induced D3-charge, equals $Q_c = \chi(Z)/24 = 972$.

\subsection{Index formulae}
\label{sec:IndexFormulae}

To count various massless string modes (i.e.\ bosonic and fermionic zero modes), it is very useful to
have index formulae relating various indices to
expressions involving characteristic classes.

For a $k$-form $\omega$ we define the fermion parity $(-)^F
\omega=(-1)^{k} \omega$. The simplest index is
\begin{equation}
 {\rm Tr}_{H^*(X)}\,(-)^F = \sum_{k} (-1)^{k} \, b^{k}(X) = \chi(X) = \int_X e(X),
\end{equation}
where $b^k(X) = {\rm dim} \, H^k(X)$ and $e(X)$ is the Euler class
defined in (\ref{eulerclassdef}).

Twisting this index by the Hodge star operator gives:
\begin{equation} \label{signatureform}
 {\rm Tr}_{H^*(X)}\, (-)^F \, * =  (-1)^d
 (b^{d}_{*+}(X) - b^{d}_{*-}(X))
 = \sigma(X) = \int_X L(X).
\end{equation}
Here $d$ is the complex dimension of $X$, $b^d_{*\pm}(X)$ is the
number of (anti-)selfdual harmonic $d$-forms on $X$, $\sigma$ is
called the signature of $X$, and $L(X)$ is the Hirzebruch $L$-genus
of the tangent bundle of $X$ as defined in (\ref{Lgenus}). This is
the Hirzebruch signature formula.

On K\"ahler manifolds $X$ there are holomorphic versions of
these index theorems, which sum over (bundle valued) $(0,p)$-forms
only. These are typically relevant to count brane moduli or fermionic zeromodes.
The simplest is the arithmetic genus / holomorphic Euler characteristic formula
\begin{equation} \label{argenus}
 {\rm Tr}_{H^{0,*}(X)} (-)^F = \sum_p (-1)^p \, h^{p,0} = \int_X {\rm Td}(X) \, ,
\end{equation}
where the Todd class ${\rm Td}(X)$ was defined in (\ref{Todddef}). This formula will allow us to check whether M5 instantons satisfy the necessary condition (\ref{holeulcond})
to contribute to the superpotential.

This can be generalized to bundle-valued forms:
\begin{equation} \label{HRR}
 {\rm Tr}_{H^{0,*}(X,V)} \, (-)^F = \sum_p (-1)^p \, h^{0,p}(V) = \int_X {\rm ch}(V) \, {\rm Td}(X).
\end{equation}
This is the Hirzebruch-Riemann-Roch theorem.

\subsection{Computing Hodge numbers} \label{sec:hodge}

Hodge numbers $h^{q,p}(X)$, or at least a set of relations between them, can be computed
using the above index theorems.

In particular, from (\ref{HRR}), taking $V = \Omega^q$, i.e.\ the space of $(q,0)$-forms on $X$, and using $H^{0,p}(X,\Omega^q) = H^{q,p}(X)$, we get a formula for the arithmetic genera $\chi_q$:
\begin{equation} \label{HRRgen}
 \chi_q := \sum_p (-1)^p \, h^{q,p}(X) =  \int_X {\rm ch}(\Omega^q) \, {\rm Td}(X) \, .
\end{equation}
${\rm ch}(\Omega^q)$ can be computed as follows. First note that $\Omega^1$ is just the holomorphic cotangent bundle $T^* X$, which is dual to the tangent bundle $TX$, so in terms of Chern roots, if $c(TX) = \prod_i (1+\lambda_i)$, we have $c(\Omega^1) = \prod_i (1-\lambda_i)$, and ${\rm ch}(\Omega^1) = \sum_i e^{-\lambda_i}$. Now $\Omega^2=\Omega^1 \wedge \Omega^1$, i.e.\ the antisymmetrization of $\Omega^1 \otimes \Omega^1$. More physically, one can think of this as the space of 2 particle states built from the 1-particle fermionic states of $\Omega^1$, where the one particle states have curvature eigenvalues $-\lambda_i$. It follows that the 2-particle states have eigenvalues $-\lambda_i-\lambda_j$, $i<j$, so ${\rm ch}(\Omega^2)=\sum_{i<j} e^{-\lambda_i - \lambda_j}$. This reasoning can be continued to higher $\Omega^q$; an efficient way to summarize the result is by the fermionic generating function:
\begin{equation}
 \sum_q {\rm ch}(\Omega^q(TX))\, y^q = \prod_i (1+y e^{-\lambda_i})\, .
\end{equation}
Combining this with (\ref{HRRgen}) and (\ref{Todddef}), we obtain the generating function for all arithmetic genera
\begin{equation}
 \chi(y) = \sum_q \chi_q \, y^q = \int_{X} \prod_{i=1}^r (1+y e^{-\lambda_i}) \frac{\lambda_i}{1-e^{-\lambda_i}}\, ,
\end{equation}
known as the \emph{Hirzebruch genus}.

This allows us to read off the following results. For $\dim_{\IC} X = 2$:
\begin{eqnarray} \label{twofoldchi0}
 \chi_0 &=& h^{0,0} - h^{0,1} + h^{0,2} = \frac{1}{12} \int_X (c_1^2 +
 c_2) \\
 \chi_1 &=&  2 \, h^{0,1} - h^{1,1} = \frac{1}{6} \int_X (c_1^2 - 5
 c_2). \label{twofoldchi1}
\end{eqnarray}
These expressions can be used to determine $h^{0,2}$ and $h^{1,1}$
once $h^{0,1}$ is known.
For $\dim_{\IC} X = 3$:
\begin{eqnarray}
 \chi_0 = h^{0,0} - h^{0,1} + h^{0,2} - h^{0,3} &=& \frac{1}{24} \int_X c_1 c_2,  \label{chi0indformula} \\
 \chi_1 = h^{0,1} - h^{1,1} + h^{1,2} - h^{0,2} &=& \frac{1}{24} \int_X (c_1 c_2 -12 \, c_3).
\end{eqnarray}
Note that the formula for $\chi_0$ allows us to rephrase the necessary condition (\ref{holeulcond}) for M5 instantons to contribute to the superpotential as $\int_{\rm M5} c_1 c_2 = 24$. Furthermore for subspaces of toric varieties, the formula (\ref{chernCI}) allows to explicitly compute this.

Finally, applied to Calabi-Yau fourfolds (so $c_1(X)=0$):
\begin{eqnarray}
 \chi_0 &=& h^{0,0} - h^{0,1} + h^{0,2} - h^{0,3} + h^{0,4} = \frac{1}{720} \int_X 3 \, c_2^2-c_4,
 \label{fourfoldchi0} \\
 \chi_1 &=& h^{0,1} - h^{1,1} + h^{1,2} - h^{1,3} + h^{0,3} = \frac{1}{180} \int_X 3 \, c_2^2 - 31 c_4,
 \label{fourfoldchi1} \\
 \chi_2 &=& 2(h^{0,2} - h^{1,2}) + h^{2,2} = \frac{1}{120} \int_X 3 \, c_2^2 + 79 \,
 c_4. \label{fourfoldchi2}
\end{eqnarray}
For Calabi-Yau fourfolds we also have $h^{0,4}=1$ (and of course
$h^{0,0}=1$) and if the holonomy is full $SU(4)$, then
$h^{0,3}=h^{0,2}=h^{0,1}=0$. The $\chi_0$ equations then becomes
trivial, and the $\chi_1$ and $\chi_2$ equations can be used to
determine $h^{1,2}$ and $h^{2,2}$ in terms of $h^{1,1}$ and
$h^{1,3}$, i.e.\ the number of K\"ahler and complex structure
moduli.

For the fourfold example (\ref{CY4eq}) discussed above, we have
$h^{1,1}=2$ K\"ahler moduli and $h^{1,3}=3878$ complex structure
moduli (obtained by direct counting of the number of polynomial
deformations modulo $GL(4,\IC)$ reparametrizations). Furthermore, using the above expressions together with (\ref{cfourfoldex}), we get
\begin{equation}
 \chi_0 = 2, \qquad \chi_1 = -3880, \qquad \chi_2 = 15564.
\end{equation}
This determines
\begin{equation} \label{fourfoldhodgenumbers}
  h^{1,1}=2, \qquad h^{1,2} = 0, \qquad h^{1,3}=3878, \qquad h^{2,2}= 15564.
\end{equation}
As we see illustrated in this example, we still need the number of complex structure moduli $h^{1,3}$ (or alternatively some other Hodge number such as $h^{1,2}$) as input. In the case at hand we obtained the correct result by counting polynomial deformations modulo coordinate redefinitions. The problem is that sometimes there are complex structure deformations which are not given by polynomial deformations. In this case, more sophisticated techniques are needed; see e.g.\ \cite{Klemm:1996ts} for CY fourfolds in particular.

\section{Statistics of flux vacua} \label{sec:statistics}

We need one last ingredient before we can start our search for explicit realizations of the moduli stabilization scenarios of sections \ref{sec:KKLT} and \ref{sec:cheese}: efficient estimates of distributions of tree level flux vacua over parameter space. This is necessary because constructing fully explicit flux vacua in typical F-theory compactifications would require specifying 20,000 or so flux quanta and finding the corresponding critical points in the 3000 or so dimensional complex structure moduli space, hoping to hit the region of parameter space we are interested in (e.g.\ weak string coupling, tiny cosmological constant, \ldots) --- an effectively intractable task.

The statistical approach to flux vacua was initiated in \cite{Douglas:2003um} building on ideas of \cite{Bousso:2000xa}, and further developed in \cite{Ashok:2003gk,Denef:2004ze,Denef:2004cf} and subsequent works.
An extensive review can be found in \cite{Douglas:2006es}, and a more pedagogical review in \cite{Denef:2007pq}.

\subsection{The Bousso-Polchinski model}

It was pointed out in \cite{Bousso:2000xa} that the freedom one has to turn on
various independent flux quanta in string theory compactifications
can lead to huge ensembles of vacua with a ``discretuum'' of low
energy effective parameters; like the continuum, the discretuum allows for fine tuning, but \emph{without} the massless moduli necessarily associated to continuously variable parameters.

This is true in particular for the cosmological constant, implying
naturally the existence of string vacua with exceedingly small
effective four dimensional cosmological constants, such as our own,
without the need to invoke any (so far elusive) dynamical mechanism
to almost-cancel the vacuum energy.

To see how this comes about, consider the potential induced by some flux $G$ characterized by
flux quanta $N^I \in \IZ$, $I=1,...,b$, of the general form we considered in section \ref{sec:basics}:
\begin{equation}
 V_N(z) \, = \, V_0(z) + \int_Z \| G \|^2 \, = \, V_0(z) + g_{IJ}(z) \, N^I
 N^J \, ,
\end{equation}
where $z$ denotes the moduli of the compactification manifold $Z$
and $g_{IJ}(z)$ is some positive definite effective metric on the
moduli space. Further on we will be interested mainly in F-theory flux vacua, but at this point we just consider the above potential as an abstract starting point for a toy model. In particular we ignore constraints such as
tadpole cancelation conditions. The bare
potential $V_0$ is taken to be negative. In the context of string theory, it will be of the order of some typically high fundamental scale, such as the string or KK scale.

\begin{figure}
\centering
\includegraphics[height=0.35\textheight]{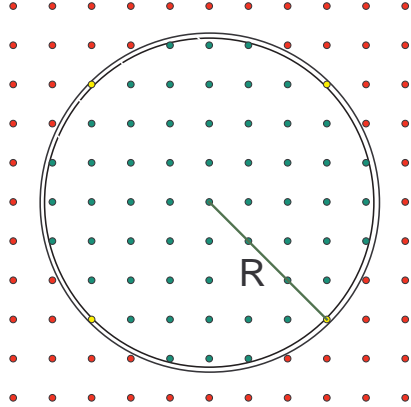}
\caption{The number of lattice points within a certain region of flux space can be estimated by the volume of this region. If the dimension is large, even thin shells can contain exponentially many lattice points.}\label{BP}
\end{figure}

Each vacuum of this model is characterized by a choice of flux
vector $N$ together with a minimum $z_*$ of $V_N(z)$. As a further drastic simplification however,
let us, following \cite{Bousso:2000xa}, simply freeze
the moduli by hand at some fixed value $z=z_0$ and ignore their
dynamics altogether. In that case $V_N$ becomes just a quadratic function of $N$, and it is
then easy to compute the
distribution of cosmological constant values. The number of vacua
with cosmological constant $\Lambda = V_N$ less than
$\Lambda_*$ is now simply given by the number of flux lattice
points in a sphere of radius squared $R^2 = |V_0| + \Lambda_*$,
measured in the $g_{IJ}$ metric. When $R$ is sufficiently large,
this is well-estimated by the volume of this $b$-dimensional ball, i.e.\
\begin{equation}
 {\rm Vol}_b(R) = \frac{1}{\sqrt{\det
g}} \, \frac{(\pi R^2)^{\frac{b}{2}}}{(\frac{b}{2})!} \ \, .
\end{equation}
This leads to the following vacuum number density as a function of $\Lambda$:
\begin{eqnarray} \label{BPdistr}
 d N_{\rm vac}(\Lambda) &\approx& \frac{1}{\sqrt{g}} \, \frac{\pi^{\frac{b}{2}} {\left(
 |V_0| + \Lambda
 \right)^{\frac{b}{2}-1}}}{(\frac{b}{2}-1)!} \, d\Lambda \\
 &\approx& \left(
 \frac{2 \pi e \, (|V_0|+\Lambda)}{\mu^4} \right)^{b/2} \, \frac{d\Lambda}{|V_0|+\Lambda}.
\end{eqnarray}
where $\mu^{4}:=(\det g)^{1/b}$ can be interpreted as the typical mass scale
of the flux part of the potential. To get the last approximate expression, we assumed large $b$ and used
Stirling's formula. Note that in particular at $\Lambda=0$, for say $|V_0|/\mu^4 \sim \CO(10)$, we
get a vacuum density $d N_{\rm vac} \sim 10^{b} \,
d\Lambda/|V_0|$. Hence for $b$ a few hundred, there will be
exponentially many vacua with $\Lambda$ in the observed range
$\Lambda \sim 10^{-120} M_p^4$, even if all fundamental scales
setting the parameters of the potential are of order $M_p^4$!

Thus, in such a model, there is no need to postulate either anomalously
large or small numbers, or an unknown dynamical mechanism, to obtain
vacua with a small cosmological constant.

However, explicitly finding the flux vectors $N^I$ which give rise to such a small cosmological constant is, even this extremely simplified setting, in general an effectively intractable problem: suitably formalized, this inversion problem can be proven to be NP-hard \cite{Denef:2006ad}!

\subsection{Distributions of F-theory flux vacua over complex structure moduli space}

\begin{figure}
\centering
\includegraphics[height=0.25\textheight]{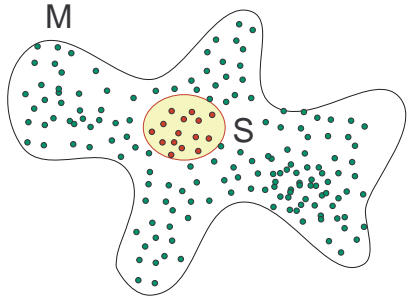}
\caption{Counting vacua.}\label{vacua}
\end{figure}

\subsubsection{Setting up the counting problem} \label{sec:setupcount}

We now turn to the problem of counting genuine F-theory flux vacua and computing their distributions over complex structure moduli space. As we have seen in section \ref{sec:superpotform}, a (tree level) F-theory flux vacuum on an elliptically fibered Calaby-Yau fourfold $Z$ is characterized by
\begin{enumerate}
 \item A choice of flux quanta $N^I$, determining the flux $G_4$ by $[G_4] = N^I \Sigma_{I}$, satisfying the tadpole cancelation condition (\ref{tpcnc})
\begin{equation} \label{tttttttt}
  \frac{1}{2} Q_{IJ} N^I N^J \, + \, N_{\rm D3} = \frac{\chi(Z)}{24} = Q_c \, , \qquad
  Q_{IJ} := \int_Z \Sigma_I \wedge \Sigma_J \, .
\end{equation}
If we require $N_{D3} \geq 0$, this imposes the bound
\begin{equation} \label{fluxbound}
 \frac{1}{2} Q_{IJ} N^I N^J \leq Q_c \, .
\end{equation}
\item A critical point $z^a=z^a_*$ in complex structure moduli space of the superpotential
\begin{equation}
 D_a W_N(z_*) = 0 \, , \qquad W_N(z):= \int_Z G_4 \wedge \Omega_4 = N^I \Pi_I(z) \, ,
\end{equation}
where $D_a W = (\partial_a + \partial_a \CK) W$, with $\CK=-\ln(\Pi_I Q^{IJ} \bar{\Pi}_J)=-\ln \int \Omega \wedge \bar \Omega$ being the K\"ahler potential.

\end{enumerate}

\noindent The number of zeros of a function $f(x)$ of one real variable is
\begin{equation} \label{onevaranalogy}
 \# \{x|f(x)=0 \} = \int dx \, \delta(f(x)) \, |f'(x)| \, .
\end{equation}
Similarly, the number of flux vacua in a given region $S$ of complex structure moduli space $\CM$ is
\begin{equation} \label{Nvacsum}
 N_{\rm vac} = \sum_N \int_{S} d^{2h} z \, \delta^{2h}(DW) \, |\det D^2 W_N| \, ,
\end{equation}
where $h=h^{3,1}(Z)$ is the complex dimension of the complex structure moduli space and the sum is over all fluxes satisfying (\ref{fluxbound}). The determinant factor ensures that each zero of $DW$ contributes $+1$ to the integral, analogous to the $|f'|$ factor in (\ref{onevaranalogy}). In vacua for which $N_{D3} > 0$, there will be residual D3 moduli at tree level. They may however be lifted after inclusion of quantum effects. This will give an additional contribution to the vacuum degeneracy, \emph{not} taken into account in (\ref{Nvacsum}). Similarly, the K\"ahler sector, left completely unfixed at tree level, may after inclusion of quantum effects give an additional vacuum degeneracy, or even destabilize the compactification altogether. At this stage however, we wish to focus exclusively on the tree level flux and complex structure sector, so (\ref{Nvacsum}) is adequate.

Note that the number of vacua without any D3-branes is just
\begin{equation} \label{tadpolesat}
 N_{\rm vac}(N_{D3}=0) = N_{\rm vac}|_{Q_c} - N_{\rm vac}|_{Q_c-1} \, .
\end{equation}

Of course (\ref{Nvacsum}) is not terribly useful yet. To make further progress, we will need to make some approximations. First, we will approximate the sum over fluxes by an integral. This is the analog of computing the number of lattice points in the Bousso-Polchinski sphere by computing its volume, and can be expected to be a good approximation in the large $Q_c$ limit. Second, we will drop the absolute value signs around the determinant factor in (\ref{Nvacsum}). This means we will be counting vacua with signs depending on the number of positive and negative eigenvalues of $D^2W$; in other words we are computing some sort of index. Strictly speaking this will only give a lower bound on the number of vacua, but in practice the index can be expected to give a good estimate of the order of magnitude of the actual number of vacua in the given region, since there is generically no particular reason for large cancelations.

Before we proceed, we will prove a counting formula in abstract generality which can be applied to many different instances of counting of flux vacua.

\subsubsection{A general asymptotic counting formula for zeros of vector field ensembles}

Consider a region $S$ in a space with real\footnote{We switch to real variables here because this makes notation more compact and because it gives a more general formula.} coordinates $x^\mu$, $\mu = 1, \ldots, m$. Let $P_{I\mu}(x)$, $I=1,\cdots,b$ be a set of real vector fields, and let $A_{IJ}$ be a nondegenerate symmetric matrix with inverse $A^{IJ}$. For a choice of integral ``flux quanta'' $N^I$ satisfying the constraint
\begin{equation} \label{ball}
 \frac{1}{2} A_{I J} N^I N^J \leq Q_c \, ,
\end{equation}
we define
\begin{equation}
 U_{N,\mu}(x):=N^I P_{I\mu}(x) \, .
\end{equation}
In applications to counting actual flux vacua, this vector field will essentially be the gradient of the superpotential. The set of ``flux vacua'' we wish to count are labeled by $(N,x_*)$ with
\begin{equation}
 U_{N,\mu}(x_*) = 0 \, .
\end{equation}
Hence, similar to (\ref{Nvacsum}), we wish to estimate
\begin{equation}
 N_{\rm vac} := \sum_N \int_{S} d^m x \, \delta^m(U_{N,\mu}) \, |\det (\partial_\mu U_{N,\nu})_{\mu\nu}| \, ,
\end{equation}
where the sum is restricted to (\ref{ball}). We will approximate this by the continuum index
\begin{equation} \label{indexxdef}
 I_{\rm vac} := \int d^b N \int_{S} d^m x \, \delta^m(U_{N,\mu}) \, \det (\partial_\mu U_{N,\nu})_{\mu\nu} \, .
\end{equation}
This can be evaluated as follows. Define a metric on $S$ by
\begin{equation} \label{auxmetricdef}
 g_{\mu\nu} := P_{I\mu} A^{IJ} P_{J\nu} \, .
\end{equation}
We assume that $g_{\mu\nu}$ is nondegenerate on $S$. We also define a covariant derivative $\nabla$ such that
\begin{equation} \label{nabladef}
 P_{I\mu} A^{IJ} \nabla_\nu P_{J\rho} \equiv 0 \, ,
\end{equation}
i.e.\ $\nabla_\nu v_\rho = \partial_\nu v_\rho - \tilde{\Gamma}^\sigma_{\nu\rho} v_\sigma$ with
\begin{equation}
 \tilde{\Gamma}^\sigma_{\nu\rho} = g^{\sigma\mu} P_{I\mu} A^{IJ} \partial_\nu P_{J\rho} \, .
\end{equation}
Then we claim
\begin{equation} \label{Ivacindex}
 I_{\rm vac} = \frac{1}{\sqrt{\det A_{I J}}} \, \frac{(2 \pi Q_c)^{\frac{b}{2}}}{(\frac{b}{2})!} \int_S e(\nabla) \, ,
\end{equation}
where $e(\nabla)$ is the Euler density derived from the connection $\nabla$:
\begin{equation} \label{eulerdensity}
 e(\nabla) = {\rm Pf} \left( \frac{\CR_{\underline{\mu}\underline{\nu}}}{2 \pi} \right)
\end{equation}
with ${\rm Pf}(\cdots)$ is the Pfaffian and $\CR_{\underline{\mu} \underline{\nu}}$ the curvature form in an orthonormal frame with respect to $g_{\mu\nu}$ (underlined indices are frame indices):
\begin{equation}
 \CR_{\underline{\mu} \underline{\nu}} = \frac{1}{2} R_{\underline{\mu}\underline{\nu}\rho\sigma}\, dx^\rho \wedge dx^\sigma , \qquad
  [\nabla_\rho,\nabla_\sigma] \, v_\nu =: {R^\mu}_{\nu \rho \sigma} \, v_\mu \, .
\end{equation}
The proof goes as follows. We define a generating function
\begin{equation} \label{generatingZ}
 Z(t) = \int d^b N e^{t \frac{1}{2} N^I A_{IJ} N^J} \int_{S} d^m x \, \delta^m(U_{N,\mu}) \, \det(\nabla_\mu U_{N,\nu})_{\mu\nu}
\end{equation}
where now the integral over $N$ is unrestricted. Trading partial derivatives for covariant derivatives in (\ref{indexxdef}) or vice versa here does not affect the result, because the difference between $\nabla_\mu U_{\nu}$ and $\partial_\mu U_{\nu}$ vanishes when $U_\nu = 0$. The index (\ref{indexxdef}) at given $Q_c$ is obtained from the generating function by the contour integral
\begin{equation} \label{IvacfromZ}
 I_{\rm vac}(Q_c) = \frac{1}{2\pi i} \int \frac{dt}{t} \, e^{-t Q_c} Z(t)
\end{equation}
where the contour runs over the imaginary axis passing the pole $t=0$ on the left. (Then if $\frac{1}{2} N^I A_{IJ} N^J - Q_c <0$, we close the contour on the right and we pick up 1 from the pole, while if $\frac{1}{2} N^I A_{IJ} N^J - Q_c >0$ we close the contour on the left and the result vanishes. This enforces the constraint (\ref{ball}).)

Furthermore we write
\begin{eqnarray}
 \delta^m(U_N) &=& \int d^m \lambda \, \, e^{2 \pi i \lambda^\mu P_{I\mu} N^I} \, , \\
 \det (\partial U_N) &=& \int d^m\psi \, d^m\chi \, \, e^{\psi^\mu \chi^\nu \nabla_\mu P_{I\nu} N^I} \, ,
\end{eqnarray}
where the second integral is over Grassmann variables. Substituting this, the integral over $N$ in (\ref{generatingZ}) becomes a simple Gaussian integral,\footnote{In general $A_{IJ}$ need not be positive or negative definite, in which case the Gaussian integral is defined by analytic continuation.} resulting in
\begin{equation}
 Z(t) =  \frac{1}{t^{b/2}} \frac{(2 \pi)^{b/2}}{\sqrt{\det (A_{IJ})}} \int_S d^m x \int d^m \lambda \, d^m\psi \, d^m\chi \,
 \, e^{-\frac{1}{2t}  f_I A^{I J} f_J}
\end{equation}
where
\begin{equation}
 f_I = 2 \pi i \lambda^\mu P_{I\mu} + \psi^\mu \chi^\nu \nabla_\mu P_{I\nu} \, .
\end{equation}
Now note that because of (\ref{nabladef}), the $\lambda$ - $\psi\chi$ cross terms obtained when expanding out $f_I A^{IJ} f_J$ all vanish. The remaining terms are proportional to
\begin{eqnarray}
 \lambda^\mu P_{I\mu} A^{IJ} \lambda^\nu P_{J\nu} &=& g_{\mu\nu} \lambda^\mu \lambda^\nu \\
 \psi^\mu \chi^\nu \nabla_\mu P_{I\nu} A^{IJ}
 \psi^\rho \chi^\sigma \nabla_\rho P_{J\sigma} &=&
 \psi^\mu \psi^\rho \chi^\nu \chi^\sigma P_{I\nu} A^{IJ}
  \nabla_{[\mu} \nabla_{\rho]} P_{J\sigma}  \nonumber \\
  &=& \mbox{$\frac{1}{2}$} \psi^\mu \psi^\rho \chi^\nu \chi^\sigma P_{I\nu} A^{IJ}
  {R^\tau}_{\sigma\mu\rho} P_{J\tau}  \nonumber \\
  &=& \mbox{$\frac{1}{2}$} \psi^\mu \psi^\rho \chi^\nu \chi^\sigma
  g_{\nu\tau} {R^\tau}_{\sigma\mu\rho} \, .
\end{eqnarray}
Performing the Gaussian integrals over $\lambda$  and $\psi$, $\chi$ then gives, using the Grassmann integral representation of the Pfaffian:
\begin{equation}
 Z(t) =  \frac{1}{t^{b/2}} \frac{(2 \pi)^{b/2}}{\sqrt{\det (A_{IJ})}} \int_S e(\nabla) \, ,
\end{equation}
with the Euler density $e(\nabla)$ defined in (\ref{eulerdensity}). Extracting $I_{\rm vac}$ from the contour integral (\ref{IvacfromZ}) finally gives
\begin{equation} \label{Ivacexpr22}
 I_{\rm vac} = \frac{1}{\sqrt{\det A_{I J}}} \, \frac{(2 \pi Q_c)^{\frac{b}{2}}}{(\frac{b}{2})!} \int_S e(\nabla) \, ,
\end{equation}
as claimed.

The prefactor can morally be thought of as giving the volume of a sphere of radius $\sqrt{2 Q_c}$ in flux space. (This is exact when $A_{IJ}$ is positive definite; if not, it is a volume in an analytically continued sense).

If $S$ is taken to be a compact, closed manifold and $e(\nabla)$ is sufficiently well-behaved, then the integral of the Euler density is a topological quantity, the Euler characteristic of the bundle for which $\nabla$ is a connection. For example when the $P_{I\mu}$ are ordinary sections of $T^* S$, then $e(\nabla) = e(T^*S) = e(TS)$, the
Euler characteristic of $S$. In this case, our counting formula reproduces the well known fact that the number of zeros of a vector field on a compact closed manifold, counted with signs, equals the Euler characteristic. However, for our formula, we actually only need the \emph{ensemble} of vector fields to be single valued; there may be monodromies acting on the individual vector fields (as will be the case typically for F-theory flux vacua). Furthermore, $S$ can be any region, and we do not just get the total number of zeros, but their actual distribution as a particular density function $e(\nabla)$.

One interesting general feature following from the expression (\ref{Ivacexpr22}) is that flux vacua will tend to cluster anomalously in singular regions where $e(\nabla)$ diverges. We will confirm this below for the example of flux vacua near conifold degenerations, where strong warping occurs.

\subsubsection{Application to F-theory flux vacua}

Although we derived (\ref{Ivacindex} thinking of the $x^\mu$, $P_{I\mu}$ as real variables, we could have thought of them as complex variables $z^a$, $\Pi_{I a}$ as well, by formally setting $x^\mu = z^\mu$ for $\mu=1,\ldots,h$ and $x^\mu = \bar{z}^{\mu-h}$ for $\mu=h+1,\ldots,2h$ and similarly $P_{I\mu} = \Pi_{I\mu}$ for $\mu=1,\ldots,h$ and $P_{I\mu} = \bar{\Pi}_{I,\mu-h}$ for $\mu=h+1,\ldots,2h$. Everything else would still have gone through, and in particular (\ref{Ivacindex}) remains true. In case $g_{\mu\nu}$ happens to be a hermitian metric, i.e.\ $g_{ab}=0=g_{\bar a \bar b}$, the Euler class can also be written in terms of a determinant, as usual for complex varieties with a hermitian metric.

With this in mind, we can immediately apply our result to counting F-theory flux vacua , taking $S$
to be a region in complex structure moduli space and
\begin{equation}
 \Pi_{Ia}(z) := e^{\CK/2} D_a \Pi_I(z) = e^{\CK/2} (\partial_a + \partial_a \CK) \Pi_I(z) \, .
\end{equation}
where the $\Pi_I(z) = \int \Sigma_I \wedge \Omega$ are the fourfold periods as in section \ref{sec:setupcount}. Furthermore we take $A_{IJ}=-Q_{IJ}$, with $Q_{IJ}$ the intersection product (\ref{tttttttt}).

With these choices, the metric (\ref{auxmetricdef}) has components
\begin{equation}
 g_{ab} = e^\CK \int D_a \Omega \wedge D_b \Omega = 0 \, , \qquad
 g_{a\bar b} = -e^{\CK} \int D_a \Omega \wedge D_{\bar b} \bar{\Omega} = \partial_a {\partial}_{\bar b} \CK \, .
\end{equation}
The first equation holds because of Griffiths transversality: The derivative $\partial_a \omega$ of a $(p,q)$-form $\omega$ with respect to the complex structure moduli produces a form of type $(p,q) + (p-1,q+1)$. Hence $\partial_a \Omega$ and therefore $D_a \Omega$ is of type $(4,0) + (3,1)$. (In fact the covariant derivative $D_a$ is defined in precisely such way that $D_a \Omega$ is exactly of type $(3,1)$.) In any case the wedge product of $(4,0)+(3,1)$ forms is zero, implying $g_{ab}=0$. The second equation is a consequence of the same Griffiths transversality and the definition of $\CK$.

Thus, interestingly, we find that the auxiliary metric (\ref{auxmetricdef}) in this case exactly coincides with the physical metric on complex structure moduli space, which appears in the low energy effective action.

As for the covariant derivative $\nabla$, this is defined in (\ref{nabladef}) by requiring
\begin{equation}
 \int (e^{\CK/2} D_a \Omega) \wedge \nabla_\mu (e^{\CK/2} D_{\bar b} \bar{\Omega}) = 0  \, ,
\end{equation}
where $\mu=c,\bar{c}$. Again using Griffiths transversality, it can easily be shown that this is satisfied for the standard Levi-Civita and K\"ahler covariant connection \cite{Denef:2004ze} on $TS \otimes \CL$ with $\CL$ the K\"ahler line bundle of which the supergravity superpotential is a section:
\begin{eqnarray}
 \nabla_a (D_b \Omega) &=& \partial_a (D_b \Omega) + (\partial_a K) (D_b \Omega) - \Gamma^c_{ab} (D_c \Omega) \, ,  \\
 \nabla_{\bar a} (D_b \Omega) &=& \partial_{\bar a} (D_b \Omega) = g_{b \bar a} \Omega \, .
\end{eqnarray}
Here $\Gamma^c_{ab}$ is the Levi-Civita connection of $g_{a\bar b}$.

Hence we conclude that the continuum index of F-theory flux vacua satisfying (\ref{fluxbound}) is
\begin{equation} \label{FtheoryIvac}
 I_{\rm vac} = \frac{1}{\sqrt{\det Q_{I J}}} \, \frac{(2 \pi Q_c)^{\frac{b}{2}}}{(\frac{b}{2})!} \int_S e(\nabla) \, ,
\end{equation}
where the euler density of $TS \otimes \CL$ can be written, using the fact that we have a complex structure, as
\begin{equation} \label{eulerforF}
 e(\nabla) = \frac{1}{\pi^h} \det(\CR + \omega \, {\bf 1}) \, .
\end{equation}
Here $\CR$ is the curvature form of the holomorphic tangent bundle to $S$ and $\omega = \frac{i}{2} \partial \bar \partial \CK$ is the K\"ahler form on $S$, which is the curvature form of $\CL$.

The F-theory flux lattice dimension is given by $b = b_4'$, where $b_4'$ is the number of 4-form fluxes with one leg on the elliptic fiber; more formally, it is the dimension of the subspace of $H^4(Z)$ orthogonal to intersections of divisors, i.e.\ satisfying (\ref{restrictedflux}). For the fourfold example (\ref{CY4eq}), we read off from (\ref{fourfoldhodgenumbers}) that $b_4 = 23322$, while (\ref{restrictedflux}), taking into account the relation (\ref{Krel}), imposes two independent constraints. Therefore $b_4' = 23320$. Furthermore, as we saw below (\ref{eulercharfourfoldex}), $Q_c = 972$. The intersection form on the full lattice $H^4(X,\IZ)$, as on any middle cohomology lattice on a compact manifold, is unimodular, i.e.\ has determinant 1. The sublattice of divisor intersections can be seen to be unimodular too using the results of (\ref{tildeintprods}), and the orthogonal complement of a unimodular lattice is unimodular. Therefore $\det Q_{IJ} = 1$, and
\begin{equation}
 I_{\rm vac} = 5 \times 10^{1786} \, \int_S e(\nabla) \, .
\end{equation}
If we let $S$ be the entire complex structure moduli space, then the integral equals the Euler characteristic of $TS \otimes \CL$.\footnote{Actually since the moduli space has singularities, the notion of Euler characteristic is ambiguous, and in particular the integral of the above Euler density need not coincide with the topological Euler characteristic. We assume it nevertheless coincides with at least one of the several natural notions of Euler characteristic for singular varieties.} Since the moduli space is some simple quotient of a projective space (namely the space of coefficients of the defining polynomial modulo coordinate redefinitions), one expects this number to be essentially order 1 compared to the exponential prefactor.

The continuum index of vacua with $N_{D3}=0$ is, analogous to (\ref{tadpolesat}):
\begin{equation}
 I_{\rm vac}(N_{D3}=0) = I_{\rm vac}|_{Q_c} - I_{\rm vac}|_{Q_c-1} \, .
\end{equation}
In fact, when the number of vacua is exponentially large, almost all flux vacua have $N_{D3}=0$ according to this estimate; this is related to the fact that for a high dimensional sphere, almost all enclosed volume is located very near its boundary. For our example:
\begin{equation} \label{D3zerofrac}
 \frac{I_{\rm vac}(N_{D3}=0)}{I_{\rm vac}} = 0.999994 \, .
\end{equation}
This illustrates we have to be particularly careful not to naively apply our low dimensional intuition to high dimensional situations.

We will discuss to what extent the continuum index does (not) give a good estimate for the actual number of vacua, counted with or without signs, in section \ref{sec:regvalidity}.

A final comment is in order. For a small domain $S$, $I_{\rm vac}$ can be quite precisely thought of as the volume in flux space of the set of $\vec{N}$ which give rise to a solution of $DW_{\vec N}=0$ located in $S$. Given the somewhat formal nature of the general computation of $I_{\rm ind}$, in particular the use of analytic continuation in evaluating the Gaussian integral $\int dN e^{-t \frac{1}{2}NQN}$, one may worry if the result we find does correctly represent this volume in flux space. In particular, given the fact that $Q_{IJ}$ is not positive or negative definite, one might worry that the actual volume is in fact infinite. However, the condition $DW=0$ effectively renders $Q_{IJ}$ positive definite, since as we saw in section \ref{sec:superpotform}, $DW=0$ implies $G_4=*G_4$, and therefore $N^I Q_{IJ} N^J = \int G \wedge G = \int G \wedge *G \geq 0$. Hence for any finite region $S$ away from singularities, $I_{\rm ind}$ will indeed be finite. Finiteness near singularities and of the actual number of IIB flux vacua has been analyzed in \cite{Ashok:2003gk,Eguchi:2005eh,Torroba:2006kt}. In \cite{Acharya:2006zw} the question of finiteness of string vacua was addressed in a much more general setting, and it was argued that, remarkably, in regimes which are in principle under control, the total number is finite as long as one stays bounded away from decompactification limits (characterized by KK modes becoming light).

\subsubsection{Application to IIB bulk flux vacua} \label{sec:bulkfluxstat}

Most of the statistical analysis in the literature has been done purely in simplified models in which one only considers the IIB bulk sector, neglecting D7 degrees of freedom. One can effectively think of these simplified models as F-theory on $Z = T^2 \times X$ with $X$ some Calabi-Yau 3-fold. Note that taken literally, these models have zero Euler characteristic, so $Q_c = 0$ and no flux vacua. However, we can still formally count solutions to $D_a W_N=0$, by choosing some $Q_c$ by hand. The number of effective F-theory fluxes is now
\begin{equation}
 b_4'(Z) = 2 \, b_3(X) \, ,
\end{equation}
namely $b_3$ RR fluxes ($G_4$ leg on $B$-cycle $T^2$) and $b_3$ NSNS fluxes ($G_4$ leg on $A$-cycle $T^2$). The intersection form is obtained from the symplectic intersection forms on $X$ and $T^2$, and again unimodular. With these substitutions, all of the above formulae remain valid. These simplified models are presumed to give estimates in some sense for the number of bulk flux vacua in the weak IIB coupling limit, although this has not been made precise. From our considerations in section \ref{weakcouplinglimitflux}, it seems plausible that this makes sense if all D7 branes are coincident with the O7-planes, although turning on bulk NSNS fluxes generically does not appear to keep the D7-branes there.

Taking $X$ to be the Calabi-Yau 3-fold arising in the weak coupling limit of the model (\ref{CY4eq}), we get $b_3(X)=300$, and therefore the continuum index for a region $S$ of the bulk moduli space (complex structures of $X$ and $T^2$), putting $Q_c = 972$, is
\begin{equation}
 I_{\rm vac} \approx 2 \times 10^{521} \, \int_S e(\nabla) \, .
\end{equation}
The restriction to these simplified models is why $10^{500}$ is such an infamous number, rather than one of the much bigger numbers one gets out of the full F-theory estimates.

\subsubsection{Toy model} \label{sec:toystat}

\begin{figure}
\centering
\includegraphics[height=0.4 \textheight]{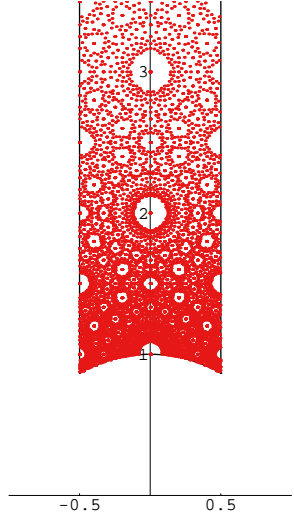}
\caption{Values of $\tau$ for rigid CY flux vacua with $Q_c = 150$.}\label{T2vacua}
\end{figure}

As a simple illustration, consider again the toy model introduced at the end of section \ref{sec:Fleeee} and further analyzed at the end of section \ref{sec:superpotform}. As we saw there, the critical points can be computed exactly, and all inequivalent flux vacua for a given $Q_c$ can be systematically enumerated \cite{Ashok:2003gk}. The exact vacua for $Q_c = 150$ are plotted in fig.\ \ref{T2vacua}. The continuum index distribution is also straightforwardly obtained \cite{Ashok:2003gk,Denef:2004ze}.  The Euler density is
\begin{equation} \label{eT2}
 e(\nabla) = \frac{i}{2 \pi} \frac{d\tau \wedge d\bar{\tau}}{(\tau-\bar{\tau})^2} \, ,
\end{equation}
hence
\begin{equation}
 I_{\rm vac} = 2 \pi Q_c^2 \, A(S) \, ,
\end{equation}
where $A(s)$ is the area of the region under consideration in the K\"ahler metric (\ref{toymetric}). Letting $S$ be the entire fundamental domain, we get $I_{\rm vac} = \pi^2 Q_c^2/6$.

\begin{figure}
\centering
\includegraphics[height=0.25\textheight]{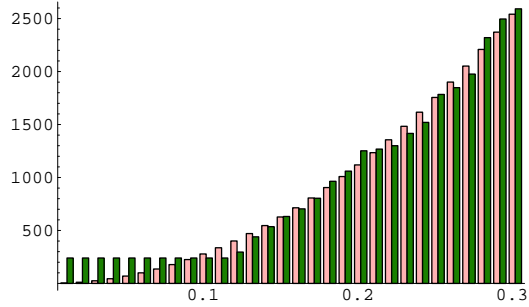}
\caption{Number of vacua in a circle of coordinate radius $R$ around $\tau = 2i$, with $R$
increasing in steps $dR = 0.01$. Pink bars give the estimated value, green bars
the actual value. The actual number starts at a nonzero value for $R=0$ because $\tau=2i$ is multiply degenerate.}\label{discs}
\end{figure}

Despite the intricate fine structure as evident from figure \ref{T2vacua} (in particular the striking ``voids'' around simple complex rational numbers), it is nevertheless true that for large $Q_c$ a disc of sufficiently large area $A$ will contain approximately $2 \pi A Q_c^2$ vacua.
This is illustrated for $Q_c=150$ in figure \ref{discs}, where
estimated and real numbers of vacua are compared in discs around the center of the largest hole $\tau = 2i$ of stepwise increasing radius.

In more complicated models there are many more fluxes and the periods are highly complex functions. As a result, flux vacua will be much more randomized than in this simple example, and continuum distributions can be expected to become good approximations already at finer scales. Some more comparisons between exact and approximate distributions can be found in \cite{DeWolfe:2004ns}.

\subsection{Regime of validity and improved estimates} \label{sec:regvalidity}

We now turn to the question when the continuum index $I_{\rm vac}$ is a good approximation for the actual number of vacua, or at least the actual discrete index of vacua. On general grounds we expect the continuum approximation to be valid in the large $Q_c$ limit, but since $Q_c$ is given to us by the topology of $Z$, we need to understand better what qualifies as ``large''.

To get an idea when the approximation certainly fails, we approximate the prefactor of (\ref{FtheoryIvac}) using Stirling's formula (and assuming $\det Q=1$) as
\begin{equation}
 \left(\frac{4 \pi e Q_c}{b_4'}\right)^{b_4'/2} \, .
\end{equation}
When $b_4' > 4 \pi e Q_c$, this is in fact exponentially small! This is related to the fact that the volume of a sphere of fixed radius goes to zero exponentially when the dimension is sent to infinity. Clearly, in this regime, the approximation breaks down badly. The reason is that in this regime a large fraction of the flux quanta will be zero or some small integer, so the continuum approximation is no longer valid.

In general we have $Q_c = \frac{\chi(Z)}{24}$ and $\chi(Z)=2+2h^{1,1}-2h^{2,1}+b_4$, so in models with $h^{1,1}, h^{2,1} \ll b_4$ (as is the case for our example and in the models listed in appendix B.4 of \cite{Klemm:1996ts}), we have $\chi(Z) \approx b_4 \approx b_4'$. Then
\begin{equation}
 \frac{4 \pi e Q_c}{b_4'} \approx \frac{\pi e}{6} \approx 1.4 \, ,
\end{equation}
so we are barely above the threshold where things go wrong badly. This indicates the continuum index $I_{\rm vac}$ may be a serious underestimate of the actual number of flux vacua in F-theory.

\begin{figure}
\centering
\includegraphics[width=\textwidth]{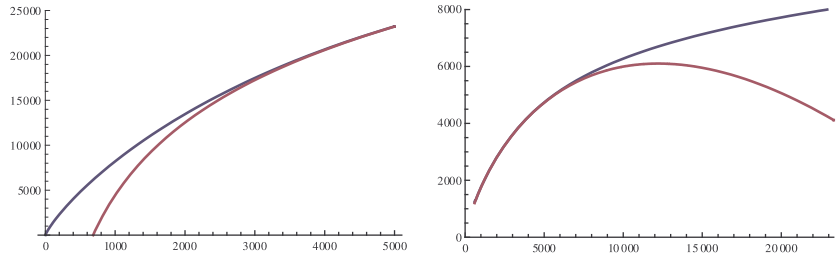}
\caption{Left: The blue (upper) line shows $\ln N(b,Q_c)$ as a function of $Q_c$ for $b=23220$, the red (lower)
line shows the continuum estimate. Right: same but as a function
of $b$ for $Q_c=972$.}\label{theta}
\end{figure}

To make this more precise, let us consider as a toy model the problem of counting the number of lattice points  $\vec n \in \IZ^b$ in a sphere of radius $\sqrt{2 Q_c}$. In the large $Q_c$ limit, this is the volume of the sphere:
\begin{equation}
 N(b,Q_c) \approx \frac{(2 \pi Q_c)^{b/2}}{(\frac{b}{2})!} \, \qquad (Q_c \to \infty) \, .
\end{equation}
The exact number can be represented as
\begin{equation}
 N(b,Q_c) = \frac{1}{2 \pi i} \int \frac{dt}{t} \, e^{-t Q_c}\, Z(t) \, , \qquad Z(t):=\sum_{\vec n} e^{t \, \vec n^2/2} = \left(\vartheta_3(e^t)\right)^{b} \, ,
\end{equation}
with $\vartheta_3(q) := \sum_{n \in \IZ} q^{n^2/2}$, and where we take the contour along the imaginary axis, passing the pole $t=0$ on the left (compare to (\ref{IvacfromZ})). For large $b$, the integral can be computed by saddle point evaluation:
\begin{equation}
 \ln N(b,Q_c) \approx S(t_*) \, , \quad \partial_t S(t_*)=0 \, , \quad S(t):=-\ln t - Q_c t + b \, \ln \vartheta_3(e^t) \, . \end{equation}
The results relevant for our usual example are shown in fig.\ \ref{theta}. We see that the continuum approximation becomes very good when $Q_c$ becomes larger than about $b/8$, but that in the regime of interest $Q_c \approx b/24$, the approximation is poor.\footnote{In the simplified bulk flux models introduced in \ref{sec:bulkfluxstat}, which have been the main focus in the literature, this problem typically does not arise, because the number of IIB bulk fluxes is usually much smaller than the total number of F-theory fluxes. In our example, the number of bulk fluxes is $b=600$, so $Q_c=972$ is well above $b/8$, and the continuum approximation is excellent.}

In fact when $Q_c \approx b/24$ and $b$ is large, we find $t_* \approx -6.18$ and
\begin{equation} \label{improvedest}
 \ln N(Q_c) \approx 8.27 \times Q_c \, , \qquad N(Q_c) \sim 10^{3.59 \times Q_c} \, .
\end{equation}
The continuum estimate on the other hand gives $N(Q_c) \sim 10^{1.84 \times Q_c}$. For our example $Q_c = 972$, so $N \sim 10^{3489}$ while the continuum estimate gives a measly $N \sim 10^{1788}$.

A natural guess for an improved estimate of the number of F-theory flux vacua would be to replace the volume factor in (\ref{FtheoryIvac}) by our toy model $N(b,Q_c)$
\begin{equation}
 I_{\rm vac}' = N(b,Q_c) \int_S e(\nabla) \, .
\end{equation}
However, one should worry that the sparseness of the typical flux vector in the ensemble will have significant effects on the distribution density as well, drastically modifying the $e(\nabla)$ density valid in the continuum approximation. In particular, one could imagine discrete effects such as clustering at enhanced symmetry loci to become more important. This has not been studied yet.

Note that as a rule of thumb, this estimate amounts to about a factor of 10 per moduli space dimension.

\subsection{More distributions}

\subsubsection{Distribution of $W$}

So far we have only discussed estimates for the total number of flux vacua and their distribution over complex structure moduli space. It is not hard to extend this to distributions of other quantities, such as the distribution over the $w:=e^{\CK/2} W$ plane. To estimate the latter in the continuum index approximation, one can insert an additional $\delta^2(e^{\CK/2} W-w)$ in the generating function $Z(t)$ for $I_{\rm vac}$, rewrite this using Lagrange multipliers, integrate out $N$ and use Griffiths transversality again, to find that the net effect of this additional insertion is (up to some constant factor) $Z(t) \to Z(t) \, t \, e^{t |w|^2}$. So effectively, this amounts to replacing $b_4' \to b_4'-2$ and $Q_c \to Q_c - |w|^2$, and we find for the combined distribution
\begin{equation}
 d I_{\rm vac} \propto N(b_4'-2,Q_c-|w|^2) \, d^2 w \, e(\nabla) \, .
\end{equation}
where $N(b,Q_c)$ is the usual sphere volume factor in the continuum approximation, or the function $N(b,Q_c)$ introduced in the previous subsection in cases where we believe this to be a better estimate. Note that at large $b_4'$, due to the exponential dependence of $N(Q_c,b)$ on $Q_c$, this distribution is approximately Gaussian on the $w$-plane, peaking at $w=0$ and cut off at $|w|^2=Q_c$. This is as one would expect if one thinks of $W$ as being the result of a random addition of a large number of complex numbers. The cutoff can be understood as well, it comes from $Q_c \geq \frac{1}{2} G^2 = |G^{4,0}|^2 + \frac{1}{2} |G^{2,2}|^2 \geq |G^{4,0}|^2 = |w|^2$, where we used that $DW=0 \Leftrightarrow G^{3,1}=0$.

The width of the Gaussian is $\sigma \sim (\partial_Q \ln N(b,Q))^{-1/2}$, which in the case of fig.\ \ref{theta} is somewhat less than one. For applications in constructions of string vacua we are however mainly interested in vacua with $|w|^2 \ll 1$. In this regime, the distribution becomes uniform on the $w$-plane:
\begin{equation} \label{smallwdistr}
 d I_{\rm vac} \sim I_{\rm vac,tot} \, d^2 w \, ,
\end{equation}
as can be expected on general grounds. In particular this implies we can expect vacua with $|w|^2$ roughly as small as $1/N_{\rm vac}$.

\subsubsection{Distribution of string coupling constants}

From the considerations in section \ref{sec:leeaweak}, we know that in Sen's weak IIB coupling limit, the fourfold complex structure moduli space $\CM_Z$ factorizes in a dilaton-axion moduli space, a threefold complex structure moduli space, and (depending on the point in the threefold complex structure moduli space), a D7 moduli space. Hence we get a corresponding factorization of the continuum index density
\begin{equation}
 dI_{\rm vac} \propto e(\nabla) = \omega_{\tau} \wedge \rho \, ,
\end{equation}
where $\rho$ is some $\tau$-independent density on the threefold complex and D7 moduli spaces, while $\omega_\tau$ is the K\"ahler form on the dilaton-axion moduli space (which proportional to the curvature form), i.e.\
\begin{equation}
 \omega_{T^2} = \frac{i}{2 \pi} \frac{d\tau \wedge d\bar{\tau}}{(\tau-\bar{\tau})^2} \, ,
\end{equation}
as in (\ref{eT2}).

This implies in particular that universally in the weak coupling limit (i.e.\ $\Im \tau$ sufficiently large), $\tau$ is uniformly distributed w.r.t.\ the standard Poincar\'e metric on the upper half plane. In terms of the string coupling constant $g_s = 1/\Im \tau$, this is simply the uniform distribution:
\begin{equation} \label{gsdistr}
 dI_{\rm vac} \propto d g_s \, .
\end{equation}
The continuum approximation for the distribution is expected to be accurate down to $g_s \sim 1/\sqrt{Q_c}$, where ``void'' effects like in the toy model might start to get important. (For example in the toy model the actual minimal value of $g_s$ is $1/Q_c$, although the continuum approximation predicts an order $1/Q_c^2$ minimum; the discrepancy can be thought of as being due to the void around $\tau=i\infty$.)


\subsubsection{Conifold clustering and distribution of warp factors}

Consider a simplified bulk flux model as described in section \ref{sec:bulkfluxstat}, with $X$ a one modulus Calabi-Yau threefold. An example is the mirror quintic, described by
\begin{equation}
 X: x_1^5 + x_2^5 + x_3^5 + x_4^5 + x_5^5 - 5 \psi x_1 x_2 x_3 x_4 x_5 = 0
\end{equation}
in $\ICP^4$, modulo phase transformations $x_i \to e^{2 \pi i k_i/5}$ leaving this equation invariant. $X$ acquires a conifold singularity when $\psi=1$. Parametrizing $z \equiv \psi-1$, the distribution near $z=0$ in the continuum approximation can be computed from (\ref{eulerforF}) to be \cite{Denef:2004ze}
\begin{equation}
 dN_{\rm vac}(z) = dI_{\rm vac}(z) \propto \frac{d^2 z}{|z|^2 \ln^2 |z|^{-1}} \propto d \biggl(\frac{1}{\ln |z|^{-1}} \biggr)\, .
\end{equation}
The distribution diverges at $z=0$, but in an integrable way, and is approximately scale invariant. As a result, there will be a sizable number of flux vacua exponentially close to the conifold point. Now recall from section \ref{sec:warping} that flux vacua close to conifold points develop a strongly warped KS-type throat. The redshift at the bottom of the throat is given by (\ref{warpzrel}): $\mu \sim |z|^{1/3}$, so the above distribution can be viewed as a distribution for warp factors. In the case of the mirror quintic, this gives, taking into account numerical factors and setting $\mu \equiv |z|^{1/3}$, about 3\% of all flux vacua has $\mu < 10^{-1}$, 0.7 \% has $\mu < 10^{-5}$, and 0.3 \% has $\mu < 10^{-12}$.

Similar to the string coupling constant, one expects the continuum distribution to be accurate for $\frac{1}{\ln |z|^{-1}}$ roughly down to $1/\sqrt{Q_c}$.

%

Having 0.3\% of flux vacua with warping $\mu < 10^{-12}$ may not sound like a terribly spectacular enhancement. Admittedly, it isn't. However in generic actual models, there are many more 3-cycles which could potentially shrink to a tiny size, and this may lead to a much higher fraction of vacua with one or more warped throats, as the following simple argument shows. Imagine we have a Calabi-Yau threefold with $b$ 3-cycles which could potentially collapse to zero size. Then the mirror quintic data suggests that a naive rough estimate for the fraction of vacua for which \emph{all} of these 3-cycles remain larger than the size to get $\mu < 10^{-12}$ is equal to something like $(1-0.003)^b \approx e^{-0.003 \times b}$. Now when $b$ becomes large, this can become a small fraction. For example if $b \sim 250$, about half of all vacua do have $\mu < 10^{-12}$, and if $b \sim 500$, this goes up to 80\%. Of course the actual numbers we used here are just for illustration purposes; but the general idea should be clear. This was studied in more detail in \cite{Hebecker:2006bn}.

The above simple argument relies on many poorly justified assumptions though, and is therefore not conclusive. No concrete model has been studied in which these ideas have been tested against actual distributions.

\subsubsection{Distribution of compactification scales}

In the KKLT scenario of moduli stabilization, the compactification radius $R$ is determined by the value of $|w|^2=e^\CK |W|^2$:
\begin{equation}
 R^4 \sim \ln |w|^{-2} \, .
\end{equation}
Therefore, from (\ref{smallwdistr}), in this scenario, the KK scale is distributed as
\begin{equation}
 dN_{\rm vac} \propto d e^{-R^4} \, .
\end{equation}
That is, large volumes are exponentially suppressed, with maximal values of order $R_{\rm max}^4 \sim \ln N_{\rm vac} \sim Q_c$.

In the large volume scenario on the other hand, we have according to (\ref{swissstab})
\begin{equation}
 R^6 \sim e^{\xi^{2/3}/g_s} \, .
\end{equation}
Thence, from (\ref{gsdistr}),
\begin{equation}
 dN_{\rm vac} \propto d\left( \frac{1}{\ln R} \right) \, .
\end{equation}
Thus, in this scenario, we have an approximate scale invariant distribution of KK scales.

\subsubsection{Distributions of nonsupersymmetric flux vacua}

The flux vacua we have been considering so far are generically nonsupersymmetric at tree level, with supersymmetry breaking scale $F^2 \sim |w|^2 m_s^4$, due to the fact that $D_T W \neq 0$. However in for example the KKLT scenario, supersymmetry gets restored by $T$-dependent quantum corrections. There could be other minima of the full effective potential where supersymmetry is still broken, by some generic $F_a = D_a W \neq 0$. Note that at tree level $D_a W \neq 0$ is forbidden by the equations of motion, so in order to get such minima, the full quantum corrected effective potential must be considered. This is in general a complicated problem.

A slightly simplified model is to consider again the flux superpotential $W$ and the supergravity potential $V = e^\CK(|DW|^2 - 3|W|^2)$, but now without including any contributions from the K\"ahler moduli --- in fact pretending there are no K\"ahler moduli whatsoever in the game. In particular solutions to $D_a W=0$ will now have negative $V$, because the covariant derivatives with respect to the K\"ahler moduli are no longer there and so no longer kill off the $-3|W|^2$ term.

By ``nonsupersymmetric flux vacua'' we mean in this model the minima of $V$ which have $F_a \sim D_a W \neq 0$. We are interested in the regime $|F| \ll 1$, that is supersymmetry breaking well below the fundamental scale. Then it was shown in \cite{Denef:2004cf}, and more intuitively explained in \cite{Dine:2005yq}, that for generic flux vacua the distribution of supersymmetry breaking scales $F$ and cosmological constants $\Lambda$ goes as
\begin{equation}
 dN_{\rm vac}(F,\Lambda) \propto F^5 dF \, d\Lambda \, .
\end{equation}
This ``favors'' high scale supersymmetry breaking. In \cite{Dine:2005yq} the possiblity was considered that other branches of the landscape could exist where low scale breaking was favored. The issue whether string theory favors high or low susy breaking remains inconclusive, and will remain so as long as we have no clue what ``favored'' means.

\subsection{Metastability, landscape population and probabilities}

\begin{figure}
\centering
\includegraphics[height=0.25 \textheight]{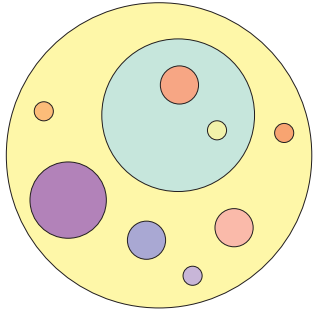}
\caption{Bubbles in bubbles.}\label{bubbles}
\end{figure}

So far we have only considered \emph{number} distributions of flux vacua over parameter space. It is of course tempting to wonder if there is any sense in which one could, as in statistical mechanics, compute \emph{probability} distributions on parameter space. This, of course, immediately runs into a heap of conceptual problems. To begin with, it already unclear of \emph{what} exactly these would be probabilities. At a vague level, one would imagine it to be the probability of finding ourselves in a particular vacuum, but making this precise, given the fact that we already found ourselves here and that there is no way to repeat the experiment, is challenging to say the least.

But regardless of the precise definition of these probabilities, it is clear that to determine them, cosmological considerations will come into play in an important way. In particular, it is necessary to consider the mechanism by which vacua actually come into existence. As we saw in section \ref{sec:Ffluxes}, fluxes are sourced by 5-brane domain walls wrapping internal cycles; the flux jumps across such domain walls. Now, if we are in a flux vacuum with positive cosmological constant, quantum fluctuations can cause nucleation of flux-changing domain wall bubbles \cite{Brown:1987dd,Brown:1988kg,Bousso:2000xa,Feng:2000if}, by a tunneling mechanism similar to Coleman-de Luccia bubble nucleation in scalar potential landscapes. If the cosmological constant inside the bubble is positive again, it will itself inflate and eventually nucleate new bubbles, and so on (see fig.\ \ref{bubbles} for a psychedelic impression of this). This is a version of eternal inflation (see e.g.\ \cite{Guth:2000ka} for a brief review, and Steve Shenker's lectures at this school).

Even if we are not interested in computing probabilities, such bubble nucleation processes are still of crucial importance to determine to what extent particular vacua are metastable.

According to \cite{Brown:1987dd,Brown:1988kg},\footnote{Their analysis does not take into account moduli dynamics and is intrinsically done in a thin-wall approximation as the membranes are taken to be infinitely thin. The resulting formulas should therefore be taken to be estimates rather than exact results for F-theory flux vacua.} the nucleation rate per unit 4d spacetime volume for a bubble with tension $T$, cosmological constant $\Lambda_o$
outside and cosmological constant $\Lambda_i$ inside is given by
\begin{equation} \label{BTrate}
 \Gamma \sim e^{-12 \pi^2 B}
\end{equation}
where (in units with $m_p \equiv 1$, which we will use in the
remainder of this section):
\begin{equation} \label{BTBformula}
 B = \frac{T \rho^3}{6}
 - \frac{1 - \sigma_i ( 1-\frac{\Lambda_i \rho^2}{3} )^{3/2}}{\Lambda_i}
 + \frac{1 - \sigma_o ( 1-\frac{\Lambda_o \rho^2}{3})^{3/2}}{\Lambda_o}.
\end{equation}
Here $\sigma_{i,o} = \mbox{sign} \left[ \pm 3\, T^2 + 4(\Lambda_o -
\Lambda_i) \right]$ and $\rho$ is the bubble radius, which must be
evaluated at the stationary\footnote{This is a minimum iff $3 T^2 -
4|\Lambda_i-\Lambda_o| > 0$.} point of $B(\rho)$:
\begin{equation} \label{rhoval}
 \rho = \frac{12 \, T}{[ 9 \, T^4 + 24 \, T^2 (\Lambda_i+\Lambda_o) +
 16 \, (\Lambda_i-\Lambda_o)^2 ]^{1/2}}.
\end{equation}
If $\Lambda_o>0$, there is always a nonzero nucleation rate. If
$\Lambda_o<0$ and the initial space is AdS, one needs in addition
$\Lambda_i < \Lambda_o - 3T^2/4$ and the argument of the square root
of (\ref{rhoval}) to be positive (which is automatic if
$\Lambda_o>0$).

Consistency of the semiclassical approximation requires $\rho \gg 1$ and therefore exponentially small decay rates, as usual with instantons. In particular any well-controlled domain wall bubble will almost tautologically give rise to an extremely small decay rate. Note that $\rho$ is infinite and the decay rate zero when $T = \frac{2}{\sqrt{3}} |\sqrt{-\Lambda_o} \pm \sqrt{-\Lambda_i}|$; this is the case when the domain wall is BPS saturated, interpolating between two superymmetric vacua.

When $T^2 \ll (\Delta\Lambda)^2/\bar{\Lambda}$ with $\bar{\Lambda} := \Lambda_i + \Lambda_o$ and
 $ \Delta \Lambda := \Lambda_o - \Lambda_i > 0$, this becomes
\begin{equation} \label{Gammalim1}
 \Gamma \sim \exp \biggl( -\frac{27 \pi^2}{2} \frac{T^4}{(\Delta \Lambda)^3} \biggr)
\end{equation}
and when $T^2 \gg \Lambda_o, \Lambda_i$:
\begin{equation} \label{Gammalim2}
 \Gamma \sim \exp \biggl( -\frac{24 \pi^2}{\Lambda_o} + \frac{64 \pi^2}{T^2} \biggr) \, .
\end{equation}
Notice that after restoring powers of $m_p$, (\ref{Gammalim1}) does not involve Newton's constant --- the result is indeed identical to the rate of bubble nucleation in the absence of gravity, in the thin wall approximation. The second expression does depend on Newton's constant. This rate is extremely suppressed for vacua with $\Lambda_o \ll m_p^4$ such as our own (although it is always larger than the Poincar\'e recurrence rate $e^{-24 \pi^2/\Lambda_0}$). Thus for the stability of a vacua, the most dangerous domain wall bubbles are those with small tension but sizable change of cosmological constant.

Decay rates such as those give here can be taken as starting point to try to find sensible probability measures on the landscape, as explained by Steve Shenker at this school.

\section{Explicit realizations of moduli stabilization scenarios} \label{sec:explconstr}

In this section we will finally get to building explicit models of moduli stabilized F-theory flux vacua, drawing on all of the techniques developed in earlier chapters. From the section on statistics, we already take that typically, there will be a fine discretuum of vacua which we can use to tune various physical parameters to our liking, and in particular generate large scale hierarchies through warping. This allows us to consider controlled regimes.

The main remaining challenge is to make sure all K\"ahler moduli are stabilized. In both the KKLT and the large volume scenarios, this hinges on the existence of suitable nonperturbative corrections to the superpotential.
The first concrete models satisfying the necessary geometrical requirements for this (in the KKLT scenario) were proposed in \cite{Denef:2004dm}, and a simpler and more explicit model was given in \cite{Denef:2005mm} and subsequently generalized in \cite{Lust:2005dy,Lust:2006zg}. Various powerful mathematical criteria for M5 instantons to have the right zeromode structure to contribute to the superpotential were systematically developed in \cite{grassi}. We will however stick to the more elementary methods we have developed in these lectures. We will in these lectures also not show explicitly the existence of a K\"ahler stabilized minimum of the effective potential, but only show that models exist where we do get the necessary structure of the K\"ahler potential and the right kind of contributions to the superpotential to make in particular the large volume scenario of section \ref{sec:cheese} work. But once these conditions are met, the existence of large volume minima of the effective potential in this scenario is guaranteed by the general analysis of \cite{Balasubramanian:2005zx,Conlon:2005ki,Conlon:2006gv,Blumenhagen:2007sm}. (However as noted in section \ref{sec:cheese} one should still check if the ``small'' K\"ahler moduli $T_{S_i}$ in string units can be made sufficiently large to trust the geometrical picture; we will not do this here.)

\subsection{The elliptic fibration over $\ICP^3$}

Let us first see if we can turn our basic example (\ref{CY4eq}) in an explicit realization of one of the moduli stabilization scenarios outlined in sections \ref{sec:KKLT} and \ref{sec:cheese}. The large volume scenario needs at least two K\"ahler moduli, so this is excluded, leaving only the KKLT scenario. To make this scenario work we need some nonperturbative contributions to the superpotential, and the existence of classical flux vacua with exponentially small $e^{\CK} |W|^2$. As explained in section \ref{sec:statistics}, according to the distribution estimates, there is certainly no shortage of the latter. Generating nonperturbative corrections to $W$ is more subtle. As noted in section \ref{sec:geomcondnonv}, in the M-theory picture, all nonperturbative effects can be thought of as being generated by holomorphic M5 instantons wrapping the elliptic fiber and a divisor in the base. A necessary condition for this instanton to contribute in the absence of fluxes is the arithmetic genus $\chi_0=1$ condition (\ref{holeulcond}). In the presence of fluxes this gets replaced by the weaker condition (\ref{fluxnecccond}): $\chi_0 \geq 1 (=\chi_{0,\rm eff})$.

In the notation of section \ref{sec:CYsubtor} where we studied this case as an example, the most general holomorphic divisor $D$ in $Z$ wrapping the elliptic fiber is given by some degree $k$ polynomial equation $P_k(\vec u)=0$ on the base, so $D = k \tK_1$, $k \in \IZ^+$. To compute $\chi_0(D)$, we use the index formula (\ref{chi0indformula}): $\chi_0 = \frac{1}{24} \int_D c_1 c_2$. Here the Chern classes $c_1$ and $c_2$ are those of the tangent bundle $TD$ of $D$, which can be computed from the Chern classes of the ambient CY fourfold using the adjunction formula (\ref{adjformula}): $c(TD)=c(TZ)/c(ND) = c(TZ)/(1+D)$. But we know $c(TZ)$ already; it is given by (\ref{cfourfoldex}). Expanding out the adjunction formula quotient, we find
\begin{equation}
 c_1(TD) = -k \tK_1 \, , \qquad c_2(TD) = \left( (k^2-10) \tK_1 + 48 \, \tK_2 \right) \tK_1 \, ,
\end{equation}
and from this, using the formula for $\chi_0$ just quoted and the intersection numbers (\ref{tildeintprods}):
\begin{equation}
 \chi_0(D) = \frac{1}{24} k \tD_1 \, c_1(TD) \, c_2(TD) = -2 k^2 \, .
\end{equation}
In particular this is always \emph{negative}, so even the weak condition $\chi_0 \geq 1$ is not satisfied.

We conclude that neither the large volume, nor the KKLT scenario for this model works.\footnote{Actually, if we tune the complex structure moduli to a locus of enhanced gauge symmetry as discussed in section \ref{sec:enhancedGS}, so $Z$ becomes singular, there could still be nonperturbative contributions associated to M5 instantons wrapping the divisors obtained by blowing up the singularity (i.e.\ going to the Coulomb branch), as explained in section \ref{sec:gauginoM5}. The blown up fourfold will have different Hodge numbers than the original $Z$, and as a result different flux lattice dimensions and D3 tadpole. Whether we still consider this to be the same model is a matter of semantics. We will consider it to be a different model here.}

\subsection{The elliptic fibration over $\CM_n$}

We consider now the CY elliptic fibration with as base manifold $B$ our example (\ref{example2}), which we denoted by $\CM_n$, the $\ICP^1$ bundle over $\ICP^2$ with twist $n \geq 0$.\footnote{The cases $n<0$ are isomorphic to $n>0$ by exchanging $u_4$ and $u_5$.} This was defined by five fields, which we will now call $u_i$, and
$U(1) \times U(1)$ gauge group, with charges

\begin{tabular}{ccccc}
 $u_1$&$u_2$&$u_3$&$u_4$&$u_5$ \\ \hline
 1&1&1&$-n$&$0$ \\
 0&0&0&1&1
\end{tabular}

\noindent and positive FI parameters $(\xi^1,\xi^2)$. Recall from (\ref{swissvol}) that the volume of $\CM_n$, $n>0$ is indeed of Swiss cheese type.

We consider again a CY elliptic fibration of the form
\begin{equation}
 Z:y^2 = x^3 + f(\vec u) \, x z^4 + g(\vec u) \, z^6 = 0
\end{equation}
over $\CM_n$, and the Calabi-Yau condition $\sum_i D_i = [Z]$ fixes the charges of the fields and polynomials to be

\begin{tabular}{cccccccccc}
 $u_1$&$u_2$&$u_3$&$u_4$&$u_5$&$x$&$y$&$z$&$f$&$g$\\ \hline
 1&1&1&$-n$&0&0&0&$n-3$&$4(3-n)$&$6(3-n)$\\
 0&0&0&1&1&0&0&$-2$&8&12 \\
 0&0&0&0&0&2&3&1&0&0
\end{tabular}

\vskip3mm
\noindent The corresponding D-term constraints are, explicitly:
\begin{eqnarray}
 |u_1|^2+|u_2|^2+|u_3|^2-n \, |u_4|^2 + (n-3) \, |z|^2  &=& \xi^1 \\
 |u_4|^2 + |u_5|^2 - 2 \, |z|^2 &=& \xi^2 \\
 2 \, |x|^2 + 3 \, |y|^2 + |z|^2 &=& \xi^3 \, .
\end{eqnarray}
In accord with the F-theory limit of vanishing elliptic fiber, we take the third FI parameter $\xi^3$ much smaller than $\xi^1,\xi^2$.

It may seem like we have constructed an infinite number of Calabi-Yau fourfolds, labeled by $n$. This is not true. We should keep in mind that we have made the implicit assumption (by using the formula $c_1 = \sum_i D_i - [Z]$) that $Z$ is smooth. If this is not the case, we should in principle first resolve the singularities before applying this formula, or use a modification of the formula appropriate for singular spaces. Now, from the $U(1)^3$ charges of the polynomials $f$ and $g$ given above, we see that if $n>3$, $f$ and $g$ become negatively charged under the first $U(1)$ and so must necessarily contain an overall factor equal to a power of $u_4$. More precisely $f(u)=u_4^k \tilde{f}(u)$, $g(u)=u_4^l \tilde{g}(u)$ where $k$ is the smallest integer  $ \geq 4(1-\frac{3}{n})$ and $l$ the smallest integer $\geq 6(1-\frac{3}{n})$. So in this case $f$, $g$ and the discriminant $\Delta = 27 \, g^2 + 4 \, f^3$ vanish as some power of $u_1$ on the divisor $D_4:u_1=0$, and hence the fourfold is singular along the locus $\Delta=0$. For $n$ not too large, the singularities are harmless in the sense that they can be resolved while preserving the $c_1=0$ condition, and moreover they have a clean physical interpretation as loci of enhanced gauge symmetry, as mentioned in section \ref{sec:enhancedGS}. For example for $n=4$, we generically have $f \sim u_4$, $g \sim u_4^2$ and $\Delta \sim u_4^3$, so from the table in section \ref{sec:enhancedGS} we read off that we get an $SU(2)$ gauge group enhancement. For $n=18$, we have $f \sim u_4^4$, $g \sum u_4^5$, $\Delta \sim u_4^{10}$ and we get an $E_8$ gauge group enhancement. For $n>18$, we fall off the table; at this point the singularity becomes so bad that it cannot be resolved preserving the CY condition. This puts a cutoff on $n$.

At any rate, we will focus on the cases without gauge symmetry enhancement, i.e.\ $n \leq 3$, for which the analysis is most straightforward.

From the charge assignments above, we read off the following relations between the divisors:
\begin{eqnarray}
 && D_1=D_2=D_3 \, , \qquad D_5 - D_4 = n \, D_1 \, , \\
 && [Z] = 3 D_x = 2 D_y = 6 D_z + (3+n) D_1 + 2 D_4 \, ,
\end{eqnarray}
where in the last line $[Z]$ is the homology class of our Calabi-Yau $Z$.

An independent set of divisors is given for instance by $D_4$, $D_5$, $D_z$. Their pullbacks to $Z$ are denoted by $\tD_4$, $\tD_5$, $\tD_z$. The first two are divisors wrapped on the elliptic fiber and a divisor in the base. The third one is a section of the elliptic fibration, i.e.\ the base itself. Using the techniques of section \ref{geometrictools}, we find the following nonzero intersection numbers between these divisors:
\begin{eqnarray}
 &&\tD_4^3 \tD_z = n^2 \, , \quad \tD_4^2 \tD_z^2 = (3-n)n \, , \quad \tD_4 \tD_z^3 = (3-n)^2 \, , \\
 &&\tD_5^3 \tD_z = n^2 \, , \quad \tD_5^2 \tD_z^2 = -(3+n)n \, , \quad \tD_5 \tD_z^3 = (3+n)^2 \, , \\
 &&\tD_z^4 = -2(n^2+24) \, .
\end{eqnarray}
This data allows us to compute volumes, characteristic classes, indices and so on. (A basis for the K\"ahler cone is given by $\tK_1 = \tD_1$, $\tK_2=\tD_5$, $\tK_3=[Z]_Z$, but we will continue to work in the above divisor basis in what follows.)

Again we need some nonperturbative contributions to $W$, associated to holomorphic M5 instantons wrapping the elliptic fiber and a divisor in $\CM_n$. The most general such divisor $D$ is given by some polynomial equation $P(\vec u) = 0$, so
\begin{equation}
 D= a D_4 + b D_5 = (a+b) D_4 + b n D_1 \,
\end{equation}
where $a+b \in \IZ^+$, $b n \in \IZ^+$.

As in the previous example, we can compute the holomorphic Euler characteristic $\chi_0$ and find (assisted by Mathematica to do the series expansions of characteristic classes and to substitute the intersection numbers):
\begin{equation}
\chi_0(D) = -\frac{1}{2} n \left( (n-3)a^2 + (n+3)b^2 \right) \, .
\end{equation}
When $n=0$ or $n \geq 3$, this is nonpositive, and therefore even the weak necessary condition $\chi_0 \geq 1$ is not satisfied.\footnote{However as we just saw when $n>3$ we need to consider more divisors, namely those obtained from resolving the enhanced gauge singularities, but we will stick to the smooth cases here.} On the other hand the diophantine equation $\chi_0(D)=1$  has infinitely many solutions for $n=1,2$. For definiteness let us specialize to
\begin{equation}
 n \equiv 1 \,
\end{equation}
from now on. Then to find divisors of arithmetic genus one, we have to solve $a^2-2b^2=1$ for $a+b,b$ nonnegative integers. This is explicitly solved as
\begin{equation} \label{absolutions}
 a = \frac{(3+2\sqrt{2})^k+(3-2\sqrt{2})^k}{2} \, , \qquad b=\frac{(3+2\sqrt{2})^k-(3-2\sqrt{2})^k}{2 \sqrt{2}} \, ,
\end{equation}
$k \geq 0$. The first few solutions are $(a,b)=\{ (1,0), (3,2), (17,12), (99,70), \cdots \}$.

In particular for $(a,b)=(1,0)$, i.e.\ $D=D_4:u_4=0$, the instanton is completely rigid and has exactly two zeromodes, i.e.\ $h^{1,0}=h^{2,0}=h^{3,0}=0$. This can be seen as follows. First, it is clear from the charge assignments of the fields that $u_4=0$ is the unique holomorphic representative in its homology class --- there are no other polynomials with the same charges as $u_1$. From the one to one correspondence between holomorphic deformations and elements of $H^{3,0}(D)$, this implies $h^{3,0}=0$.  Furthermore, from the Lefschetz hyperplane theorem, it follows that $b_1(D)=b_1(Z)=0$, and therefore $h^{1,0}=0$. Hence $1=\chi_0=1+h^{2,0}$, and therefore also $h^{2,0}=0$.

Thus, flux or no flux, $D$ will always contribute to the superpotential, and given (\ref{swissvol}) this is moreover exactly the kind of contribution we need for the large volume scenario to work! Note also that unlike in the KKLT scenario, we only need one instanton correction to stabilize all K\"ahler moduli.\footnote{Given the infinite set of solutions (\ref{absolutions}) it is possible that we have more instanton contributions, but these will be exponentially suppressed compared to $(a,b)=(1,0)$.}

For completeness we give some further topological data for this model, obtained in a way similar to what we did for the example of the elliptic fibration over $\ICP^3$. The Hodge data of $Z|_{n=1}$ is
\begin{equation}
 h^{1,1}=3 \, , h^{2,1}=0 \, , h^{3,1} = 3397 \, , h^{2,2} = 13644 \, .
\end{equation}
This implies in particular $b_4=20440$ and a curvature induced D3 tadpole
\begin{equation}
 Q_c = \frac{\chi(Z)}{24} = 852 \, .
\end{equation}
According to the estimate (\ref{improvedest}), this yields a discretuum of about $10^{3000}$ flux vacua.

Following section \ref{sec:orlimitgen}, we find that the IIB weak coupling limit is a $v \to -v$ orientifold of the Calabi-Yau hypersurface:
\begin{equation}
 X: v^2 = h(\vec u) \, ,
\end{equation}
in a toric variety with fields $(u_1,\ldots,u_5,v)$ and the following charge assignments:

\begin{tabular}{ccccccc}
 $u_1$&$u_2$&$u_3$&$u_4$&$u_5$&$v$&$h$ \\ \hline
 1&1&1&$-1$&$0$&2&4 \\
 0&0&0&1&1&2&4
\end{tabular}

Computing the third Chern class in the usual way, we find $\chi(X)=-260$, so (using $h^{1,1}=2$) $h^{2,1}=132$. This also determines the number $\xi$ defined in (\ref{Kahlercorr}): $\xi \approx 0.315$.

Thus we conclude that in this model, the large volume scenario can indeed be realized.

\vskip15mm \noindent {\bf Acknowledgements} \vskip3mm

I would like to thank the organizers of this excellent school, Costas Bachas, Laurent Baulieu,
Michael Douglas, Elias Kiritsis, Eliezer Rabinovici, Pierre Vanhove and Paul Windey, for providing me
the opportunity to teach on this topic. I am very much indebted to my collaborators, Andres Collinucci, Michael Douglas, Mboyo Esole, Bogdan Florea, Antonella Grassi, Shamit Kachru and Greg Moore, whose insights and work are reflected throughout these lectures. Thanks also to the schools' students and fellow lecturers for questions and discussions and providing such a pleasant environment, and to the company responsible for brewing Desperados. Finally special thanks to Alessandro Tomasiello for discussions related to the contents of these lectures, and to Andres Collinucci, Jonathan Ruel and Erik Plauschinn for helpful comments on the manuscript.

\end{document}